\documentclass[a4paper,preprint,superscriptaddress,preprintnumbers,nofootinbib]{revtex4}

\usepackage[dvipdf,dvips,dvipdfmx]{graphicx}
\usepackage{amsmath}
\usepackage{amssymb}
\usepackage{float}
\usepackage{wrapfig}
\usepackage{bm}
\usepackage{color}

\def\lsim{\mathrel{\mathpalette\@versim<}}
\def\gsim{\mathrel{\mathpalette\@versim>}}

\DeclareMathOperator{\tr}{tr}
\DeclareMathOperator{\Tr}{Tr}
\DeclareMathOperator{\STr}{STr}

\newcommand{\T}{\text{T}}
\newcommand{\wt}{\widetilde}

\newcommand{\BB}{\text{BB}}
\newcommand{\BF}{\text{BF}}
\newcommand{\FB}{\text{FB}}
\newcommand{\FF}{\text{FF}}

\newcommand{\mc}{\mathcal}

\newcommand{\mO}{\mathcal{O}}

\newcommand{\al}[1]{\begin{align}#1\end{align}}

\newcommand{\bp}{\begin{pmatrix}}
\newcommand{\ep}{\end{pmatrix}}
\newcommand{\nn}{\nonumber\\}
\newcommand{\paren}[1]{\left(#1\right)}
\newcommand{\sqbr}[1]{\left[#1\right]}
\newcommand{\br}[1]{\left\{#1\right\}}

\newcommand{\p}{\partial}

\newcommand{\df}{\text{d}}

\newcommand{\bs}[1]{\boldsymbol}

\newcommand{\pmat}[1]{\begin{pmatrix}#1\end{pmatrix}}
\newcommand{\bmat}[1]{\begin{bmatrix}#1\end{bmatrix}}
\newcommand{\fn}[1]{\!\left(#1\right)}

\newcommand{\Slash}[1]{{\ooalign{\hfil/\hfil\crcr$#1$}}}

\newcommand{\ft}{\tilde} 

\newcommand{\bara}{\overleftarrow}

\usepackage{etoolbox}
\let\bbordermatrix\bordermatrix
\patchcmd{\bbordermatrix}{8.75}{4.75}{}{}
\patchcmd{\bbordermatrix}{\left(}{\left[}{}{}
\patchcmd{\bbordermatrix}{\right)}{\right]}{}{}

\begin{document}

\title{Asymptotic safety of higher derivative quantum gravity non-minimally coupled with a matter system}

\author{Yuta \surname{Hamada}}
\email{yhamada@wisc.edu}
\affiliation{Department of Physics, University of Wisconsin, Madison, WI 53706, USA}
\affiliation{KEK Theory Center, IPNS, KEK, Tsukuba, Ibaraki 305-0801, Japan}

\author{Masatoshi \surname{Yamada}}
\email{m.yamada@thphys.uni-heidelberg.de}
\affiliation{Institut f\"ur Theoretische Physik, Universit\"at Heidelberg, Philosophenweg 16, 69120 Heidelberg, Germany}
\preprint{
}

\begin{abstract}
We study asymptotic safety of models of the higher derivative quantum gravity with and without matter.
The beta functions are derived by utilizing the functional renormalization group, and non-trivial fixed points are found.
It turns out that all couplings in gravity sector, namely the cosmological constant, the Newton constant, and the $R^2$ and $R_{\mu\nu}^2$ coupling constants, are relevant in case of higher derivative pure gravity.
For the Higgs-Yukawa model non-minimal coupled with higher derivative gravity, we find a stable fixed point at which the scalar-quartic and the Yukawa coupling constants become relevant.
The relevant Yukawa coupling is crucial to realize the finite value of the Yukawa coupling constants in the standard model.
\end{abstract}
\maketitle

\section{Introduction}
One of the important problems in elementary particle physics is the construction of quantum gravity.\footnote{The necessity of the quantum gravity is indirectly supported by an experiment~\cite{Page:1981aj}.}
In perturbation theory at one-loop level the quantized Einstein-Hilbert action can be renormalizable only without cosmological constant~\cite{Hooft:1974bx}.
However, the perturbation theory for the systems coupled to matter does not work at one-loop level~\cite{Hooft:1974bx,Deser:1974cz,Deser:1974cy,Deser:1974zzd,Deser:1974xq}.
At two-loop level the pure gravity system becomes perturbatively non-renormalizable~\cite{Goroff:1985sz,vandeVen:1991gw}.
Although the inclusion of the higher derivative terms such as $R^2$ and $R_{\mu\nu}R^{\mu\nu}$ helps the theory to be perturbatively renormalizable~\cite{Stelle:1976gc}, the ghost problem arises, that is, the norm of some states becomes negative~\cite{Stelle:1977ry}.\footnote{
See also \cite{Narain:2011gs,Narain:2012nf}.
}
These facts may indicate that approaches beyond perturbation theory are needed.
 
Asymptotic safety is a general feature of ultraviolet (UV) completeness in quantum field theory. 
The first evidence that quantum gravity may be asymptotically safe was reported in \cite{Hawking:1979ig}.
It is crucial for the scenario of asymptotic safety that a theory has a non-trivial UV fixed point at which the beta functions of the theory vanish.
If there exists a UV fixed point, the continuum limit $k \to \infty$ can be taken (UV complete). 
Further, if the number of relevant couplings is finite, the theory can be renormalizable, that is, the low energy physics is predicted.
Since perturbation theory is valid only around  the vicinity of the trivial (Gaussian) fixed point, generally a non-perturbative methods are required to see asymptotic safety.\footnote{
See e.g.,~\cite{Percacci:2007sz,Percacci:2011fr,Nagy:2012ef} for details.
}

Although the $\epsilon$ expansion method in $2+\epsilon$ dimension has been applied in order to find the non-trivial fixed point~\cite{Hawking:1979ig,Kawai:1989yh}, this method fails for $\epsilon>1$.
A powerful method to investigate an asymptotically safe theory is the functional renormalization group (FRG) which originated from Kadanoff's and Wilson's renormalization group~\cite{Kadanoff:1966wm,Wilson:1973jj}.\footnote{
There are review papers~\cite{Bagnuls:2000ae,Berges:2000ew,Aoki:2000wm,Polonyi:2001se,Pawlowski:2005xe,Gies:2006wv,Rosten:2010vm} on the FRG.
}
Since the FRG method does not depend on any asymptotic expansion by in the spacetime dimension and coupling constant, we can analyze strongly coupled systems in arbitrary dimension.
In particular, the approximation in the FRG is systematically improved by including the higher order operators.
This feature is adequate for studying the stability of the fixed point structure of the system.

Analyses using FRG with the background field method have been performed for various truncations of the quantum gravitational systems: for Einstein-Hilbert truncation~\cite{Reuter:1996cp,Dou:1997fg,Souma:1999at,Reuter:2001ag,Lauscher:2001rz,Lauscher:2001cq,Lauscher:2001ya,Litim:2003vp,Percacci:2004sb,Bonanno:2004sy,Fischer:2006fz,Manrique:2011jc,Harst:2012ni,Falls:2014tra,Falls:2015qga,Gies:2015tca,Biemans:2016rvp,Falls:2017cze};
for gravity with matters and gauge fields~\cite{Percacci:2002ie,Percacci:2003jz,Percacci:2005wu,Narain:2009fy,Narain:2009gb,Zanusso:2009bs,Daum:2009dn,Daum:2010bc,Vacca:2010mj,Harst:2011zx,Eichhorn:2011pc,Folkerts:2011jz,Eichhorn:2012va,Dona:2012am,Dona:2013qba,Henz:2013oxa,Dona:2014pla,Percacci:2015wwa,Labus:2015ska,Oda:2015sma,Dona:2015tnf,Eichhorn:2016esv,Henz:2016aoh,Eichhorn:2016vvy,Biemans:2017zca,Christiansen:2017gtg}; for higher derivative gravity with $f\fn{R}$-type truncation~\cite{Lauscher:2002sq,Codello:2007bd,Machado:2007ea,Hindmarsh:2012rc,Dietz:2012ic,Dietz:2013sba,Falls:2013bv,Falls:2014zba,Eichhorn:2015bna,Ohta:2015efa,Ohta:2015fcu,Schroder:2015xva,Falls:2016wsa,Falls:2016msz}; $R_{\mu\nu}R^{\mu\nu}$~\cite{Granda:1998wn,Codello:2006in,Benedetti:2009rx,Benedetti:2009gn,Benedetti:2009iq,Ohta:2012vb,Ohta:2013uca,Groh:2011vn}; and the Goroff--Sagnotti term~\cite{Gies:2016con};
see also \cite{Niedermaier:2006wt,Niedermaier:2006ns,Reuter:2007rv,Codello:2008vh,Litim:2008tt,Litim:2011cp,Reuter:2012id} for review papers.\footnote{
The background, gauge and cutoff scheme dependences have been investigated in~\cite{Falkenberg:1996bq,Souma:2000vs,Morris:2016spn,Percacci:2016arh,Ohta:2017dsq}; see also~\cite{Wetterich:2016ewc}
The ghost interactions have been discussed in~\cite{Eichhorn:2010tb,Eichhorn:2009ah,Groh:2010ta}.
Asymptotically safe gravity in viewpoint of the fractal space-time structure has been studied in \cite{Lauscher:2005qz,Reuter:2011ah,Rechenberger:2012pm,Reuter:2012xf}.
Analyses with lattice simulation have been performed in~\cite{Egawa:1996ff,Egawa:1996fu,Horata:2002uf,Horata:2003hm,Egawa:2003gk,Ambjorn:2004qm,Ambjorn:2005db,Ambjorn:2009ts,Laiho:2011ya,Ambjorn:2012jv,Laiho:2016nlp}.
}
Also studies based on the vertex expansion have been performed in~\cite{Donkin:2012ud,Christiansen:2012rx,Christiansen:2014raa,Christiansen:2015rva,Meibohm:2015twa,Meibohm:2016mkp,Christiansen:2016sjn,Denz:2016qks}. 
Further, a scenario was suggested in \cite{Aad:2012tfa,Chatrchyan:2012xdj} which provides a prediction of a Higgs mass with 129\,GeV~\cite{Shaposhnikov:2009pv,Bezrukov:2012sa}.
The asymptotically safe gravity scenario could have the possibility to solve problems in elementary particle physics such as the gauge hierarchy problem~\cite{Wetterich:2016uxm} and the $\text{U}\fn{1}$ triviality problem~\cite{Christiansen:2017gtg}.
These studies have encouraged the scenario of asymptotically safe gravity as a strong candidate for quantum gravity.

In this paper, we study higher derivative gravity coupled without and with matter fields using the FRG.\footnote{
See related studies~\cite{Buchbinder:1992rb,Elizalde:1994gv,Elizalde:1995at,Narain:2016sgk}.
}
In refs.~\cite{Codello:2006in,Benedetti:2009rx,Benedetti:2009gn}, higher derivative gravity has been studied.
We reanalyze higher derivative gravity with different ghost and gauge fixing actions from \cite{Codello:2006in,Benedetti:2009rx,Benedetti:2009gn} and investigate the gauge dependence of the fixed points and the critical exponents.
In the higher derivative gravity coupled to matter fields, the Higgs-Yukawa model is employed for the matter sector as the minimal toy model of the standard model.
Ref.~\cite{Oda:2015sma} studied the Higgs-Yukawa model coupled to gravity without the higher derivative terms.
It has been shown that this model can become asymptotically safe, and especially fermionic fluctuations make the scalar mass and the non-minimal coupling between the scalar field and the Ricci scalar $\phi^2 R$ irrelevant around the UV fixed point.\footnote{
Ref.~\cite{Oda:2015sma} studied the truncated effective action spanned by six couplings, namely, the Newton coupling, the cosmological constant, the scalar mass-squared, the quartic scalar coupling and the Yukawa coupling.
Among them, the Newton constant and the cosmological constant becomes relevant. The others are irrelevant.
The fact that the quartic scalar coupling and the Yukawa coupling within asymptotically safe gravity with matter become irrelevant was reported in \cite{Percacci:2003jz} and \cite{Zanusso:2009bs}, respectively.
}

Besides the asymptotic safety scenario, the Higgs-Yukawa model non-minimally coupled to gravity has been studied as a toy model of Higgs inflation.\footnote{
See also \cite{Myrzakulov:2016tsz}
}
Recent result from the Planck satellite~\cite{Ade:2015xua} put strong constraints on inflationary parameters, that is, the tensor to scalar ratio and the spectral index of scalar perturbation.
Among a lot of inflation models, the predictions of Higgs inflation~\cite{Bezrukov:2007ep,Hamada:2013mya,Hamada:2014iga,Bezrukov:2014bra,Hamada:2014wna,Hamada:2014raa} and the Starobinsky~\cite{Starobinsky:1980te} model are close to the best fit values.
In addition to the support from observation, these models are attractive because they do not introduce extra degrees of freedom except for standard model particles and gravity.
Usually, these models require large coupling in the gravity sector~\cite{Hamada:2014iga}, whose validity should be discussed in the context of a UV completed theory such as asymptotic safety.\footnote{
Models for inflation with the feature of asymptotic safety have been discussed in~\cite{Weinberg:2009wa,Bonanno:2010bt,Tye:2010an,Bonanno:2010mk,Hindmarsh:2011hx,Cai:2011kd,Bonanno:2012jy,Hong:2011ws,Fang:2012ca,Copeland:2013vva,Xianyu:2014eba,Nielsen:2015una,Bonanno:2015fga,Saltas:2015vsc}.
}

This paper is structured as follows:
The effective action and the set-up to derive the beta function is given in section~\ref{models}.
The RG equations for the effective action and their numerical analysis are shown in Sect.~\ref{flow equations section} and \ref{numericalan}, respectively.
Section~\ref{Summary and Discussion} is devoted to summary and discussion.
In the appendix~\ref{frgtreatment}, we explain the basic concepts of the FRG and the fixed point structure.
We list several formulas of variations to compute beta functions in the appendix~\ref{variations of action}.
In the appendix~\ref{hkeapp}, the formula of the heat kernel expansion, which is used to evaluate the functional trace, is shown.
We show derivations of beta functions in the appendix~\ref{explicit derivation of beta functions}.
The fixed point values and the critical exponents obtained in this analysis are listed in the appendix~\ref{FPCX}.

\section{Effective action}\label{models}
In \cite{Oda:2015sma}, the Higgs-Yukawa model with Ricci scalar $R$ and non-minimal coupling between the scalar field and the Ricci scalar has been analyzed.
It has been shown that the fermionic fluctuation turns the mass of the scalar field and the non-minimal coupling irrelevant in that truncation.
Note that a comparison of \cite{Dona:2014pla} with \cite{Meibohm:2015twa} highlights major differences between the effect of fermion fluctuations in a single-metric and a bimetric treatment.
The results in \cite{Oda:2015sma} are obtained within a single-metric approximation and could therefore change significantly if the difference between the background metric and the full metric is resolved. 
In refs.~\cite{Benedetti:2009rx,Falls:2013bv} pure higher derivative gravity has been studied and it has been shown that the $R^2$ term becomes relevant.

In this paper, we investigate the effects of the higher derivative gravity on the Higgs-Yukawa system.
Here, we briefly summarize the structure of this section.
In the next subsection, we introduce the effective actions for the the Higgs-Yukawa model non-minimally coupled to higher derivative gravity.
We employ the ghost and gauge fixing actions with a higher derivative operator~\cite{Ohta:2016jvw} in order to simplify the kinetic terms of graviton; see Eq.~\eqref{gaugefixedaction}.
In the subsection~\ref{YD}, the York decomposition~\cite{York:1973ia} is briefly explained.
We list explicit forms of the two-point functions using the Lichnerowicz Laplacians in the subsection~\ref{modelandmanipulations}.
Note that since the ghost and gauge fixing actions with a higher derivative operator are used, the structures of the two-point function for graviton differs from ones given in e.g.~\cite{Narain:2009fy,Oda:2015sma} but instead are same as \cite{Ohta:2016jvw}.
The two-point functions for fermion are same as \cite{Zanusso:2009bs,Oda:2015sma}.

\subsection{Model}
The effective action of higher derivative gravity with matter interactions in four dimensional Euclidean spacetime dimension is given by
\al{
\Gamma_k&= \int \df^4x \sqrt{g}\Bigg\{
V\fn{\Phi^2}-F\fn{\Phi^2}R +a R^2 +b R_{\mu\nu}R^{\mu\nu}
		+\frac{1}{2} g^{\mu \nu}\,\p _\mu{ \Phi}\,\p _{\nu} \Phi \nn
&\quad
		+\sum_{i=1}^{N_f}{\bar \Psi_i}{\Slash D}\Psi_i	
		+y \sum_{i=1}^{N_f} \Phi {\bar \Psi}_i\Psi_i
		\Bigg\}
				+S_{\rm gf}	+S_{\rm gh},
		\label{originaleffectiveaction}
}
where $\Phi$ and $\Psi$ are the scalar and fermion fields, respectively, $S_\text{gf}$ and $S_\text{gh}$ are the gauge fixing and ghost terms, respectively, the covariant derivative $\Slash D$ in the kinetic term of the fermions is ${\Slash D} \Psi={\Slash\p}\Psi +\gamma^\mu\hat\Gamma_\mu\Psi$ where $\hat\Gamma_\mu$ is the spin connection.
In this paper, we employ the local potential approximation, that is, the corrections to the field renormalization factors in the kinetic terms are neglected, which means $\eta=0$ in \eqref{explicit beta general}.
We assume that the effective action is invariant under the $Z_2$ transformation ($\Phi\to -\Phi$, $\Psi\to \gamma^5\Psi$ and $\bar \Psi\to -\bar \Psi\gamma^5$) and CP transformation which prohibits $\Phi \bar \Psi i\gamma^5 \Psi$.
Moreover, we assume the $SU\fn{N_f}$ flavor symmetry.

\eqref{originaleffectiveaction} can be rewritten as
\al{
\Gamma_k
	=	\int\df^4x \sqrt{g}&\Bigg\{ V\fn{\Phi ^2} 
	-F \fn{\Phi ^2} R
	+ \left(a + \frac{1}{4}b \right)R^2
	+ \frac{b}{4} R_{\mu\nu\rho\sigma}R^{\mu\nu\rho\sigma}
	-\frac{b}{4} E \nn
&\quad		+\frac{1}{2} g^{\mu \nu}\,\p _\mu{ \Phi}\,\p _{\nu} \Phi 
		+\sum_{i=1}^{N_f}{\bar \Psi_i}{\Slash D}\Psi_i
		+y  \sum_{i=1}^{N_f}\Phi {\bar \Psi}_i\Psi_i
		\Bigg\}
		+S_{\rm gf}	+S_{\rm gh},
		\label{effectiveaction}
}
where $E= R^2 -4R_{\mu\nu}R^{\mu\nu}+R_{\mu\nu\rho\sigma}R^{\mu\nu\rho\sigma}$ is the integrand of the Gauss--Bonnet term which is topological in four dimensional spacetime, and then it does not contribute to the beta function.

To calculate the beta function, we use the background method and then split the fields as
\al{
 g_{\mu\nu}
	&=	\bar g_{\mu\nu}+h_{\mu\nu},&
\Phi
	&=	\phi+\varphi,&
\Psi
	&=	\psi+\chi,&
	\label{backgroundsplit}
}
where the matter background are constant.
We assume the Einstein metric as our background, i.e., 
\al{
\bar R_{\mu\nu}=\frac{\bar R}{4}\bar g_{\mu\nu}.
}
The potentials $V\fn{\phi}$ and $F\fn{\phi}$ are expanded into the polynomial of $\phi^2$:
\al{
V\fn{\phi}&= \sum_{n=0}^\infty\lambda_{2n} \phi^{2n},&
F\fn{\phi}&= \sum_{n=0}^\infty\xi_{2n} \phi^{2n}.&
}
The coupling constants $\lambda_0=\Lambda_\text{c.c}$, $\lambda_2=m^2/2$, $\lambda_4=\lambda/4$, $\xi_0=M_\text{pl}^2/2$ and $\xi_2=\xi$ are the cosmological constant, the scalar mass-square, scalar quartic coupling constant, the Planck mass (inverse Newton constant) and non-minimal coupling constant, respectively.
Pure gravity can be obtained by taking $\lambda_2\to \infty$, $N_f=0$, $y\to 0$, $\xi_{2n}\to 0$ and $\lambda_{2(n+1)}\to0$ for $n\geq 1$.

Using the Kugo--Ojima formulation~\cite{Kugo:1979gm}, the gauge-fixing and the ghost actions for the diffeomorphisms are given as~\cite{Ohta:2016jvw}
\al{
S_{{\rm gf}}+S_{{\rm gh}}	
&= \int \df^4x \sqrt{\bar g}\,{\bf \delta_{B}} \left[{\bar C}_\mu Y^{\mu\nu}\left( \Sigma_\nu -\frac{\alpha}{2} B_\nu \right)\right] \nn
&=\int \df^4x \sqrt{\bar g}\bigg[
-\frac{\alpha}{2}{\hat B}_\mu Y^{\mu\nu} {\hat B}_\nu +\frac{1}{2\alpha}\Sigma_\mu Y^{\mu\nu} \Sigma_\nu 
-{\bar C}_\mu Y^{\mu\rho}{\bar \Delta}^\text{ghost}_{\rho\nu}C^\nu
\bigg],
					\label{gaugefixedaction}
}
respectively, where $\bar \Box:= \bar g^{\mu\nu}\bar\nabla_\mu\bar\nabla_\nu$ and $ \bar\Delta^\text{ghost}_{\mu\nu}:= \bar g_{\mu\nu}\bar\Box + \frac{1-\beta}{2}{\bar\nabla}_\mu{\bar\nabla}_{\nu}+{\bar R}_{\mu\nu}$; ${\bf \delta_{B}} $ is the Grassmann-odd BRST transformation; $B_\mu$ is the bosonic auxiliary field (Nakanishi-Lautrup field); $C_\mu$ and ${\bar C}_\mu$ are the ghost and anti-ghost fields for the diffeomorphisms, respectively;
\al{
Y^{\mu\nu}&:= \bar g^{\mu\nu}\bar\Box + \rho_1\bar\nabla_\mu \bar\nabla_\nu -\rho_2\bar\nabla_\nu \bar\nabla_\mu;&
\Sigma_\mu		
	&:= \bar\nabla^\nu h_{\nu \mu}-\frac{\beta +1}{4} \bar\nabla_\mu h&
}
with $h:=\bar g^{\mu\nu}h_{\mu\nu}$; $\hat B_\mu=B_\mu + \Sigma_\mu/\alpha$; and $\alpha$, $\beta$, $\rho_1$ and $\rho_2$ are gauge parameters.
Note here that whereas $B_\mu$ is not the dynamical field in the Einstein gravity case where $Y^{\mu\nu}=\bar g^{\mu\nu}$ and then it is integrated out.
We use a dynamical $B_\mu$ in case of the higher derivative gravity and call it $B$ ghost.
We note also that in previous works on higher derivative gravity the Nielsen--Kallosh (NK) ghost being a Grassmann-odd and corresponding to the contribution $(\det Y^{\mu\nu})^{1/2}$ has been introduced within the path-integral formalism and the Faddeev-Popov (FP) ghost has been given as ${\bar C}^\mu\bar\Delta^\text{ghost}_{\mu\nu}C^\nu$.
Then the total contributions from the ghost fields are given by $(\det Y^{\mu\nu})^{1/2}\cdot (\det\bar\Delta^\text{ghost}_{\mu\nu})$.
However, it is unclear why the differential operators for the NK and the FP ghosts differ from each other.
In contrast, the contributions from the ghost fields given in \eqref{gaugefixedaction} become $(\det Y^{\mu\nu})^{-1/2}\cdot (\det Y^{\mu\nu})(\det\bar \Delta^\text{ghost}_{\mu\nu})=(\det Y^{\mu\nu})^{1/2}\cdot (\det\bar\Delta^\text{ghost}_{\mu\nu})$ which agrees with the case of the path-integral formalism.

\subsection{York decomposition}\label{YD}
The graviton fluctuation $h_{\mu\nu}$ is decomposed as~\cite{York:1973ia}
\al{\label{metricdecomposition}
h_{\mu\nu}	=	h_{\mu\nu}^\perp		+\bar\nabla _\mu\ft\xi_{\nu}	+\bar\nabla _{\nu}\ft\xi_\mu	
+\paren{\bar\nabla_\mu \bar\nabla_{\nu}	-\frac{1}{4} \bar g_{\mu\nu}\bar\Box}\ft\sigma
 	+\frac{1}{4} \bar g_{\mu\nu}h,
}
where $h^\perp_{\mu\nu}$ is transverse and traceless tensor field with spin 2, thus satisfies $\bar\nabla^{\mu}h^\perp_{\mu\nu}=0$ and $\bar g^{\mu\nu}h^\perp_{\mu\nu}=0$;
 $\ft \xi_\mu$ satisfying $\bar\nabla^\mu\tilde \xi_{\mu}=0$ is the transverse vector field with spin 1; and $\ft \sigma$ and $h:={\bar g}^{\mu\nu}h_{\mu\nu}$ are the scalar fields with spin 0.

The ghost fields are also decomposed into the transverse and scalar components:
\al{
{\hat B}_\mu	
	&=	B_\mu^\perp			+\bar\nabla_\mu \tilde B,&
C_\mu	
	&=	C_\mu^\perp			+\bar\nabla_\mu \tilde C,&
{\bar C}_\mu
	&=	{\bar C}_\mu^\perp+\bar\nabla_\mu \tilde {\bar C},&
	\label{ghostdecomposition}
}
where $\tilde B$, $\tilde C$ and $\tilde {\bar C}$ are spin-0 scalar fields, and $B^\perp_\mu$, $C^\perp_\mu$ and $\bar C_\mu ^\perp$ are spin-1 transverse vector fields that satisfy $\bar\nabla^\mu B^\perp_\mu=\bar\nabla^\mu C^\perp_\mu=\bar\nabla^\mu {\bar C}^\perp_\mu=0$.

Due to the decompositions \eqref{metricdecomposition} and \eqref{ghostdecomposition}, the following contributions come out in the path integral (see e.g.~\cite{Machado:2007ea}):\footnote{
Here the Einstein metric is imposed.
In more general metric, there are mixing terms such as $\tilde \xi_\mu {\bar R}^{\mu\nu}\bar \nabla_\nu \tilde \sigma$.
}
\al{
J:=\int {\mc D}\Omega \exp\bigg[ 
-2\tilde \xi^\mu\left(-{\bar \Box}-\frac{\bar R}{4} \right) \tilde \xi^\mu 
-\frac{3}{4}\tilde \sigma \left( -{\bar \Box} \right)\left( -{\bar \Box}-\frac{\bar R}{3} \right) \tilde \sigma
-{\tilde B} \left( -{\bar \Box} \right) \tilde B
-\tilde {\bar {C}} \left( -{\bar \Box} \right) \tilde C
 \bigg],
 \label{jacobian}
}
where $\int {\mc D}\Omega :=\int {\mc D}\tilde \xi\, {\mc D} \tilde \sigma \, {\mc D} \tilde B\,  {\mc D} \tilde{\bar{C}}\,  {\mc D} \tilde C$.
To remove them, we redefine the fluctuations $\tilde \xi$, $\tilde \sigma$, $\tilde B$, $\tilde {\bar C}$ and $\tilde C$ as
\al{
\xi_\mu
	&=	\sqrt{-{\bar \Box}-\frac{\bar R}{4}}\,\ft\xi_\mu,	&
{\sigma}	
	&=	\sqrt{-{\bar \Box}-\frac{\bar R}{3}}\sqrt{-{\bar \Box}}\,\ft\sigma,	&
{B}	
	&=	\sqrt{-{\bar \Box}}\,\ft B,&
{C}
	&=	\sqrt{-{\bar \Box}}\,\ft C,&
{\bar C}	
	&=   \ft {\bar C}\sqrt{-{\bar \Box}}.&
}
The Jacobian from this field redefinition exactly cancels Eq.~\eqref{jacobian} (see e.g. \cite{Dou:1997fg}), and then the term \eqref{jacobian} does not contribute to the beta functions.
Hereafter we consider the two-point functions in field bases without the tilde.

\subsection{Two-point functions}\label{modelandmanipulations}
For the background fields $\Xi:=\paren{\bar g_{\mu\nu},\phi,\psi,\bar \psi}$ and the fluctuations $\Upsilon:=\paren{h_{\mu\nu},\varphi,\chi, \bar \chi,C_\mu,\bar C_\mu,B_\mu}$, the effective action is written as $\Gamma_k[\Xi;\Upsilon]$ and is expanded as
\al{
\Gamma_k[\Xi;\Upsilon]=\Gamma_k[\Xi] + \Gamma_k^{(1)}[\Xi;\Upsilon]+
 \Gamma_k^{(2)}[\Xi;\Upsilon]+\mO\fn{\Upsilon^3},
}
where $\Gamma_k^{(n)}[\Xi;\Upsilon]$ contains the terms of order $\Upsilon^n$.

To derive the beta functions for the Higgs-Yukawa model, we need to evaluate the $\Gamma_k^{(2)}$ terms 
\al{
\Gamma_k^{(2)}[\Xi;\Upsilon]=\frac{1}{2}\frac{\delta^2 \Gamma_k[\Xi]}{\delta \Upsilon_i\delta \Upsilon_j}\Upsilon_i \Upsilon_j,
\label{secondvariations}
}
where $S_{\rm gf}$ and $S_{\rm gh}$ are given in Eqs.~\eqref{gaugefixedaction}
The explicit calculation of \eqref{secondvariations} is given in appendix~\ref{variations of action}.
The second variation of the effective action, i.e., the Hessian, becomes
\al{
&\Gamma_k^{(2)}[\Xi;\Upsilon]
=\frac{1}{2}\int \df^4x \sqrt{\bar g}
\bigg[
h_{\mu\nu}^\perp \Gamma_k ^{(\perp\perp)}h^{\mu\nu}{}^\perp
+\xi_\mu \Gamma_k ^{(\xi\xi)}\xi^\mu
+S^T\Gamma_k^{(SS)}S
+ B_\mu^\perp \Gamma_k ^{(B^\perp B^\perp)} B^\perp{}^\mu
+ B \Gamma_k ^{(BB)} B\nn
&\quad+ \xi_\mu \Gamma_k ^{(\xi \chi)}\chi
+\bar \chi \Gamma_k ^{(\bar \chi \xi)}\xi_\mu
+S^T \Gamma_k ^{(S \chi)}\chi
+\bar \chi \Gamma_k ^{(\bar \chi S )}S
+\bar \chi\Gamma_k ^{(\bar \chi \chi)} \chi
+\bar C_\mu^\perp \Gamma_k ^{(\bar C^\perp C^\perp)} C^\perp{}^\mu
+\bar C \Gamma_k ^{(\bar C C)} C
\bigg],
} 
where $S:=(\sigma,h ,\varphi)^T$ denotes the scalar fields with spin 0 and the York decomposition~\eqref{metricdecomposition} was employed.
We show the explicit forms of the Hessian below.
For bosonic fields, we have
\al{
\Gamma_{\BB}=
\pmat{
\Gamma_k ^{(\perp\perp)} & 0 & 0 & 0 & 0\\
0 & \Gamma_k ^{(\xi\xi)}& 0 & 0 & 0\\
0 & 0 & \Gamma_k^{(SS)} & 0 & 0\\
0 & 0 & 0 & \Gamma_k^{(B^\perp B^\perp)} & 0\\
0 & 0 & 0 & 0 & \Gamma_k^{(BB)}
},
} 
where each component is given by
\al{
\Gamma_k ^{(\perp\perp)}
&=\frac{F}{2}\bar\Delta_{L2}
 -a{\bar R}\left(\bar \Delta_{L2} -\frac{\bar R}{2} \right)
+\frac{b}{2}\left( \bar\Delta_{L2}^2-\frac{3\bar R}{2}\bar\Delta_{L2}+\frac{\bar R^2}{2}\right) 
  -\frac{V+Y}{2},\\
\Gamma_k ^{(\xi\xi)}    
&= \frac{F}{2}{\bar R}-(V+Y) -\frac{1}{\alpha}\left(\bar \Delta_{L1}-\frac{\bar R}{2} \right)\left( \bar\Delta_{L1}-\frac{1-\rho_2}{4}{\bar R} \right),\\
\Gamma_k^{(SS)}
&=
\pmat{
\Gamma_k^{(\sigma\sigma)} & \Gamma_k^{(\sigma h)} &\Gamma_k^{(\sigma\varphi)} \cr
\Gamma_k^{(h \sigma)} & \Gamma_k^{(h h)} &\Gamma_k^{(h\varphi)} \cr
\Gamma_k^{(\varphi \sigma)} & \Gamma_k^{(\varphi h)} &\Gamma_k^{(\varphi\varphi)}
},\\
\Gamma_k^{(B^\perp B^\perp)}
&=  \alpha\left( \bar\Delta_{L1} -\frac{1-\rho_2}{4}{\bar R} \right),\\
\Gamma_k^{(B B)}
&=\alpha\left( (1+\rho_1-\rho_2)\bar\Delta_{L0}-\frac{1-\rho_2}{4}{\bar R}\right),
}
with 
\al{
\Gamma_k^{(\sigma\sigma)}
&=
-\frac{3F}{16}\left( \bar\Delta_{L0}  - {\bar R} \right)+\frac{9a}{8}\bar\Delta_{L0}^2 + \frac{3b}{8}\bar\Delta_{L0}^2   -\frac{3\paren{V +Y}}{8}\nn
&\quad -\frac{9}{16\alpha} \left( \bar\Delta_{L0} -\frac{\bar R}{3}\right)\left[(1+\rho_1-\rho_2)\bar\Delta_{L0} -\frac{1-\rho_2}{4}{\bar R} \right]
,\\
\Gamma_k^{(\sigma h)} &=\Gamma_k^{(h \sigma)} 
=\left(-\frac{3F}{16}+\frac{9a}{8}\bar\Delta_{L0} +\frac{3b}{8}\bar\Delta_{L0} \right) \sqrt{\bar\Delta_{L0}}\sqrt{\bar\Delta_{L0}-\frac{\bar R}{3}}\nn
&\quad -\frac{3\beta}{16\alpha}\left[(1+\rho_1-\rho_2)\bar\Delta_{L0} -\frac{1-\rho_2}{4}{\bar R} \right]\sqrt{\bar\Delta_{L0}}\sqrt{\bar\Delta_{L0} -\frac{\bar R}{3}}
,\\
\Gamma_k^{(\sigma\varphi)}&=\Gamma_k^{(\varphi \sigma)} 
= -\frac{3\phi F'}{2}\sqrt{\bar\Delta_{L0}-\frac{{\bar R}}{3}}\sqrt{\bar\Delta_{L0}},\\
\Gamma_k^{(h h)}
&=-\frac{3F}{16}\bar\Delta_{L0}+\frac{9a}{8}\left( \bar\Delta_{L0} -\frac{\bar R}{3} \right)\bar\Delta_{L0} 
+\frac{3b}{8}\left( \bar\Delta_{L0} -\frac{\bar R}{3} \right)\bar\Delta_{L0}
+\frac{V+Y}{8}\nn
&\quad -\frac{\beta^2}{16\alpha}\bar\Delta_{L0}\left[(1+\rho_1-\rho_2)\bar\Delta_{L0} -\frac{1-\rho_2}{4}{\bar R} \right]
,\\
\Gamma_k^{(h\varphi)}
&=\Gamma_k^{(\varphi h)}
=-\frac{3\phi F'}{2} \left( \bar\Delta_{L0} + \frac{{\bar R}}{3}\right) + \phi V'+{y{\bar \psi}\psi\over 2},\\
\Gamma_k^{(\varphi\varphi)}
&=\bar\Delta_{L0}+(2V' +4\phi^2 V'')-{\bar R}(2F' +4\phi^2 F'').
}
Here the prime $'$ denotes the derivative with respect to $\varphi^2$, i.e.
\al{
V'&=\frac{\p V}{\p \varphi^2},&
V''&=\frac{\p^2 V}{\p (\varphi^2)^2},&
F'&=\frac{\p F}{\p \varphi^2},&
F''&=\frac{\p^2 F}{\p (\varphi^2)^2},&
}
and we have defined the Lichnerowicz Laplacians (see e.g.~\cite{Christensen:1978md}) with the Einstein metric
\al{
\bar\Delta_{L0}S&:=-\bar\Box S,\\
\bar\Delta_{L\frac{1}{2}}\psi&:= -\bar D^2 \psi +\frac{\bar R}{4}\psi,\\
\bar\Delta_{L1} \xi_\mu&:=-\bar\Box \xi_\mu +\frac{\bar R}{4}\xi_\mu\\
\bar\Delta_{L2}h_{\mu\nu} &:=-\bar\Box h_{\mu\nu} + \frac{\bar R}{2} h_{\mu\nu}-2\bar R_{\mu~\nu}^{~\alpha~\beta}h_{\alpha \beta}.
}
More general forms of these Laplacians are represented in \eqref{scalar Lich}--\eqref{tensor Lich}.

For the fermionic fields, the Hessian becomes
\al{
\Gamma_\FF = 
\pmat{
\Gamma^{(\bar \chi \chi)}_k & 0 & 0 \\
0 & \Gamma_k^{\bar C^\perp C^\perp} & 0 \\
0 & 0 & \Gamma_k^{\bar C C} \\
},
}
with
\al{
\Gamma_k^{(\bar \chi \chi)} 
&=\pmat{
0 & -({\bara{\Slash {\bar D}}}^\T +y\phi)\\
{ \Slash {\bar D}} +y \phi & 0
},\\
\Gamma_k^{\bar C^\perp C^\perp}
&=\pmat{
0 &{ \left( \bar\Delta_{L1}-\frac{\bar R}{2} \right)\left( \bar\Delta_{L1} -\frac{1-\rho_2}{4}{\bar R} \right)} \\
-{\left( \bar\Delta_{L1}-\frac{\bar R}{2} \right)\left( \bar\Delta_{L1} -\frac{1-\rho_2}{4}{\bar R} \right)}  & 0\\
},\\
\Gamma_k^{\bar C C}
&=
\pmat{
0 & {\scriptstyle \frac{3-\beta}{2} \left( \bar\Delta_{L0} -\frac{\bar R}{3-\beta} \right)\left[ (1+\rho_1-\rho_2)\bar\Delta_{L0} -\frac{1-\rho_2}{4}{\bar R}  \right]}\\
{\scriptstyle -\frac{3-\beta}{2}  \left( \bar\Delta_{L0} -\frac{\bar R}{3-\beta} \right)\left[ (1+\rho_1-\rho_2)\bar\Delta_{L0} -\frac{1-\rho_2}{4}{\bar R}  \right]} & 0
},
}
where $\T$ on the derivative operator is the transposition acting on the spinor space and the over-left-arrow denotes that the derivative acts on the operator from the right-hand side.

The parts with both bosonic and fermionic fields are given by
\al{
\Gamma_\BF
	&=\pmat{
	\frac{\overrightarrow \delta}{\delta h^\perp_{\mu\nu}}\Gamma_k \frac{\overleftarrow \delta}{\delta \chi}
&	\frac{\overrightarrow \delta}{\delta h^\perp_{\mu\nu}}\Gamma_k \frac{\overleftarrow \delta}{\delta \bar \chi^\T}\\
	\frac{\overrightarrow \delta}{\delta \xi_{\mu}}\Gamma_k \frac{\overleftarrow \delta}{\delta \chi}
&	\frac{\overrightarrow \delta}{\delta \xi_{\mu}}\Gamma_k \frac{\overleftarrow \delta}{\delta \bar \chi^\T}\\
	\frac{\overrightarrow \delta}{\delta \sigma}\Gamma_k \frac{\overleftarrow \delta}{\delta \chi}
&	\frac{\overrightarrow \delta}{\delta \sigma}\Gamma_k \frac{\overleftarrow \delta}{\delta \bar \chi^\T}\\
	\frac{\overrightarrow \delta}{\delta h}\Gamma_k \frac{\overleftarrow \delta}{\delta \chi}
&	\frac{\overrightarrow \delta}{\delta h}\Gamma_k \frac{\overleftarrow \delta}{\delta \bar \chi^\T}\\
	\frac{\overrightarrow \delta}{\delta \varphi}\Gamma_k \frac{\overleftarrow \delta}{\delta \chi}
&	\frac{\overrightarrow \delta}{\delta \varphi}\Gamma_k \frac{\overleftarrow \delta}{\delta \bar \chi^\T}\\
	}
=	\pmat{
				0&0\cr
				-\frac{1}{4}\sqrt{-\bara {\bar D}^2}\paren{\bar\psi\gamma^\mu}	&	-\frac{1}{4}\sqrt{-\bara {\bar D}^2}\paren{\gamma^\mu\psi}^\T\cr
				-\frac{3}{16}\paren{\bara{\bar D_{\mu}}}\paren{\bar\psi\gamma^\mu} 	&	-\frac{3}{16}\paren{\bara{\bar D_{\mu}}}\paren{\gamma^\mu\psi}^\T\cr
				\frac{y}{2}\phi {\bar \psi} 	-\frac{3}{16}\paren{\bara{\bar D_{\mu}}}\paren{\bar\psi\gamma^\mu}	&	-\frac{y}{2}\phi \psi^\T	-\frac{3}{16}\paren{\bara{\bar D_{\mu}}}\paren{\gamma^\mu \psi}^\T\cr
				y\bar\psi		&	-y\psi^\T
				}	,\\
\Gamma_\FB
	&=\pmat{
	\frac{\overrightarrow \delta}{\delta \chi^\T}\Gamma_k \frac{\overleftarrow \delta}{\delta h^\perp_{\mu\nu}}
&	\frac{\overrightarrow \delta}{\delta \chi^\T}\Gamma_k \frac{\overleftarrow \delta}{\delta \xi{\mu}}
&	\frac{\overrightarrow \delta}{\delta \chi^\T}\Gamma_k \frac{\overleftarrow \delta}{\delta \sigma}
&	\frac{\overrightarrow \delta}{\delta \chi^\T}\Gamma_k \frac{\overleftarrow \delta}{\delta h}
&	\frac{\overrightarrow \delta}{\delta \chi^\T}\Gamma_k \frac{\overleftarrow \delta}{\delta \varphi}\\
	\frac{\overrightarrow \delta}{\delta \bar \chi}\Gamma_k \frac{\overleftarrow \delta}{\delta h^\perp_{\mu\nu}}
&	\frac{\overrightarrow \delta}{\delta \bar \chi}\Gamma_k \frac{\overleftarrow \delta}{\delta \xi_\mu}
&	\frac{\overrightarrow \delta}{\delta \bar \chi}\Gamma_k \frac{\overleftarrow \delta}{\delta \sigma}
&	\frac{\overrightarrow \delta}{\delta \bar \chi}\Gamma_k \frac{\overleftarrow \delta}{\delta h}
&	\frac{\overrightarrow \delta}{\delta \bar \chi}\Gamma_k \frac{\overleftarrow \delta}{\delta \varphi}
	}\nn
	&=	\pmat{
	0	&	\frac{1}{4}\paren{{\bar \psi}\gamma^\mu}^\T\sqrt{-{\bar D}^2}	&	\frac{3}{16}\paren{{\bar \psi}\gamma^\mu}^\T {\bar D}_{\mu}	&	-\frac{y}{2}\phi {\bar \psi}^\T+\frac{3}{16}\paren{{\bar \psi}\gamma^\mu}^\T {\bar D}_{\mu}	&	-y{\bar \psi}^\T \cr
	0	&	\frac{1}{4}\paren{\gamma^\mu \psi}\sqrt{-{\bar D}^2}	&	\frac{3}{16}\paren{\gamma^\mu\psi}{\bar D}_{\mu}	&	\frac{y}{2}\phi \psi	+\frac{3}{16}\paren{\gamma^\mu\psi}{\bar D}_{\mu}	&	y\psi
 }.
			}
Note that we have neglected the terms which do not contribute to the truncated effective action \eqref{originaleffectiveaction}.

\begin{figure}
\begin{center}
\fbox{\includegraphics[width=15cm]{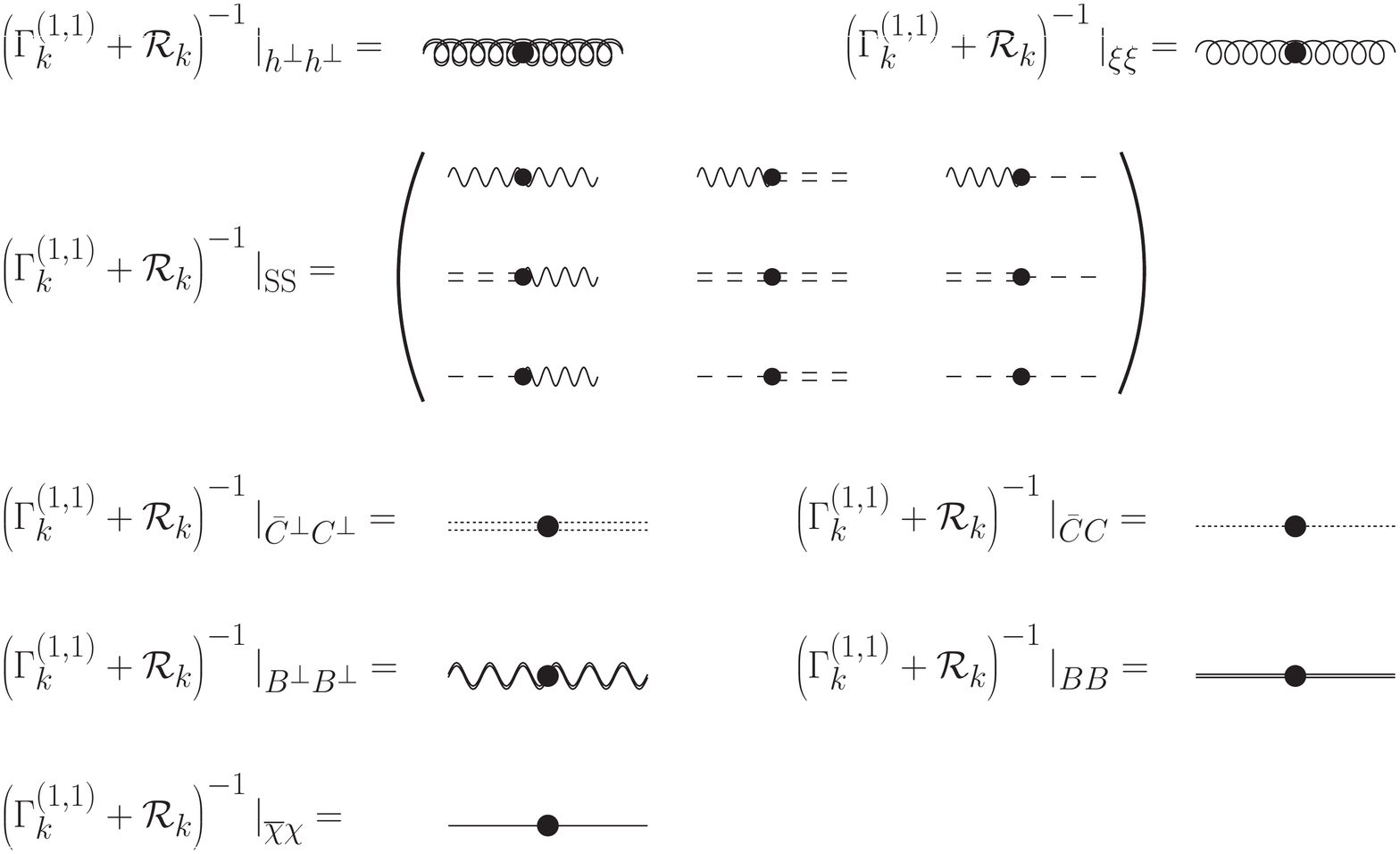}}
\end{center}
\caption{The Feynman diagrams of propagators.}
\label{feynman diagrams of propagators}
\end{figure}

We next give the cutoff function $\mathcal R_k$.
The cutoff functions are employed so that the Lichnerowicz Laplacians in the Hessian are replaced as $\bar\Delta_{Ln}\to P_n\fn{\bar\Delta_{Ln}}=\bar\Delta_{Ln}+R_k\fn{\bar\Delta_{Ln}}$, namely
\al{
\mathcal R_\BB
&=\pmat{
{\mathcal R_k^{(h^\perp h^\perp)} }& 0 & 0 & 0 &0 \\
0 & {\mathcal R_k^{(\xi \xi)}} & 0 &0 & 0 \\
0 & 0 & {\mathcal R_k^{(SS)}} &0 & 0\\
0 & 0 & 0 & {\mathcal R_k^{(B^\perp B^\perp)}} & 0\\
0 & 0 & 0 & 0 & {\mathcal R_k^{(BB)}}
},&
\mathcal R_\FF
&=\pmat{
{\mathcal R_k^{(\bar \chi \chi)} } & 0\\
0 & {\mathcal R_k^{\text{ghost}}}
},&
}
where
\al{
{\mathcal R_k^{(h^\perp h^\perp)} }
&=\Gamma_k^{(h^\perp h^\perp)}[P_k\fn{\bar\Delta_{L2}}]-\Gamma_k^{(h^\perp h^\perp)}[\bar\Delta_{L2}],\\
{\mathcal R_k^{(\xi \xi)}}
&=\Gamma_k^{(\xi \xi)}[P_k\fn{\bar\Delta_{L1}}]-\Gamma_k^{(\xi \xi)}[\bar\Delta_{L1}],\\
{\mathcal R_k^{(SS)}}&=
\pmat{
{\mathcal R}_k^{(\sigma\sigma)} & {\mathcal R}_k^{(\sigma h)} &{\mathcal R}_k^{(\sigma\varphi)} \cr
{\mathcal R}_k^{(h \sigma)} & {\mathcal R}_k^{(h h)} &{\mathcal R}_k^{(h\varphi)} \cr
{\mathcal R}_k^{(\varphi \sigma)} & {\mathcal R}_k^{(\varphi h)} &{\mathcal R}_k^{(\varphi\varphi)}\\
}
=\Gamma_k^{(SS)}[P_k\fn{\bar\Delta_{L0}}]-\Gamma_k^{(SS)}[\bar\Delta_{L0}],\\
{\mathcal R}_k^{B^\perp B^\perp}&=\Gamma_k^{(B^\perp B^\perp)}[P_k\fn{\bar\Delta_{L1}}]-\Gamma_k^{(B^\perp B^\perp)}[\bar\Delta_{L1}],\\
{\mathcal R}_k^{\bar BB}&=\Gamma_k^{({\bar B}B)}[P_k\fn{\bar\Delta_{L0}}]-\Gamma_k^{({\bar B}B)}[\bar\Delta_{L0}],\\
{\mathcal R_k^{(\bar \chi \chi)} }&=
\pmat{
0 & -\bara{\Slash{\bar D}}^\T\left( \sqrt{1+\frac{R_k\fn{\bar\Delta_{L\frac{1}{2}}}}{\bar\Delta_{L\frac{1}{2}}}} -1\right)\\
\left( \sqrt{1+\frac{R_k\fn{\bar \Delta_{L\frac{1}{2}}}}{\bar\Delta_{L\frac{1}{2}}}} -1\right) {\Slash {\bar D}} & 0
},\\
{\mathcal R_k^{\text{ghost}}}&=
\pmat{
0 & -{\mathcal R}_k^{\bar C^\perp C^\perp}\fn{\bar\Delta_{L1}}^T & 0 & 0\\
{\mathcal R}_k^{\bar C^\perp C^\perp}\fn{\bar\Delta_{L1}}& 0 & 0 & 0\\
0 & 0 & 0 & -{\mathcal R}_k^{\bar CC}\fn{\bar\Delta_{L0}}^T \\
0 & 0 & {\mathcal R}_k^{\bar CC}\fn{\bar\Delta_{L0}} & 0
},
}
with 
\al{
{\mathcal R}_k^{\bar C^\perp C^\perp}\fn{\bar\Delta_{L1}}&=\Gamma_k^{(\bar C^\perp C^\perp)}[P_k\fn{\bar\Delta_{L1}}]-\Gamma_k^{(\bar C^\perp C^\perp)}[\bar\Delta_{L1}],\\
{\mathcal R}_k^{\bar C C}\fn{\bar\Delta_{L0}}&=\Gamma_k^{({\bar C}C)}[P_k\fn{\bar\Delta_{L0}}]-\Gamma_k^{({\bar C}C)}[\bar\Delta_{L0}].
}
In this paper, we use the optimized cutoff function~\cite{Litim:2001up} for $R_k\fn{p^2}$, namely,
\al{
R_k\fn{p^2}=\fn{k^2-p^2}\theta\fn{k^2-p^2},
}
where $\theta\fn{x}$ is the step function.
For $p^2<k^2$, $P_n\fn{\bar\Delta_{Ln}}=k^2$.
We note here that for the fermionic field $\mathcal R_k^{(\bar \chi \chi)}$ the Type-II cutoff function should be employed in order to obtain the correct sign of the femionic fluctuation in $R$ term~\cite{Dona:2012am}.
In Fig.~\ref{feynman diagrams of propagators}, we show the Feynman diagrams of the propagators.

\section{Flow equations}\label{flow equations section}
Using the Hessian matrices shown in the last section we can derive the beta functions.
The Wetterich equation now is reduced as
\al{
\p_t\Gamma_k	
	&=	\frac{1}{2}{\rm Tr}\left. \frac{\p_t{\mathcal R}_k}{\Gamma _k^{(1,1)}	
					+{\mathcal R}_k}\right|_{h^\perp h^\perp}
							+\frac{1}{2}{\rm Tr}\left. \frac{\p_t{\mathcal R}_k}{\Gamma_k^{(1,1)}
									+{\mathcal R}_k}\right|_{\xi \xi}
						+\frac{1}{2}{\rm Tr}\left. \frac{\p_t{\mathcal R}_k}{\Gamma _k^{(1,1)}
									+{\mathcal R}_k}\right|_{\rm SS}\nn
							&\quad +\frac{1}{2}{\rm Tr}\left. \frac{\p_t{\mathcal R}_k}{\Gamma _k^{(1,1)}
									+{\mathcal R}_k}\right|_{ B^\perp B^\perp}
								+\frac{1}{2}{\rm Tr}\left. \frac{\p_t{\mathcal R}_k}{\Gamma _k^{(1,1)}
								+{\mathcal R}_k}\right|_{BB}\nn
							&\qquad -\left. \Tr \frac{\p_t {\mathcal R} _k}{\Gamma_k^{(1,1)}+{\mathcal R}_{k}}\right|_{\bar\chi \chi}
							-\left. \Tr \frac{\p_t {\mathcal R} _k}{\Gamma_k^{(1,1)}+{\mathcal R}_{k}}\right|_{{\bar C}^\perp C}
							-\left. \Tr \frac{\p_t {\mathcal R} _k}{\Gamma_k^{(1,1)}+{\mathcal R}_{k}}\right|_{\bar C C},
				\label{betafunctions}
}
where we defined the dimensionless scale $t:=\ln\fn{k/\Lambda}$ and the derivative
\al{
\p_t
	&:=	k{\p\over\p k}.
}
The functional traces are evaluated by using the heat kernel techniques shown in appendix~\ref{hkeapp}.
The explicit calculations are presented in appendix~\ref{explicit derivation of beta functions}.

In this work, the non-minimal potential $F\fn{\phi^2}$ and the scalar potential $V\fn{\phi^2}$ are expanded around the symmetric phase $\langle \phi \rangle =0$ and truncated as
\al{
k^2{\tilde F}\fn{\phi^2}&=F\fn{\phi^2}=\xi_0 + \xi_2 \phi^2
= k^2 (\tilde \xi_0 + \tilde \xi_2 \tilde \phi^2),\\
k^4{\tilde V}\fn{\phi^2}&=V\fn{\phi^2}=\lambda_0 + \lambda_2\phi^2+\lambda_4\phi^4
=k^4(\tilde \lambda_0 + \tilde \lambda_2\tilde \phi^2+\tilde \lambda_4\tilde \phi^4),
}
where dimensionless variables are introduced as $\tilde \phi=\phi/k$, $\tilde \xi_{2n}=\xi_{2n}/k^{2-2n}$ and $\tilde \lambda_{2n}=\lambda_{2n}/k^{4-2n}$.
The other coupling constants $a$, $b$ and $y$ are dimensionless, thus they are written as $\tilde a=a$, $\tilde b=b$ and $\tilde y=y$.
We consider the theory space spanned by maximally eight coupling constants $g=\{ \tilde \lambda_0, \tilde \xi_0,\tilde a, \tilde b, \tilde \xi_2, \tilde \lambda_2, \tilde \lambda_4, \tilde y\}$.

\subsection{Comparison with previous works}
Here we compare our calculation to the previous work.
First, the pure four derivative gravity case ($\lambda_2\to \infty$, $\lambda_{2n}\to 0$ for $n\geq 2$ and $F\fn{\phi^2}\to 0$) is considered.
The standard form of the action for higher derivative gravity is given by
\al{
\Gamma=k^{d-4}\int \df^dx \Bigg[
\frac{1}{\kappa^2}(R-2\Lambda_\text{cc} ) + \frac{1}{\xi}R^2 + \frac{1}{2\lambda}C^2 - \frac{1}{\rho}E
\Bigg]+S_\text{gh}+S_\text{gf},
\label{pert effective action}
}
where the generic background metric is imposed.
The perturbative one-loop contributions to the beta functions for $d\to 4$ are obtained as~\cite{deBerredoPeixoto:2004if}
\al{
k\frac{\df }{\df k}\Gamma &=\frac{k^{d-4}}{(4\pi)^2}
\int \df^dx\sqrt{g}\Bigg[
\frac{133}{20}C^2 -\frac{196}{45}E 
+\left( \frac{10\lambda^2}{\xi^2}-\frac{5\lambda}{\xi} +\frac{5}{36} \right)R^2
\nn
& \quad 
+\left( \frac{\xi}{12\lambda} -\frac{13}{6}-\frac{10\lambda}{\xi} \right)\frac{\lambda}{\kappa^2}R
+\left( \frac{56}{3} -\frac{2\xi}{9\lambda} \right) \frac{\lambda \Lambda_\text{cc}}{\kappa^2}
+\left( \frac{\xi^2}{72} +\frac{5\lambda^2}{2} \right)\frac{1}{\kappa^4}
\Bigg].
\label{higher perturbative}
}
The Gauss--Bonnet term and the squared Weyl tensor are 
\al{
E&= R_{\mu\nu\rho\sigma}R^{\mu\nu\rho\sigma}-4R_{\mu\nu}R^{\mu\nu}+R^2,\\
C^2&= R_{\mu\nu\rho\sigma}R^{\mu\nu\rho\sigma}-\frac{4}{d-2}R_{\mu\nu}R^{\mu\nu} +\frac{2}{(d-1)(d-2)}R^2,
}
and then one can recast \eqref{pert effective action} as
\al{
\Gamma=\mu^{d-4}\int \df^dx \Bigg[
\lambda_0 -\xi_0 R + aR^2 + bR_{\mu\nu}R^{\mu\nu} + z R_{\mu\nu\rho\sigma}R^{\mu\nu\rho\sigma}
\Bigg] +S_\text{gh}+S_\text{gf},
}
where
\al{
\lambda_0&= -\frac{2\Lambda_\text{cc}}{\kappa^2},&
\xi_0&=- \frac{1}{\kappa^2},&\nn
a&=-\frac{1}{\rho} +\frac{1}{\xi} +\frac{1}{\lambda(d-1)(d-2)},&
b&=-\frac{2}{(d-2)\lambda}+\frac{4}{\rho},&
z&=\frac{1}{2\lambda}-\frac{1}{\rho}.&
}
The beta functions for the higher derivative terms in $d=4$ become
\al{
\beta_a&=\frac{1}{(4\pi)^2}
\left( \frac{1}{3}\frac{133}{20} -\frac{196}{45} -\frac{5(8z^2+12bz+y^2+32az-4ab-24a^2)}{12(b+4z)^2} \right)\nn
&=\frac{1}{(4\pi)^2}\frac{90a^2-23b^2-338z^2+15ab-120az-199bz}{9(b+4z)^2},\nn
\beta_b&=\frac{1}{(4\pi)^2}\left( -2 \frac{133}{20} -4\left(-\frac{196}{45} \right) \right)
=\frac{1}{(4\pi)^2}\frac{371}{90},\nn
\beta_z&= \frac{1}{(4\pi)^2}\left( \frac{133}{20} -\frac{196}{45} \right)
=\frac{1}{(4\pi)^2}\frac{413}{180}.
\label{uni4}
}
It is known that they are universal, i.e. do not depend on the gauge parametrization and cutoff scheme.
Moreover, 
\al{
\beta_{\lambda_0}&= \frac{1}{4\pi^2},&
\beta_{\xi_0}&=-\frac{15a-14b}{48\pi^2 b^2},&
\label{uni1}
}
are also universal~\cite{Ohta:2016jvw}.

When the Einstein metric is imposed, one obtains the combination
\al{
\beta\big|_{R^2\,\text{term}}= \beta_a +\frac{1}{4}\beta_b\Big|_{z\to 0}
=\frac{1200a^2 +200ab-183b^2}{1920\pi^2 b^2}.
\label{uni3}
}
Our beta functions deriven in this work agree with \eqref{uni4}--\eqref{uni3}.
Note that when using the results from the perturbation theory \eqref{higher perturbative}, we define 
\al{
\theta&=\frac{\lambda}{\rho},&
\omega&=-\frac{3\lambda}{\xi},&
}
we have
\al{
(4\pi)^2k\frac{\df \theta}{\df k}&=-\frac{133}{10}\theta\lambda+\frac{196}{45}\lambda,\nn
(4\pi)^2k\frac{\df \lambda}{\df k}&=-\frac{133}{10}\lambda^2,\nn
(4\pi)^2k\frac{\df \omega}{\df k}&=-\frac{133}{10}\omega\lambda-\left(
\frac{10}{3}\omega^2+5\omega+\frac{5}{12}
\right)\lambda.
}
These beta functions have a UV stable fixed point~\cite{deBerredoPeixoto:2004if}:
\al{
\lambda^*&=0,&
\theta^*&=0.32749,&
\omega^*&=-0.02286.&
}
Thanks to the contributions from the quadratic and quartic divergences taken into account by the FRG computations, the values of fixed point for the Newton and cosmological constants become finite~\cite{Codello:2006in,Groh:2011vn}.
As a result, the theory is asymptotically safe at the non-trivial UV fixed point rather than asymptotically free~\cite{Codello:2006in}.

We next consider the Einstein--Hilbert limit, i.e., $V\fn{\phi^2}=\lambda_0$, $F\fn{\phi^2}=\xi_0$, $a\to 0$ and $b\to 0$.
In this case the beta functions for $\lambda_0$ and $b$ in our work become
\al{
\beta_{\lambda_0}&=\frac{1}{16\pi^2},&
\beta_{b}&=\frac{53}{720\pi^2}.&
}
They agree with~\cite{Ohta:2016jvw}.
Note that $\beta_{\xi_0}$ depends on the gauge parametrization.

Next, we consider the beta functions for the matter couplings.
Setting the coupling constants as $\tilde \lambda_0\to 0$, $\tilde \xi_2\to 0$, $\tilde \xi_0=1/(16\pi g)$, $\tilde a\to 0$, $\tilde \beta \to 0$ and the gauge parameters as $\alpha=0$, $\beta=1$ and $\rho_1=\rho_2=0$, the beta functions become
\al{
\p_t {\tilde \lambda_0}&=-\frac{N_f}{8\pi^2}+\frac{5+8 \tilde{\lambda }_2}{32 \pi ^2 (1+2 \tilde{\lambda}_2)},\\
\p_t {\tilde \lambda_2}&=-2{\tilde \lambda}_2 + \frac{N_f {\tilde y}^2}{8\pi^2}
- \frac{3{\tilde \lambda}_4}{8\pi^2(1+2{\tilde \lambda}_2)^2}
+2 g\tilde{\lambda}_2\frac{ (2 +\tilde{\lambda }_2)}{\pi(1+2 \tilde{\lambda}_2)^2},\\
\p_t {\tilde \lambda_4}&=- \frac{N_f {\tilde y}^4}{8 \pi ^2} +\frac{9 \tilde{\lambda }_4^2}{2 \pi ^2 (1+2 \tilde{\lambda}_2)^3}
+4 g\tilde{\lambda }_4 \frac{1-8 \tilde{\lambda }_2+28 \tilde{\lambda}_2^2+24 \tilde{\lambda }_2^3}{\pi (1+2 \tilde{\lambda }_2)^3}
   +{\mathcal O}\fn{g^2},\\
\p_t{\tilde\xi_0}&=\p_t\left(\frac{1}{16\pi g}\right)
=\frac{N_f}{48\pi^2}+\frac{1}{96\pi^2(1+2\lambda_2)}+\frac{-10g+3\pi}{24g\pi^2},\\
\p_t{\tilde \xi_2}&=\frac{N_fy^2}{48\pi^2}-\frac{\tilde{\lambda }_4}{8 \pi ^2 (1+2 \tilde{\lambda}_2)^2}
+\frac{g (18 \tilde{\lambda }_2+97 \tilde{\lambda}_2^2+104 \tilde{\lambda }_2^3)}{3 \pi (1+2\tilde{\lambda }_2)^2},\\
\p_t {\tilde y}&=
-\frac{\tilde{y}^3(1+\tilde{\lambda }_2) }{8 \pi ^2(1+2 \tilde{\lambda }_2)^2}
+g\tilde{y}\frac{ 2395+1388 \tilde{\lambda}_2+1260 \tilde{\lambda }_2^2}{1120 \pi  (1+2\tilde{\lambda}_2)^2}.
}
The contributions from matter fluctuations agree with \cite{Zanusso:2009bs,Oda:2015sma,Eichhorn:2016esv}.
Although the gauge fixing and ghost actions are different from them in \cite{Zanusso:2009bs,Oda:2015sma,Eichhorn:2016esv}, the gravitational contributions are similar.

\begin{figure}
\begin{center}
\includegraphics[width=15cm]{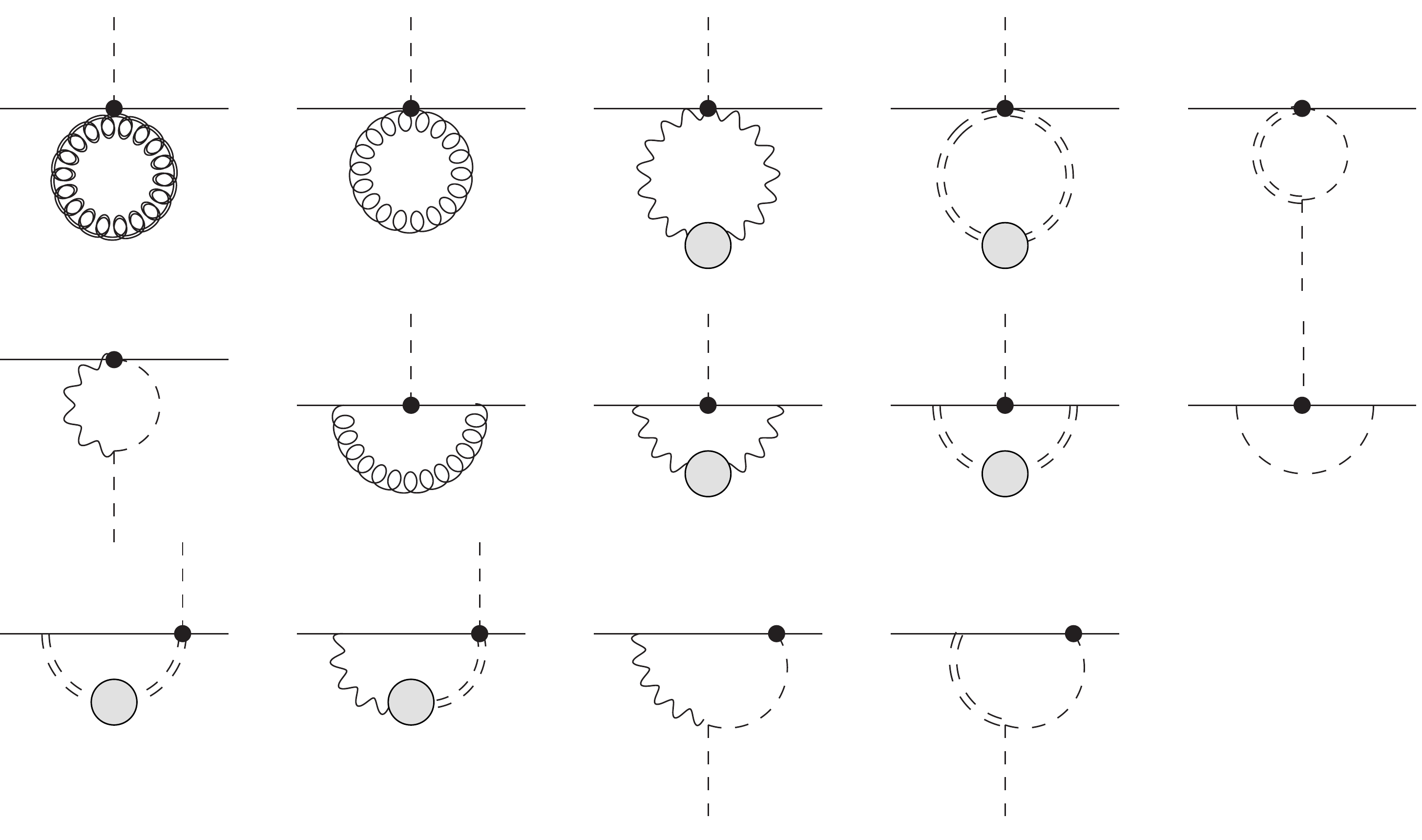}
\put(-440,250){(I)}
\put(-350,250){(II)}
\put(-250,250){(III)}
\put(-160,250){(IV)}
\put(-70,250){(V)}
\put(-440,160){(VI)}
\put(-350,160){(VII)}
\put(-250,160){(VIII)}
\put(-160,160){(IX)}
\put(-70,160){(X)}
\put(-440,80){(XI)}
\put(-350,80){(XII)}
\put(-250,80){(XIII)}
\put(-160,80){(XIV)}
\end{center}
\caption{The loop corrections to the Yukawa coupling. The Feynman diagrams for the propagators are shown in Fig.~\ref{feynman diagrams of propagators}}
\label{feynman diagrams of yukawa vertex}
\end{figure}
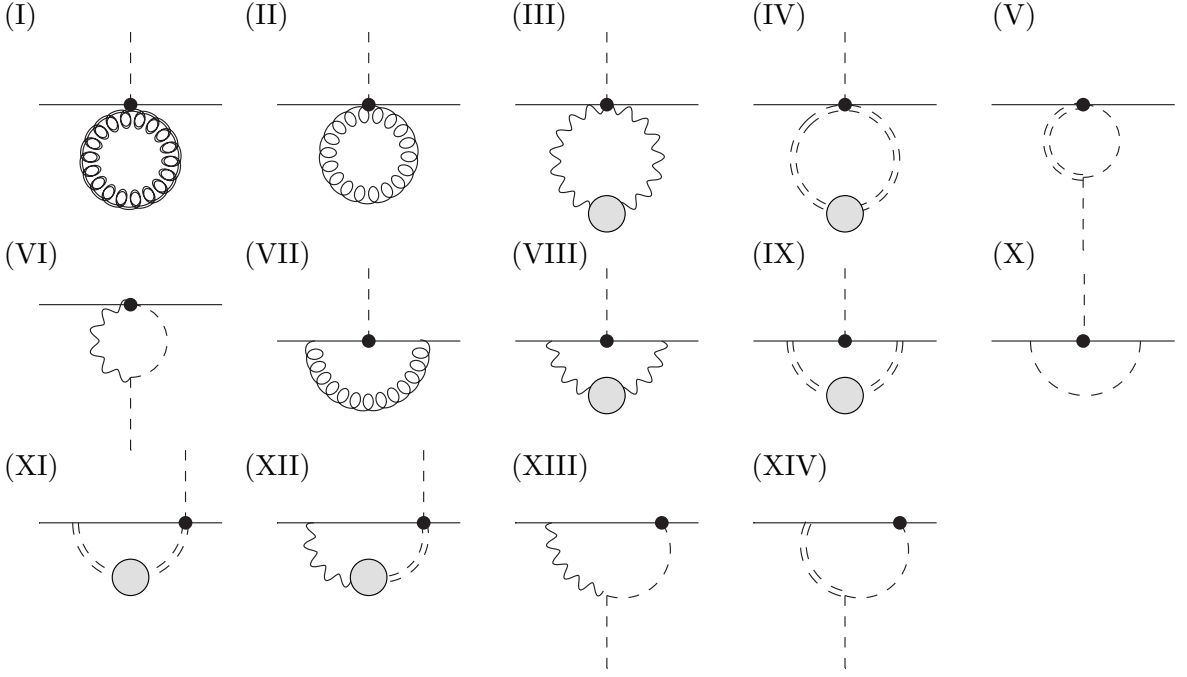

The contributions of higher derivative gravity to the Yukawa coupling constant is considered and compared with~\cite{Eichhorn:2017eht}.
To this end, we set the gauge parameters as $\alpha=0$, $\beta=0$, $\rho_1=\rho_2=0$ and the coupling constants as $\tilde \lambda_2\to \infty$, $\tilde \xi_2\to 0$, $\tilde \lambda_4\to 0$, $\tilde \xi_0=1/(16\pi g)$, $\tilde \lambda_0={\Lambda_\text{cc}/(16\pi g)}$, $\tilde a= {\bar a}/(16\pi g)$ and $\tilde b = {\bar b}/(16\pi g)$, and then obtain
\al{
\p_t y=\beta_y&=
\frac{g {\tilde y}}{3360 \pi}
\Bigg(
\frac{5600 (3 {\bar b} +2)}{({\bar b} -2{\Lambda_\text{cc}}+1)^2}
-\frac{13440 (9 {\bar a} +3 {\bar b} -1)}{(18 {\bar a} +6 {\bar b} +4 {\Lambda_\text{cc}}-3)^2}\nn
&\quad-\frac{315 (216 {\bar a} +72 {\bar b} +16 {\Lambda_\text{cc}}-27)}{(18 {\bar a} +6 {\bar b} +4{\Lambda_\text{cc}}-3)^2}
   +\frac{576(630 {\bar a}+210{\bar b} +28 {\Lambda_\text{cc}}-75)}{(18{\bar a} +6 {\bar b} +4
   {\Lambda_\text{cc}}-3)^2}
   \Bigg).
   }
The first, second, third and last terms correspond to the contribution from the diagrams (I), (III)(IV), (VIII)(IX) and (XI)(XII) shown in Fig.~\ref{feynman diagrams of yukawa vertex}, respectively.
The first term being the contribution from the transverse graviton ($h^\perp_{\mu\nu}$) agrees with \cite{Eichhorn:2017eht}.
As the choice of gauge in \cite{Eichhorn:2017eht} differs from ours, the remaining terms depend on the gauge.
\subsection{Structures of beta functions and stability matrix}
As discussed in appendix~\ref{frgtreatment}, the fixed point $g_*$ is defined by $\beta_i\fn{g_*}=0$ for all coupling constants which span the truncated theory space.
Here we show the explicit beta functions of the gravitational coupling constants in the limits $\tilde \xi_2=\tilde \lambda_4=\tilde y=0$, with vanishing anomalous dimension (i.e., $\p_t g_i=0$ on the right-hand side), and $\alpha\to 0$ and $\beta=1$.
\al{
\p_t {\tilde \lambda}_0
&=-4 \tilde{\lambda }_0
+\frac{1-N_f}{8 \pi ^2}
+\frac{1}{32 \pi ^2(1+2 \tilde {\lambda}_2)}
+\frac{1}{48\pi^2}\frac{5(\tilde{b}+2 \tilde{\lambda}_0)}{\tilde{b}-\tilde{\lambda }_0+\tilde{\xi }_0}
-\frac{1}{24\pi^2}\frac{3\tilde{a}+\tilde{b}-\tilde{\lambda }_0}{-6\tilde{a}-2\tilde{b}-\tilde{\lambda }_0+\tilde{\xi }_0},\\
 \p_t {\tilde \xi}_0&=
 -2 \tilde{\xi }_0
-\frac{N_f-20}{48 \pi ^2}
-\frac{1}{96 \pi ^2 (1+2 \tilde{\lambda }_2)}
-\frac{1}{192\pi^2}
\Bigg(
\frac{5 (4 \tilde{a}-2 \tilde{b}-15 \tilde{\lambda}_0)}{\tilde{b}-\tilde{\lambda }_0+\tilde{\xi}_0}
  +\frac{10 (4\tilde{a}+3 \tilde{b}) (\tilde{b}+2
   \tilde{\lambda }_0)}{(\tilde{b}-\tilde{\lambda}_0+\tilde{\xi }_0)^2}\Bigg)\nn
&\qquad 
  -\frac{1}{196\pi^2}\Bigg(\frac{-12 \tilde{a}-4 \tilde{b} +3 \tilde{\lambda}_0}{-6 \tilde{a}-2 \tilde{b}- \tilde{\lambda }_0- \tilde{\xi }_0}
  +\frac{ 8(3 \tilde{a}+\tilde{b})( 3\tilde{a}+ \tilde{b} - \tilde{\lambda}_0)}{(-6 \tilde{a}-2 \tilde{b}- \tilde{\lambda }_0- \tilde{\xi }_0)^2}
 \Bigg),
\\
\p_t \tilde c
&=
-\frac{6N_f-997}{11520 \pi ^2}
+\frac{1}{1280 \pi ^2 (1+2 \tilde{\lambda }_2)}\nn
&\quad
   -\frac{180
   \tilde{a}+115 \tilde{b}-41 \tilde{\lambda }_0}{384 \pi ^2
   (\tilde{b}-\tilde{\lambda }_0+\tilde{\xi
   }_0)}
   +\frac{5 (-4 \tilde{a} \tilde{b}-76
   \tilde{a} \tilde{\lambda }_0+16 \tilde{a}^2-53 \tilde{b}
   \tilde{\lambda }_0-10 \tilde{b}^2)}{384 \pi ^2
   (\tilde{b}-\tilde{\lambda }_0+\tilde{\xi
   }_0)^2}
   +\frac{5 (4 \tilde{a}+3
   \tilde{b})^2 (\tilde{b}+2 \tilde{\lambda
   }_0)}{192 \pi ^2 (\tilde{b}-\tilde{\lambda}_0+\tilde{\xi }_0)^3}\nn
&\quad
   -\frac{210 \tilde{a}+70 \tilde{b}-13\tilde{\lambda }_0}{11520 \pi ^2 (-6 \tilde{a}-2\tilde{b}- \tilde{\lambda }_0+ \tilde{\xi}_0)}
  +\frac{(3 \tilde{a}+\tilde{b}) (12 \tilde{a}+4 \tilde{b}-3 \tilde{\lambda}_0)}{192 \pi ^2 (-6 \tilde{a}-2 \tilde{b}-\tilde{\lambda }_0+\tilde{\xi }_0)^2}
 -\frac{(3 \tilde{a}+\tilde{b})^2 (3 \tilde{a}+ \tilde{b}- \tilde{\lambda }_0)}{6 \pi ^2 (-6 \tilde{a}-2 \tilde{b}- \tilde{\lambda }_0+ \tilde{\xi }_0)^3}
 ,  \\
\p_t {\tilde b}&=
\frac{7 (N_f+118)}{1440 \pi ^2}
+\frac{1}{720 \pi ^2 (1+2 \tilde{\lambda }_2)}
+\frac{1}{9\pi^2}
\Bigg( 
 \frac{5\tilde{\lambda }_0}{\tilde{b}-\tilde{\lambda}_0+\tilde{\xi }_0}
\Bigg)
+\frac{1}{360\pi^2}
\left(
\frac{\tilde{\lambda }_0}{-6\tilde{a}-2 \tilde{b}-\tilde{\lambda}_0+ \tilde{\xi }_0}
   \right),
    }
where $\tilde c:={\tilde a} +\frac{\tilde b}{4}$.
The fixed point is given by solving the coupled equations $\p_t {\tilde \lambda}_0= \p_t {\tilde \xi}_0=\p_t \tilde c =\p_t {\tilde b}=0$.
For pure gravity the limits $\tilde \lambda_2\to \infty$ and $N_f\to 0$ have to be taken.
The numerical calculation is performed in the next section.

Next, we show the diagonal parts of the stability matrix in the matter sector at the Gaussian-matter fixed point,
\al{
\frac{\p \beta_{\tilde \lambda_2}}{\p \tilde \lambda_2}\bigg|_{\tilde \lambda_2\to 0}
&=-2 - \frac{1}{384 \pi^2}
\left(
   -\frac{40 (3\tilde{b}+2 \tilde{\xi}_0)}{(\tilde{b}-\tilde{\lambda }_0+\tilde{\xi}_0)^2}
+\frac{16 (9 \tilde{a}+3 \tilde{b}-\tilde{\xi}_0)}{(5 \tilde{a}+2 \tilde{b}+\tilde{\lambda}_0- \tilde{\xi }_0)^2}
\right),\\
\frac{\p \beta_{\tilde \xi_2}}{\p \tilde \xi_2}\bigg|_{\tilde \xi_2 \to 0}
&=-\frac{1}{1152 \pi^2}\Bigg(
\frac{30 \br{\tilde{\xi }_0 (-8 \tilde{a}+9 \tilde{b}+5
   \tilde{\lambda }_0)-2 (\tilde{b}+\tilde{\lambda}_0) (12 \tilde{a}+4 \tilde{b}+5 \tilde{\lambda}_0)+5 \tilde{\xi }_0^2}}{(\tilde{b}-\tilde{\lambda }_0+\tilde{\xi}_0)^3}\nn
&\quad   +\frac{6 \sqbr{\tilde{\xi }_0
   \br{\tilde{\lambda }_0-30 (3\tilde{a}+\tilde{b})}+2 (6 \tilde{a}+2 \tilde{b}-\tilde{\lambda }_0) (24 \tilde{a}+8\tilde{b}+\tilde{\lambda }_0)+ \tilde{\xi}_0^2}}{(6 \tilde{a}+2 \tilde{b}+  \tilde{\lambda }_0- \tilde{\xi}_0)^3}
\Bigg),\\
\frac{\p \beta_{\tilde \lambda_4}}{\p \tilde \lambda_4}\bigg|_{\tilde \lambda_4 \to 0}
&=\frac{1}{384 \pi^2}\left(
   \frac{40 (3\tilde{b}+2 \tilde{\xi}_0)}{(\tilde{b}-\tilde{\lambda }_0+\tilde{\xi}_0)^2}
-\frac{16 (9 \tilde{a}+3 \tilde{b}-\tilde{\xi}_0)}{(6 \tilde{a}+2\tilde{b}+\tilde{\lambda}_0-\tilde{\xi }_0)^2}
\right),
\label{criticalquartic}
\\
\frac{\p \beta_{\tilde y}}{\p \tilde y}\bigg|_{\tilde y \to 0}
&=\frac{1}{53760 \pi^2}
\Bigg(
\frac{5600 (3\tilde{b}+2 \tilde{\xi}_0)}{(\tilde{b}-\tilde{\lambda }_0+\tilde{\xi}_0)^2}
+\frac{3(32760 \tilde{a}+10920 \tilde{b}+1596 \tilde{\lambda }_0-4015 \tilde{\xi }_0)}{(6\tilde{a}+2 \tilde{b}+\tilde{\lambda }_0-\tilde{\xi}_0)^2} 
\Bigg),
\label{criticalyukawa}
}
where the matter coupling constants and $\p_t g_i$ are set to zero.\footnote{
In the next section, we numerically take into account the contributions of $\p_t g_i$.
}
The first term in the parentheses for each beta function corresponds to the transverse graviton loop contribution which is the physical mode and then is dominant.

Let us naively estimate the value of the critical exponent.
We will perform a numerical analysis in the next section.
In the beta function of the scalar mass, the first term is its canonical dimension and the scalar mass becomes relevant at the Gaussian fixed point where all coupling constant vanish $g_i^*=0$.
When we have $3\tilde{b}+2 \tilde{\xi}_0>0$ at a non-trivial fixed point, the transverse graviton loop contribution tends to make the critical exponent of the scalar mass negative.
The critical exponents for the quartic scalar and Yukawa coupling constants also tend to become negative due to the gravitational fluctuations.
On the other hand, the critical exponent for the non-minimal coupling constant tends to become positive.

\section{Numerical Analysis}\label{numericalan}
\subsection{Fixed point structure and critical exponent}
The fixed points and the critical exponents are investigated numerically.
In this section we employ the Landau gauge $\alpha=0$ and the other gauge parameters are set to $\beta=1$.
As will be seen, the beta functions do not depend on $\rho_1$ and $\rho_2$ in the Landau gauge.
We investigate the following cases:
\begin{itemize}
\item Einstein--Hilbert (EH) truncation; $g=\{ {\tilde \xi}_0, {\tilde \lambda}_0\}$,
\item EH $+R^2$; $g=\{ {\tilde \xi}_0, {\tilde \lambda}_0,{\tilde a}\}$,
\item EH $+R^2+R_{\mu\nu}R^{\mu\nu}$; $g=\{ {\tilde \xi}_0, {\tilde \lambda}_0,{\tilde a}, {\tilde b}\}$,
\item EH--scalar; $g=\{ {\tilde \xi}_0, {\tilde \lambda}_0, {\tilde \lambda_2}, {\tilde \xi}_2, {\tilde \lambda}_4\}$,
\item EH--Higgs-Yukawa (HY); $g=\{ {\tilde \xi}_0, {\tilde \lambda}_0, {\tilde \lambda_2}, {\tilde \xi}_2, {\tilde \lambda}_4,{\tilde y}\}$,
\item Full theory space~\eqref{originaleffectiveaction}; $g=\{ {\tilde \xi}_0, {\tilde \lambda}_0, {\tilde a}, {\tilde b}, {\tilde \lambda_2}, {\tilde \xi}_2, {\tilde \lambda}_4,{\tilde y}\}$.
\end{itemize} 
The values of fixed points in the these truncations are shown in table~\ref{FPalpha0}.
We find a Gaussian-matter fixed points for the system with matter, that is, the fixed points for matter couplings become the Gaussian fixed point $g_i^*=0$.
In the last row of table~\ref{FPalpha0}, we show the value of a combination ${\tilde \lambda}_0^*/ {\tilde \xi}_0^*{}^2$ which is dimensionless and then has a weak dependence pn cutoff and gauge choices.
The values of the critical exponents at the fixed point given in table~\ref{FPalpha0} are listed in table~\ref{criticalalpha0}.

\begin{figure}
\begin{center}
\includegraphics[width=10cm]{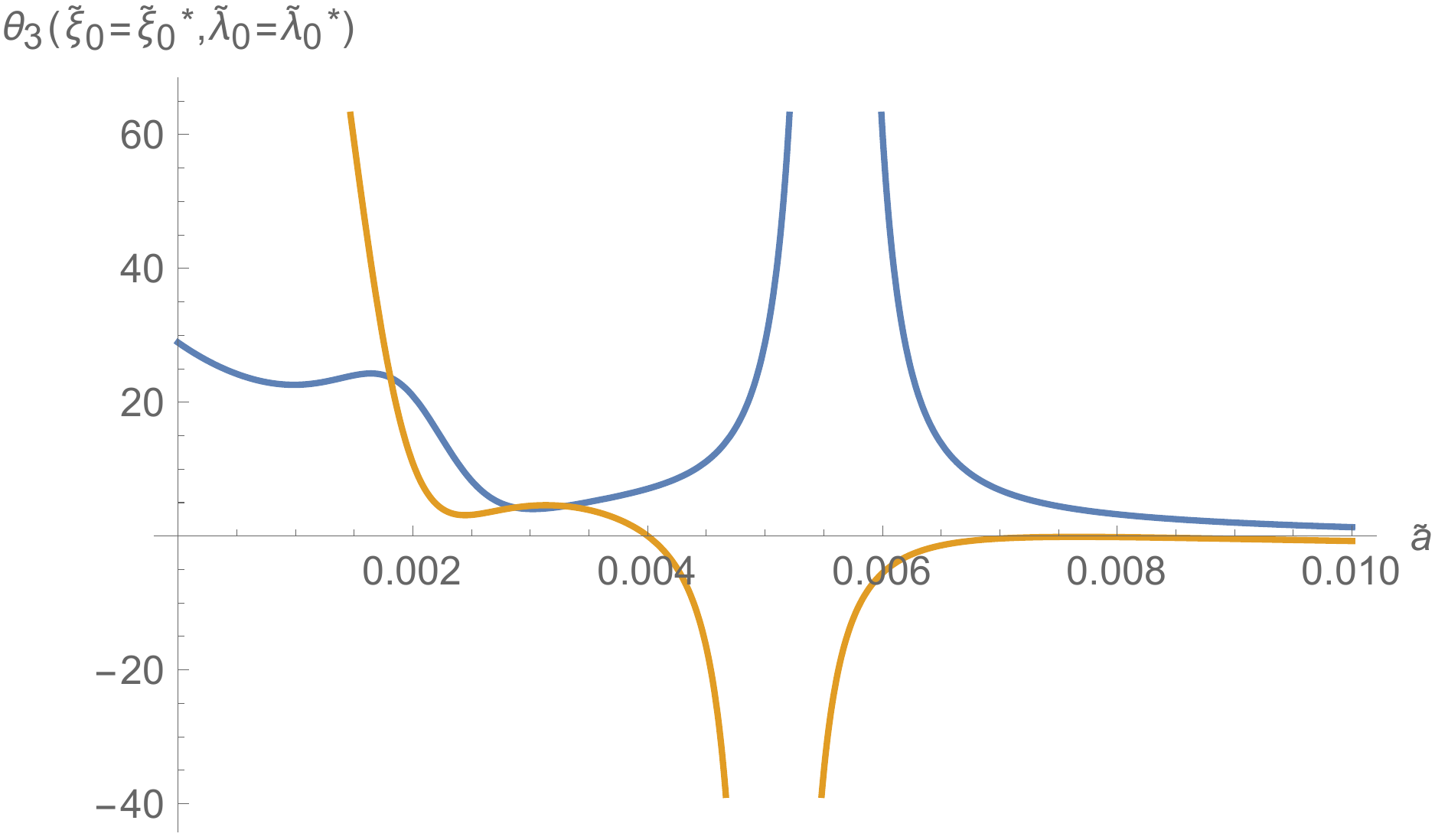}
\end{center}
\caption{The behavior of $\theta_3$ as a function of $\tilde a$. The blue and orange lines correspond to the ``EH $+R^2$" (i) and (ii), respectively. }
\label{theta3}
\end{figure}
For higher derivative pure gravity truncations, i.e, ``EH $+R^2$" and ``EH $+R^2+R_{\mu\nu}R^{\mu\nu}$", two fixed points are found. 
At the fixed point (i) in the EH $+R^2$ truncation, the critical exponents are positive and then all couplings are relevant.
This result agrees with~\cite{Lauscher:2002sq,Falls:2013bv}.
In contrast, at the fixed point (ii) one of the critical exponents becomes negative.
This is because its fixed point is located around a pole of $\theta_3$.
In Fig.~\ref{theta3}, we show the behavior of $\theta_3$ with $\tilde \xi_0=\tilde \xi_0^*$ and  $\tilde \lambda_0=\tilde \lambda_0^*$ as a function of $\tilde a$.
For EH $+R^2$ (i) and (ii), there is a pole at $\tilde a\simeq 0.00560$ and $\tilde a\simeq 0.00507$, respectively.
The value $\tilde a^*\simeq 0.004603$ in EH $+R^2$ (ii) is near the pole.
Therefore, the fixed point (ii) may be an artifact of the truncation of theory space.
The fixed point (ii) in the EH $+R^2+R_{\mu\nu}R^{\mu\nu}$ truncation may correspond to the result given in~\cite{Benedetti:2009rx} since the there are three positive critical exponents and one irrelevant one and they do not have imaginary part.
However, the value of $\theta_3$ is larger than that of \cite{Benedetti:2009rx}.
To see the stability of the critical exponents, we investigate their gauge dependence in the next subsection.

The result of the scalar--gravity system (``EH--scalar") agrees with~\cite{Percacci:2003jz}.
In the present work for the Higgs--Yukawa system (``EH--HY"), the scalar mass and the non-minimal coupling become relevant whereas the previous study~\cite{Oda:2015sma} reported that they become irrelevant.
This is because the gauge fixing and ghost actions~\eqref{gaugefixedaction} are different.
Nevertheless, the real parts of $\theta_5$ and $\theta_6$ become smaller than the ones in the scalar--gravity system.
Therefore, the fermionic fluctuation tends to make the critical exponents of the scalar mass and the non-minimal coupling small, but their magnitude depends on the gauge choice.
For the full theory space~\eqref{originaleffectiveaction}, we find three sets of fixed point for the present gauge parameters.
The cosmological constant $\tilde \lambda_0^*$ in one of them is negative (Full (iii)).
At this point we cannot conclude which fixed point is reasonable.

Here we consider the reason why the critical exponents obtained in ``Full (ii)" have large.
In the Landau gauge the propagator for the transverse and longitudinal gravitons has the following pole structure:
\al{
{\mathcal T}_\text{T}&:={\tilde \lambda _0- \tilde \xi _0-\tilde \kappa },&
{\mathcal T}_\text{L}&:={\tilde\lambda _0-\tilde\xi _0+6\tilde \omega+2\tilde \kappa},&
\label{pole structure}
}
respectively.
At the fixed point of ``Full (ii)", their values become  ${\mathcal T}_\text{T}\simeq 4.65\times 10^{-3}$ and ${\mathcal T}_\text{L}\simeq 4.78\times 10^{-5}$.
${\mathcal T}_\text{L}$ is smaller than ${\mathcal T}_\text{T}$ at the values of the fixed point.
In other words, the values of the fixed point are located at near the pole of the longitudinal graviton propagator.
In particular, it seems that the critical exponents for the quartic and the Yukawa couplings are affected by the values of the longitudinal graviton propagator. 
Note that at the fixed point of ``Full (i)", values of \eqref{pole structure} become ${\mathcal T}_\text{T}\simeq 2.73\times 10^{-3}$ and ${\mathcal T}_\text{L}\simeq 3.56\times 10^{-4}$.
Then we see that the critical exponents in ``Full (i)" tend to become smaller than the ones in ``Full (ii)".
The fixed point being close to the pole in the propagator is not suitable and indicate that this fixed point might be a gauge artifact.
The eigenvalues, i.e. the critical exponents, are shown in table.~\ref{FPalpha0}.

\begin{table}
  \begin{center}
    \begin{tabular}{|c||c|c|c|c|c|} \hline
Truncation & $\tilde \xi_0^*\times 10^{2}$ & $\tilde \lambda_0^*\times 10^{3}$ & $\tilde a^*\times 10^{2}$ & $\tilde b^*\times 10^{2}$ & ${\tilde \lambda}_0^*/ {\tilde \xi}_0^*{}^2$  \\ \hline \hline
EH & $2.529$ & $4.559$ & ---  & --- & $7.13$ \\
EH $+R^2$ (i) & $2.254$ & $3.392$ & $0.2312$  & --- & $6.68$ \\
EH $+R^2$ (ii) & $2.192$ & $5.800$ & $0.4603$  & --- & $12.1$ \\
EH $+R^2+R_{\mu\nu}^2$ (i) & $1.115$ & $4.331$ & $0.5021$ & $-1.109$ & $34.9$ \\
EH $+R^2+R_{\mu\nu}^2$ (ii) & $1.695$ & $5.666$ & $1.186$ & $-2.993$ & $19.7$ \\
EH $+R^2+R_{\mu\nu}^2$ (iii) & $2.690$ & $-9.349$ & $2.943$ & $-2.720$ & $-12.9$ \\ 
EH--scalar & $2.60$ & $5.76$ & ---  & --- & $8.53$ \\
EH--HY & $2.07$ & $1.17$ & ---  & --- & $2.74$ \\ 
Full (i) & $0.773$ & $2.74$ & $0.346$ & $-0.772$ & $45.9$ \\
Full (ii) & $0.973$ & $3.26$ & $0.479$ & $-1.11$ & $34.5$ \\
Full (iii) & $2.099$ & $-9.841$ & $2.512$ & $-2.141$ & $-22.3$\\ 
Full (iv) & $9.893\times10^{-4}$ & $-6.697$ & $4.771\times10^{-2}$ & $-2.427\times10^{-2}$ & $-6.842\times10^7$\\ 
\hline
    \end{tabular}
  \end{center}
\caption{The values of fixed points for $\alpha=0$ and $\beta=1$. ``EH", ``HY" and ``Full" denote the Einstein--Hilbert truncation, the Higgs-Yukawa term and the theory space \eqref{originaleffectiveaction}, respectively.}
\label{FPalpha0}
\end{table}

\begin{table}
  \begin{center}
    \begin{tabular}{|c||c|c|c|c|c|c|c|c|} \hline
      Truncation & $\theta_1$ & $\theta_2$  & $\theta_3$ & $\theta_4$ & $\theta_5$  & $\theta_6$ & $\theta_7$ & $\theta_8$ \\ \hline \hline
EH & $2.74 + 1.43i$ & $\theta_1^*$ & ---  & --- &  ---   &  ---   & ---  &  ---   \\
EH $+R^2$ (i)  & $2.54 + 1.43i$ & $\theta_1^*$ & $12.2$ & --- & --- & --- & --- & --- \\
EH $+R^2$ (ii)  & $2.70 + 1.75i$ & $\theta_1^*$ & $-36.8$ & --- & --- & --- & --- & --- \\
EH $+R^2+R_{\mu\nu}^2$ (i) & $2.95 + 1.06i$  & $\theta_1^*$ & $8.76 + 1.61i$ & $\theta_3^*$ &  --- & --- & --- & ---\\
EH $+R^2+R_{\mu\nu}^2$ (ii) & $2.08$  & $8.12$ & $72.1$ & $-2.52$ &  --- & --- & --- & --- \\
EH $+R^2+R_{\mu\nu}^2$ (iii) & $3.63+ 4.72i$ & $\theta_1^*$  & $2.04$ & $44.0$ & --- & ---& --- & ---\\
EH--scalar & $2.77+ 1.56i$ & $\theta_1^*$ & --- & --- & $0.766 + 1.56i$ & $\theta_5^*$ & $-1.61$ & ---\\
EH--HY & $2.63 + 1.24i$ & $\theta_1^*$ & --- & --- & $0.631 + 1.24i$ & $\theta_5^*$ & $-1.38$ & $-0.746$ \\
Full (i) & $3.54 + 1.70i$ & $\theta_1^*$ & $5.69$ & $15.7$ & $0.622 + 4.34i$ & $\theta_5^*$  & $20.4$ & $-26.6$ \\
Full (ii) & $4.03$ & $7.92$ & $81.7$ & $-2.07$ & $1.92 + 3.33i$ & $\theta_5^*$  & $84.0$ & $-1405$ \\
Full (iii) & $4.09+ 3.84i$ & $\theta_1^*$ & $1.95$ & $34.6$ & $13.6$ & $0.487$  & $2.74$ & $2.49$ \\
Full (iv) & $2.32$ & $4.12$ & $4.74$ & $12.8$ & $2.72$ & $-90.3$  & $0.980$ & $-4.76\times10^2$ \\ \hline
\end{tabular}
\end{center}
\caption{The values of critical exponents for $\alpha=0$ and $\beta=1$. ``EH", ``HY" and ``Full" denote the Einstein--Hilbert truncation, the Higgs-Yukawa term and the theory space \eqref{originaleffectiveaction}, respectively.}
\label{criticalalpha0}
\end{table}

\subsection{Gauge dependence}
In the previous subsection, we have found sets of non-trivial UV fixed point.
Here we investigate the gauge dependence for the fixed point and the critical exponent.
If the fixed point is not an artifact of the approximation, the structure of the fixed point should be stable against the change of gauge parameters.

Let us first explicitly look at the gauge dependence of the beta functions in the Landau gauge $\alpha\to 0$.
We see that in the Wetterich equation $\mathcal X$ and ${\mathcal Y}$ defined in \eqref{xmath} and \eqref{ymath} within the spin 1 and 0 fields cancel out:
\al{
{\mathcal A}&:=\frac{1}{2}{\rm Tr}\left. \frac{\p_t{\mathcal R}_k}{\Gamma_k^{(1,1)}
									+{\mathcal R}_k}\right|_{\xi \xi}
+\frac{1}{2}{\rm Tr}\left. \frac{\p_t{\mathcal R}_k}{\Gamma _k^{(1,1)}
									+{\mathcal R}_k}\right|_{ B^\perp B^\perp}
-\left. \Tr \frac{\p_t {\mathcal R} _k}{\Gamma_k^{(1,1)}+{\mathcal R}_{k}}\right|_{{\bar C}^\perp C^\perp}\nn
&={\tilde \p}_t
\Bigg[
\frac{1}{2}{\rm Tr}\log \fn{\Gamma_k^{(1,1)}+{\mathcal R}_k} \bigg|_{\xi \xi}
+\frac{1}{2}{\rm Tr}\log \fn{{\Gamma _k^{(1,1)}+{\mathcal R}_k}}\bigg|_{ B^\perp B^\perp}
-{\rm Tr}\log \fn{{\Gamma_k^{(1,1)}+{\mathcal R}_{k}}}\bigg|_{{\bar C}^\perp C^\perp}
\Bigg]\nn
&={\tilde \p}_t
\Bigg[
\frac{1}{2}{\rm Tr}\log \fn{{\mathcal X}\fn{P_k}\left( P_k\fn{\bar\Delta_{L1}}-\frac{\bar R}{2}\right)}
+\frac{1}{2}{\rm Tr}\log \fn{{\mathcal X}\fn{P_k}} \nn
&\quad
-{\rm Tr}\log \fn{{\mathcal X}\fn{P_k}\left( P_k\fn{\bar\Delta_{L1}}-\frac{\bar R}{2}\right)}
\Bigg]_\text{spin 1}\nn
&=
-\frac{1}{2}{\tilde \p}_t{\rm Tr}\log \fn{ P_k\fn{\bar\Delta_{L1}}-\frac{\bar R}{2}},
\label{spin 1 contributions}
}
and
\al{
{\mathcal B}&:=\frac{1}{2}{\rm Tr}\left. \frac{\p_t{\mathcal R}_k}{\Gamma _k^{(1,1)}
									+{\mathcal R}_k}\right|_{\rm SS}
+\frac{1}{2}{\rm Tr}\left. \frac{\p_t{\mathcal R}_k}{\Gamma _k^{(1,1)}
								+{\mathcal R}_k}\right|_{BB}	
-\left. \Tr \frac{\p_t {\mathcal R} _k}{\Gamma_k^{(1,1)}+{\mathcal R}_{k}}\right|_{\bar C C}	\nn
&={\tilde \p}_t
\Bigg[
\frac{1}{2}{\rm Tr} \log \fn{{\Gamma _k^{(1,1)}+{\mathcal R}_k}}\bigg|_{\rm SS}
+\frac{1}{2}{\rm Tr}\log\fn{{\Gamma _k^{(1,1)}+{\mathcal R}_k}}\bigg|_{BB}	
-\Tr\log\fn{{\Gamma_k^{(1,1)}+{\mathcal R}_{k}}}\bigg|_{\bar C C}
\Bigg]\nn
&= {\tilde \p}_t
\Bigg[
\frac{1}{2}\br{{\rm Tr} 
\log\fn{{\mathcal Y}\fn{P_k}\left( \sqrt{P_k\fn{\bar \Delta_{L0}}-\frac{\bar R}{3}} +\beta\sqrt{P_k\fn{\bar\Delta_{L0}}} \right)}+...}\nn
&\quad
+\frac{1}{2}{\rm Tr}\log\fn{{\mathcal Y}\fn{P_k}}
-\Tr\log\fn{{\mathcal Y}\fn{P_k}\left(P_k\fn{\bar \Delta_{L0}}-\frac{\bar R}{3-\beta}\right)}
\Bigg]_{\text{spin 0}}\nn
&={\tilde \p}_t
\Bigg[
\frac{1}{2}{\rm Tr} \log\left( \sqrt{P_k\fn{\bar \Delta_{L0}}-\frac{\bar R}{3}} +\beta\sqrt{P_k\fn{\bar \Delta_{L0}}} \right)
\left( P_k\fn{\bar \Delta_{L0}}-\frac{\bar R}{3-\beta}\right)^{-2}+...
\Bigg],
}
where $...$ represents the contribution which does not depend on $\rho_1, \rho_2$, but depends on $\beta$. 
That is, the dependences of the gauge parameters $\rho_1$ and $\rho_2$ do not appear and there is only the $\beta$ dependence in the spin 0 contributions.
Setting $V\fn{\phi^2}=\xi_2=y=a=b=0$, the corrections to the operators $V\fn{\phi^2}$ and $\bar R_{\mu\nu\rho\sigma}\bar R^{\mu\nu\rho\sigma}$ do not depend on the gauge parameter $\beta$.
For $\beta=0$, $h$ disappears in the gauge fixing action \eqref{actionrewritten}.
Alternatively, taking $\beta\to \pm \infty$, $\sigma$ disappears, which is called ``unimodular physical gauge"~\cite{Percacci:2015wwa}.
Besides, it has been discussed in \cite{Wetterich:2016qee,Wetterich:2016ewc,Wetterich:2016vxu} that the choice $\beta=-1$ is ``physical gauge fixing".

\begin{figure}
\begin{center}
\includegraphics[width=8cm]{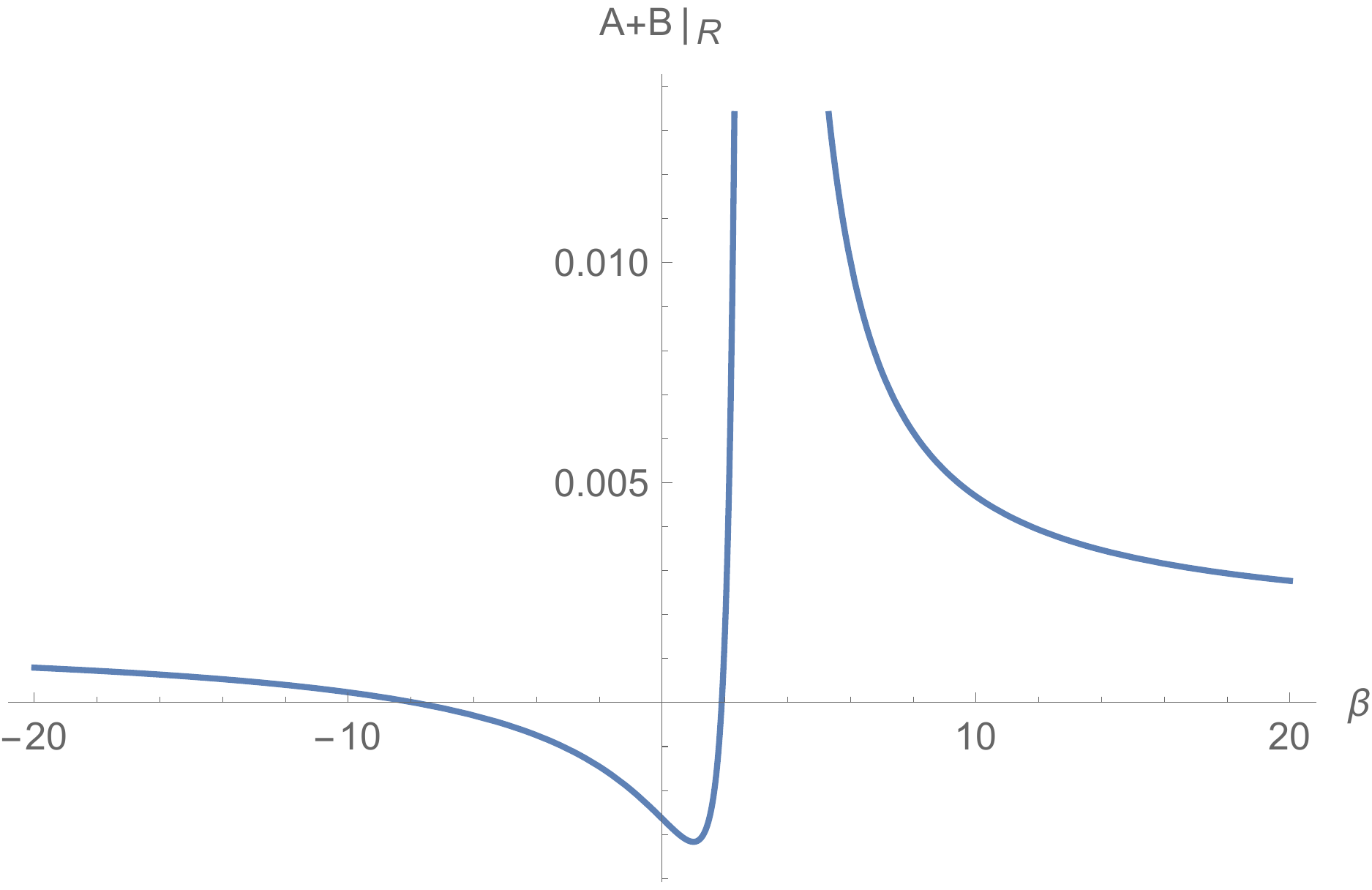}
\includegraphics[width=8cm]{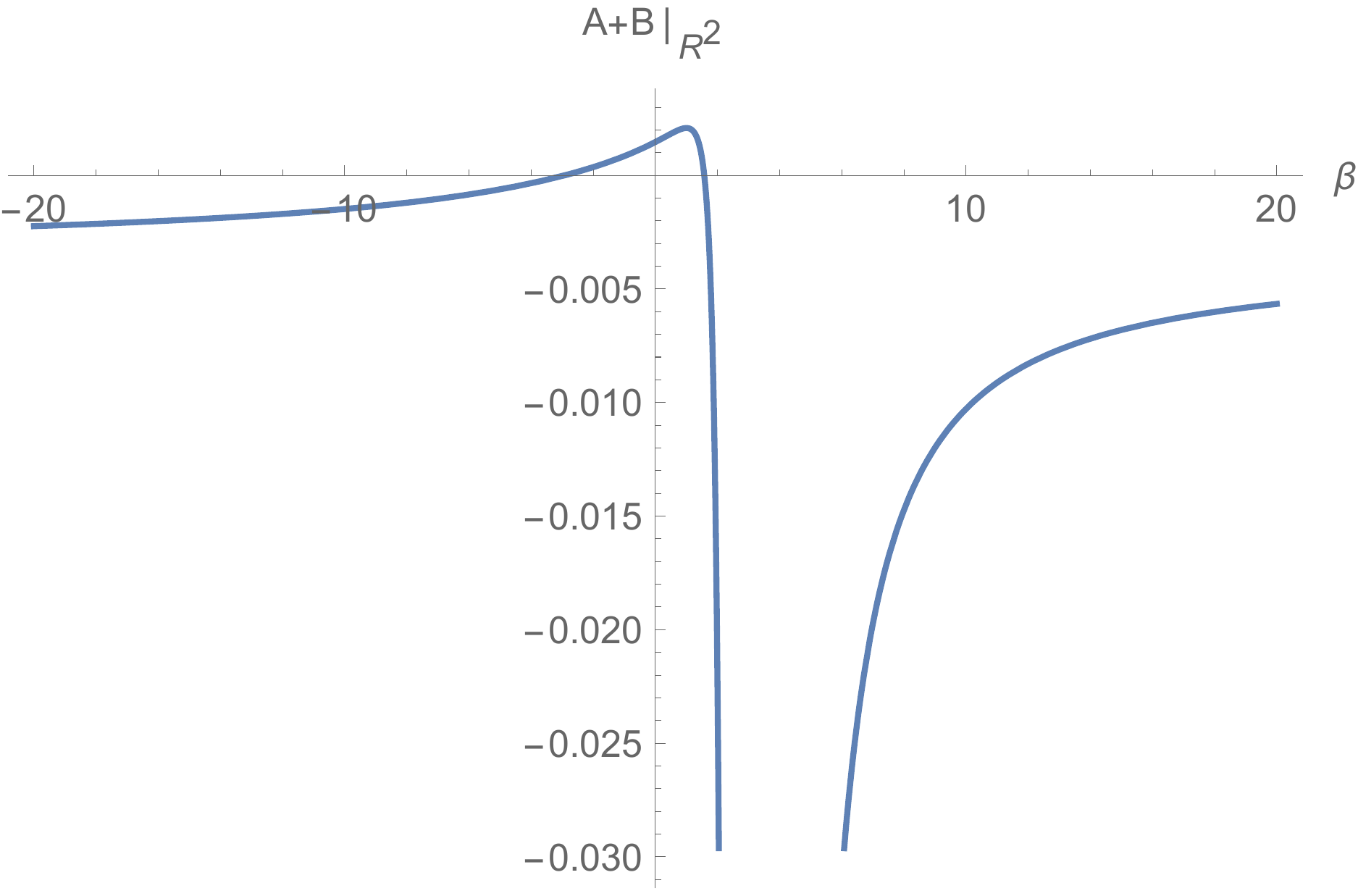}
\end{center}
\caption{$\beta$-dependence of \eqref{depA} (left) and \eqref{depB} (right).}
\label{deps}
\end{figure}
The corrections to the operators $\bar R$ and $\bar R^2$ depend on $\beta$ and become 
\al{
{\mathcal A}+{\mathcal B}\bigg|_{\bar R}
&=\frac{\beta ^2+6 \beta -15}{64 \pi ^2(\beta -3)^2},
\label{depA}
\\
{\mathcal A}+{\mathcal B}\bigg|_{\bar R^2}
&=\frac{131 \beta ^4-372 \beta ^3-366 \beta ^2+3852 \beta-4509}{3840 \pi ^2 (\beta -3)^4  }.
\label{depB}
}
Their gauge dependences on $\beta$ are shown in Fig.~\ref{deps}.
One can see that there is pole at $\beta=3$~\cite{Eichhorn:2016esv}.
The gauge parameters have to avoid to be chosen values near the poles.

The numerical values of the transverse graviton loop contributions are
\al{
{\mathcal S}^\text{TT}_F&:= -\frac{25}{96\pi^2}\approx -0.0263857,
\label{physicV}
\\
{\mathcal S}^\text{TT}_c&:= -\frac{41}{768\pi^2}\approx -0.00540907.
\label{physicC}
} 
The contributions with the gauge dependences should not be larger than \eqref{physicV} and \eqref{physicC}.
We see that ${\mathcal A}\simeq {\mathcal S}^\text{TT}_F$ at $\beta\simeq 2.4$ and ${\mathcal B}\simeq {\mathcal S}^\text{TT}_c$ at $\beta\simeq 1.7$.
Then, $\beta<1.5$ should be chosen. 

\begin{figure}
\begin{center}
\includegraphics[width=8cm]{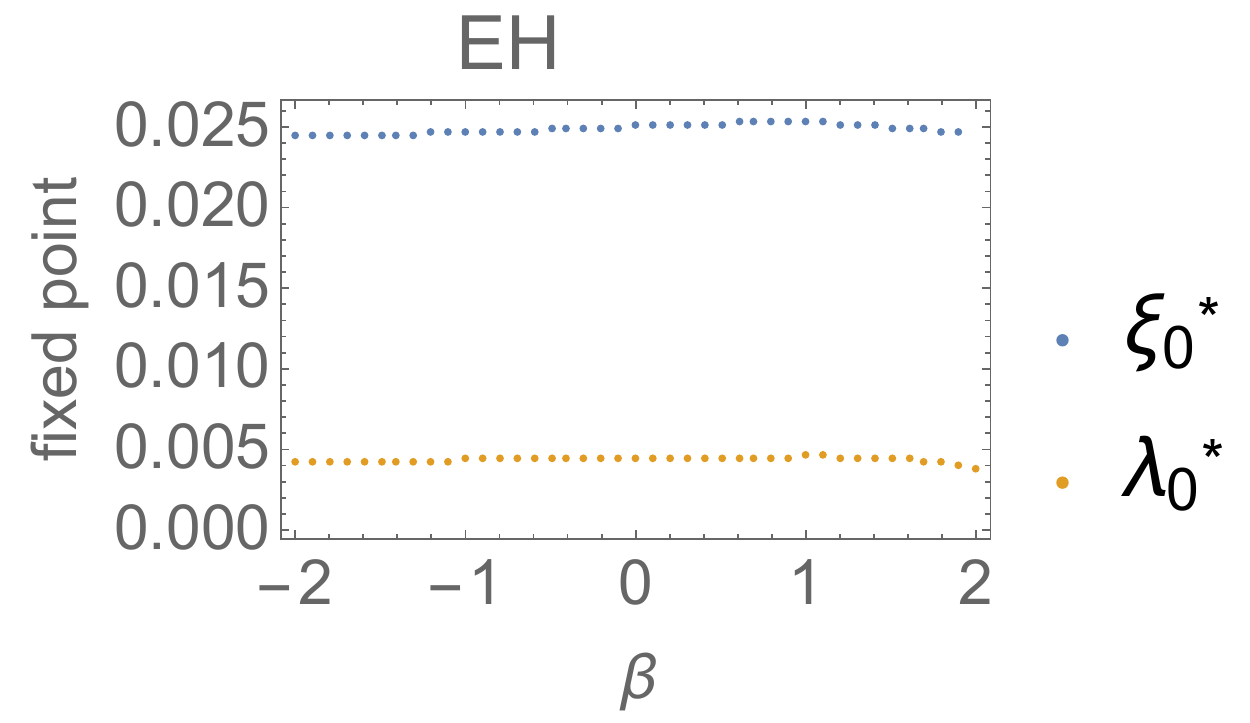}
\includegraphics[width=8cm]{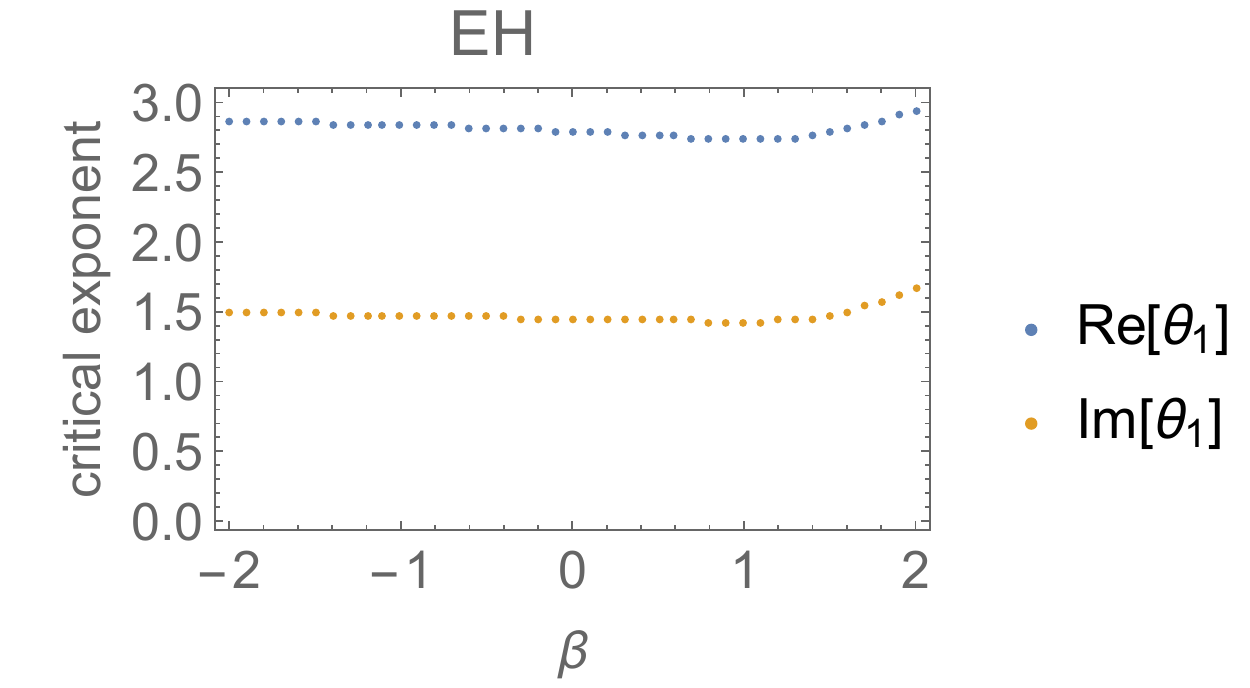}
\end{center}
\caption{The dependences of the stable fixed point and the stable critical exponent on $\beta$ in the ``EH" truncation.}
\label{fpcrpureEH}
\end{figure}

\begin{figure}
\begin{center}
\includegraphics[width=8cm]{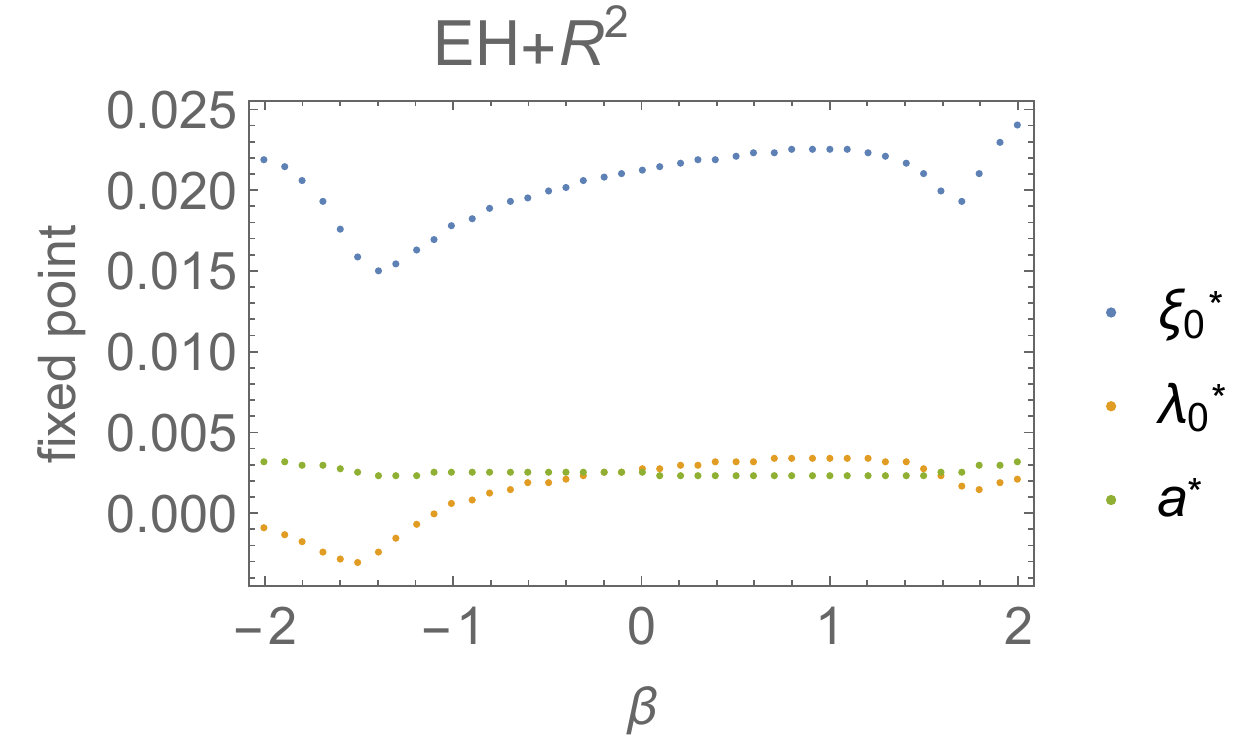}
\includegraphics[width=8cm]{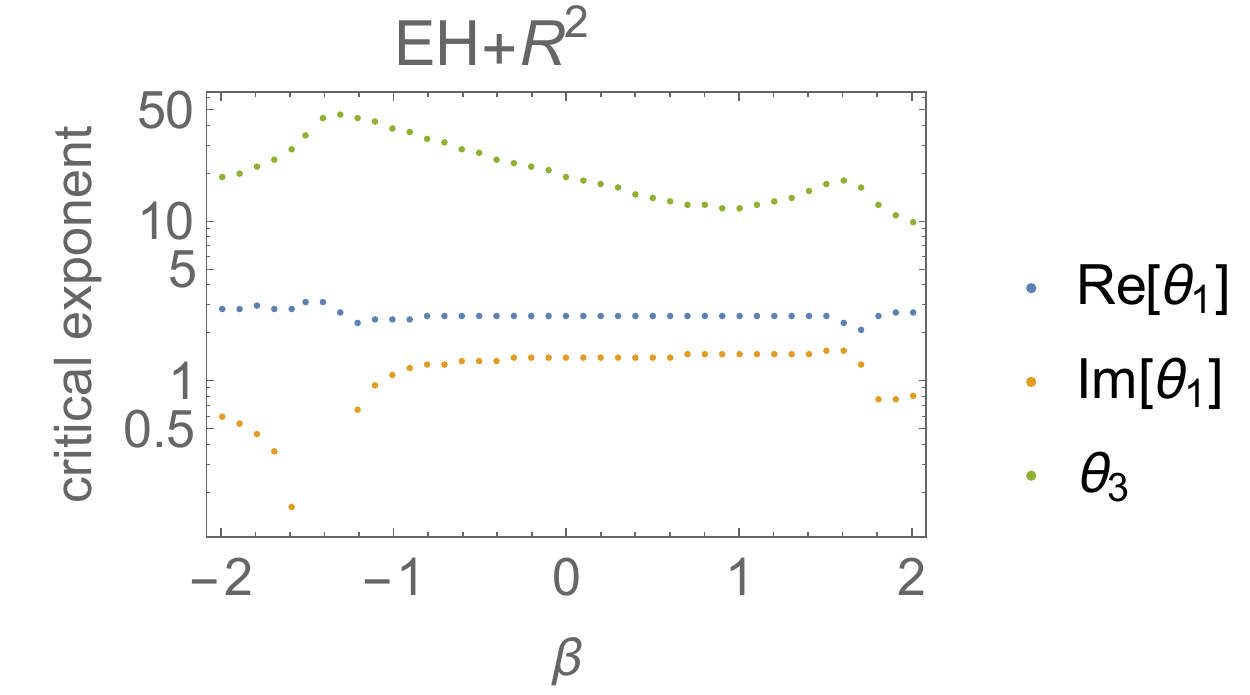}
\end{center}
\caption{The dependences of the stable fixed point and the stable critical exponent on $\beta$ in the ``EH $+R^2$" truncation.}
\label{fpcrpureR2}
\end{figure}

\begin{figure}
\begin{center}
\includegraphics[width=8cm]{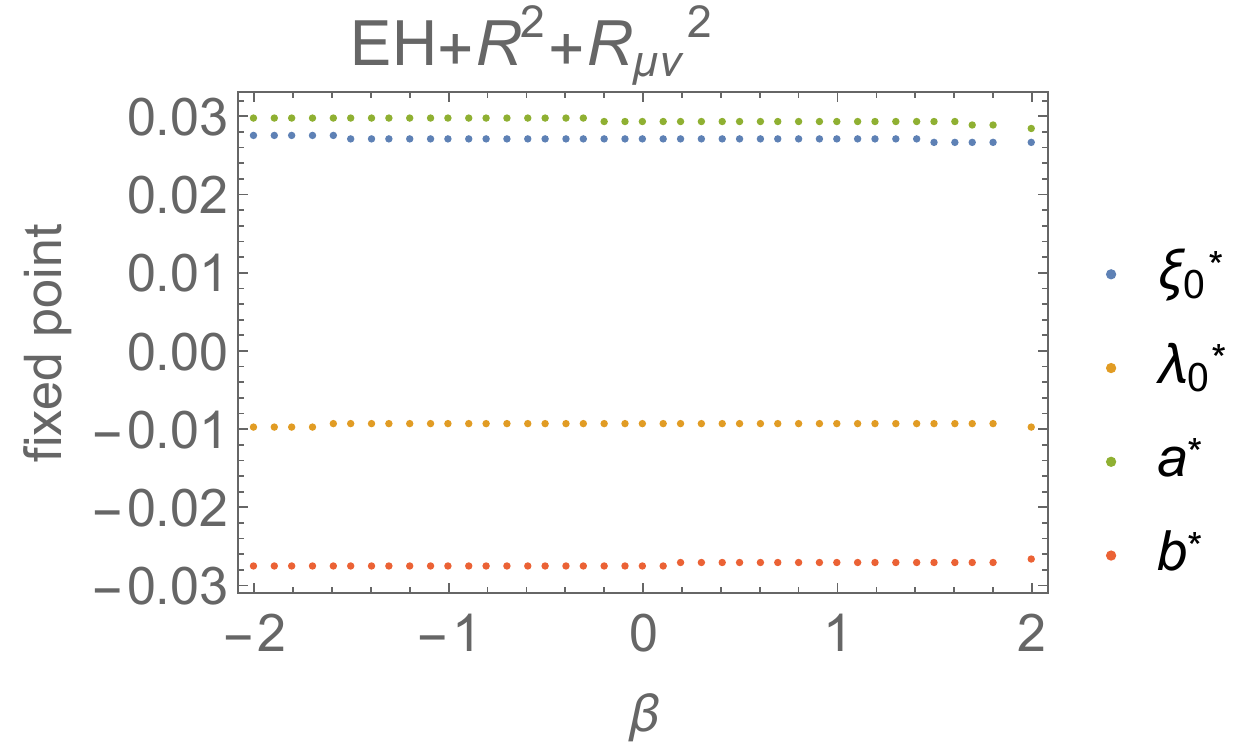}
\includegraphics[width=8cm]{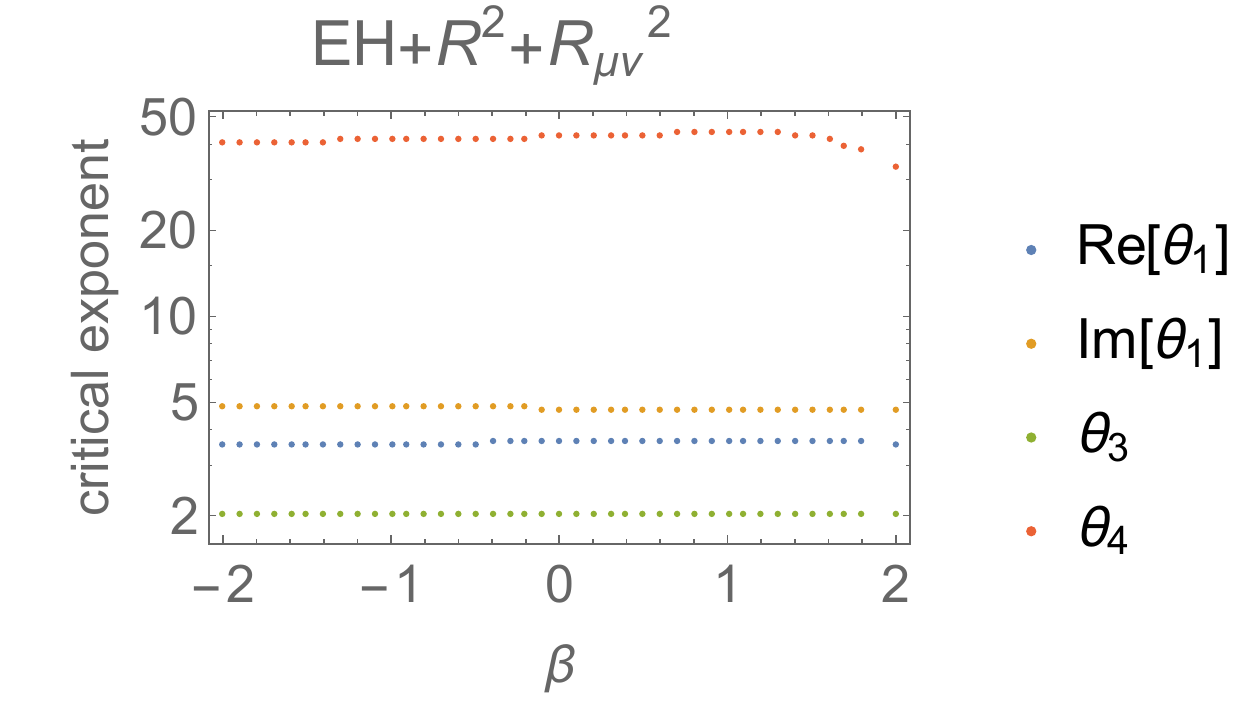}
\end{center}
\caption{The dependences of the stable fixed point and the stable critical exponent on $\beta$ in the ``EH $+R^2+R_{\mu\nu}^2$" truncation.}
\label{fpcrpureRRmunu}
\end{figure}

\begin{figure}
\begin{center}
\includegraphics[width=8cm]{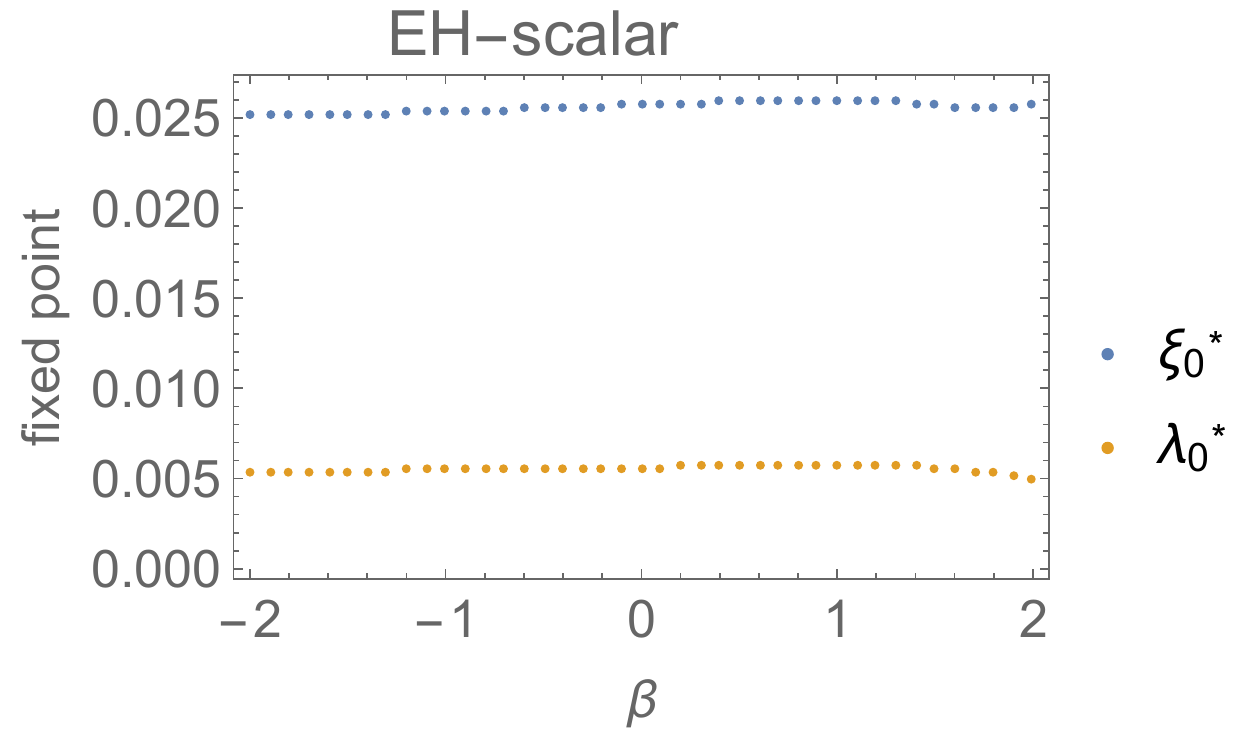}
\includegraphics[width=8cm]{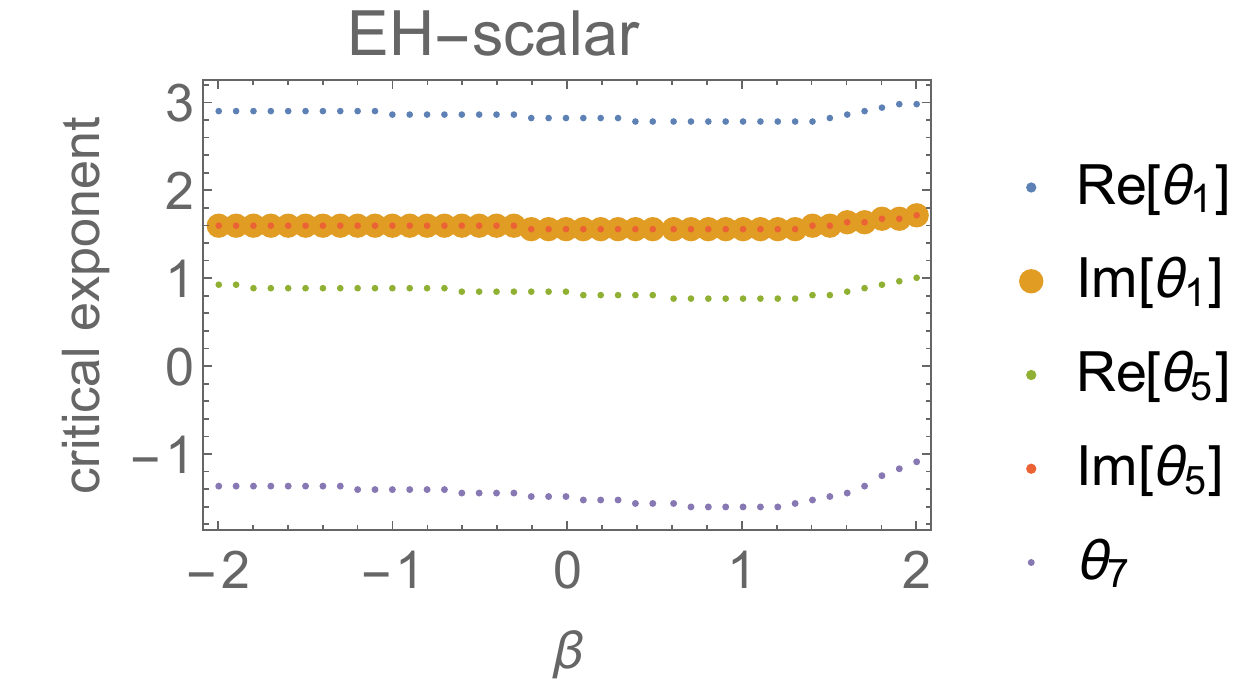}
\end{center}
\caption{The dependences of the stable fixed point and the stable critical exponent on $\beta$ in the ``EH--scalar" truncation.}
\label{fpcrEHscalar}
\end{figure}

\begin{figure}
\begin{center}
\includegraphics[width=8cm]{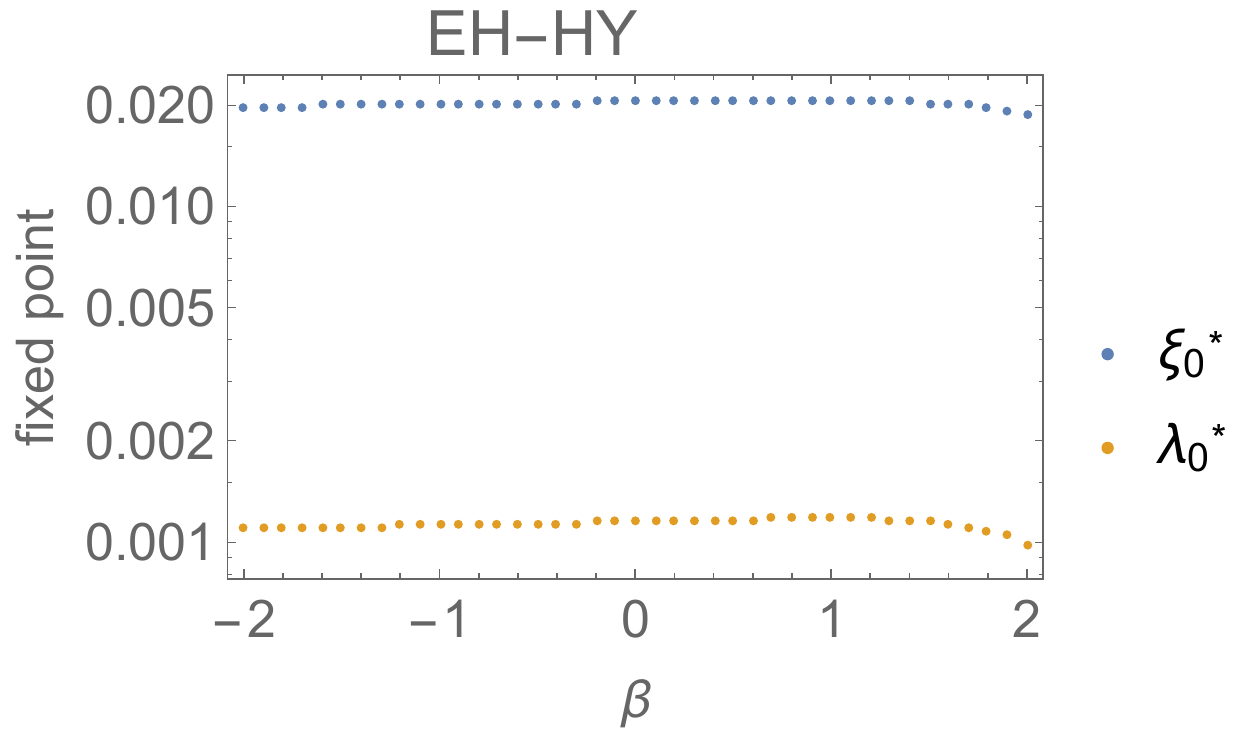}
\includegraphics[width=8cm]{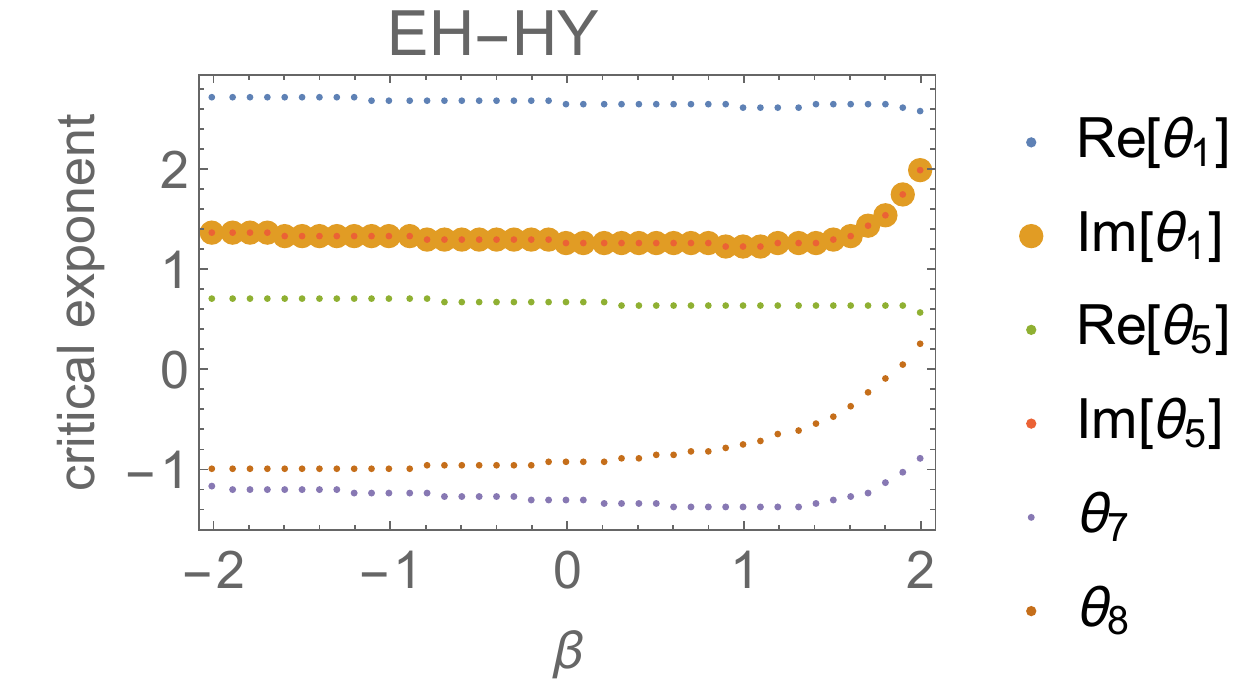}
\end{center}
\caption{The dependences of the stable fixed point and the stable critical exponent on $\beta$ in the ``EH--HY" truncation.}
\label{fpcrEHHY}
\end{figure}

\begin{figure}
\begin{center}
\includegraphics[width=8cm]{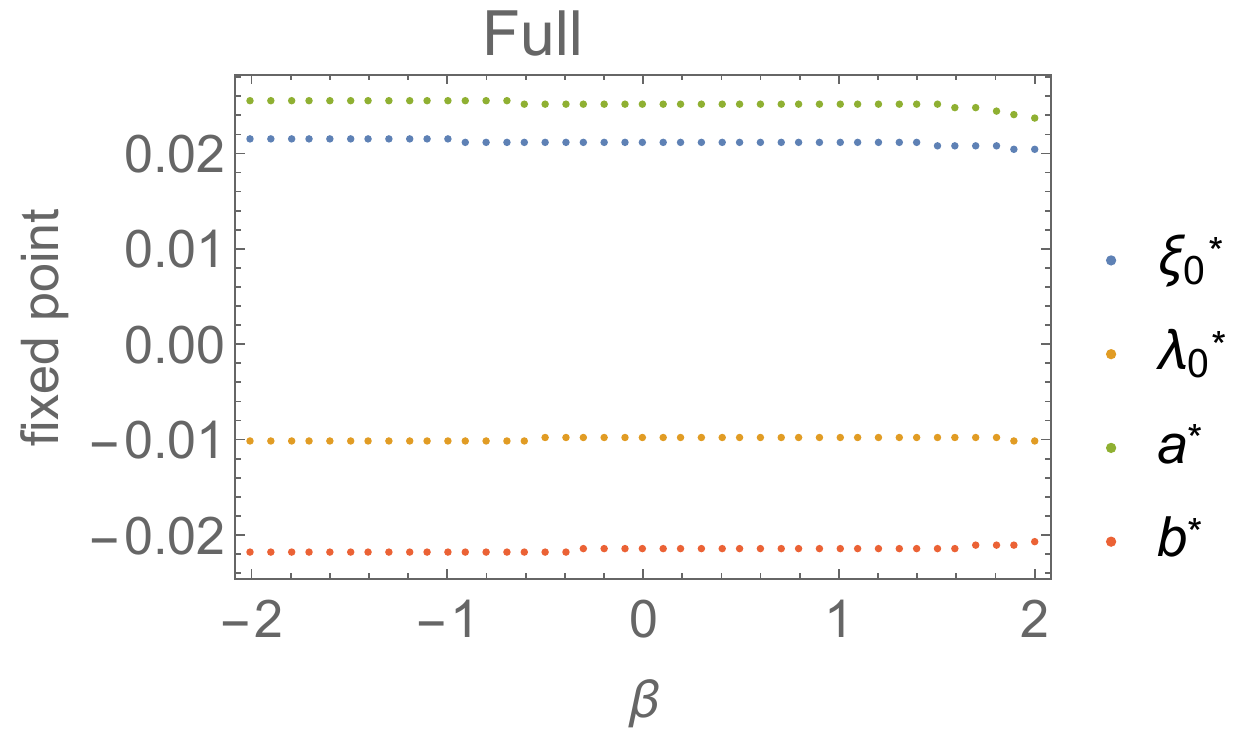}
\includegraphics[width=8cm]{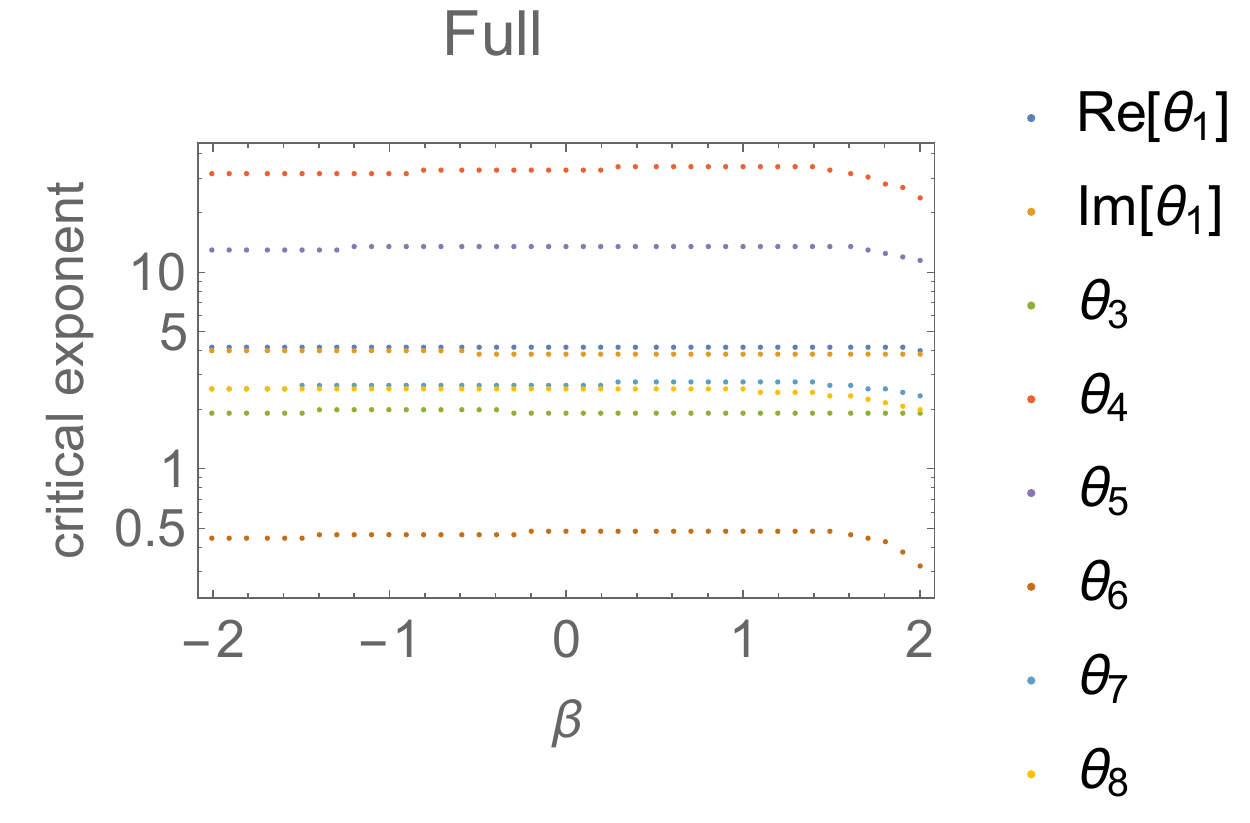}
\end{center}
\caption{The dependences of the stable fixed point and the stable critical exponent on $\beta$ in the ``full" theory space.}
\label{fpcrfull}
\end{figure}

Next, we numerically investigate the gauge dependences of the fixed point and the critical exponent.
First, we vary the value of $\beta$.
In Fig.~\ref{fpcrpureEH}--\ref{fpcrfull}, we show the stable fixed points and the critical exponents for the each truncation.
The fixed point structures for $\beta=0, -1,\pm2$ are summarized in table~\ref{FPbeta0}, \ref{criticalbeta0}, \ref{FPbeta-1}, \ref{criticalbeta-1} \ref{FPbeta2}, \ref{criticalbeta2}, \ref{FPbeta-2} and \ref{criticalbeta-2} in Appendix~\ref{FPCX}.

One can see that the result of EH is stable under the change of $\beta$.
Both the cosmological and Newton constants are relevant.
The other solutions with $\beta=-1$ in table~\ref{FPbeta-1} are artifact.
Similarly, EH$+R^2$ has a rather stable fixed point where all three couplings are relevant.
The stable fixed point with four relevant couplings is found in the case of EH$+R^2+R_{\mu\nu}^2$.
However, this feature is not consistent with Refs.~\cite{Benedetti:2009rx,Benedetti:2009gn}, where one of the couplings is irrelevant.
This disagreement would come from the difference of the treatment of ghost action:
The ghost action~\eqref{gaugefixedaction} is based on the BRST formalism, whereas \cite{Benedetti:2009rx,Benedetti:2009gn} employ the Nielsen-Kallosh ghost. 
Besides, Refs.~\cite{Benedetti:2009rx,Benedetti:2009gn} imposes a ``mode by mode" cancelation between the gauge-degrees of freedom in the metric and the ghost sector.

One of the critical exponents in gravity sector has large value ($\theta_4 \sim 30$), which means that the system has to be extended.

As for EH--scalar and EH--HY, we can find a stable fixed points.
The quartic coupling of the scalar is irrelevant in EH--scalar, and the quartic and Yukawa couplings are irrelevant in EH--HY. 
Although the Yukawa coupling becomes relevant for $\beta=2$, this may not be reliable because this point is close to the pole of ghost propagator at $\beta=3$, see \eqref{depA}, \eqref{depB} and \eqref{Eq:C contribution}.
Note that the value of $\theta_1, \theta_2, \theta_5, \theta_6$ decreases in EH--HY compared with EH--scalar. This can be interpreted as the effect of the fermion loop. 
In the ``Full" result, the number of sets of fixed points changes while varying the gauge parameter $\beta$.
The fixed point, at which all coupling constants become relevant, is stable for varying the gauge parameter $\beta$. 
Therefore, this fixed point is suitable for the present truncated system; see Fig.~\ref{fpcrpureRRmunu}.
In this case, all couplings of the gravity sector become relevant as with the EH$+R^2+R_{\mu\nu}^2$ truncation.

In the ``Full" result with stable critical exponents, the quartic and the Yukawa coupling become relevant.
This is because of the large negative fixed point value of $b$.
See Fig.~\ref{fpcrfull} where the dependences of the stable fixed point and the stable critical exponent on $\beta$ is shown.
For the critical exponents of the quartic coupling~\eqref{criticalquartic} the Yukawa coupling~\eqref{criticalyukawa}, the main contribution comes from the transverse and traceless mode which has $3{\tilde b}+2{\tilde \xi_0}$ in the numerator.
Substituting the values of fixed point given in e.g., table~\ref{FPalpha0}, we have $3{\tilde b}+2{\tilde \xi_0} \approx -0.02225$.
Then the sign of their critical exponents changes to positive.

To compare with~\cite{Eichhorn:2017eht}, we calculate
\al{
g&=\frac{1}{16\pi \tilde \xi_0}\approx 0.947802,&
\Lambda_\text{cc}&=16\pi \tilde \xi_0 \tilde \lambda_0 \approx -0.234421,&
\label{aaron1}
\\
{\bar a}&=16\pi \tilde \xi_0 \tilde a\approx 1.20105,&
{\bar b}&=16\pi \tilde \xi_0 \tilde b \approx -1.02001,&
\label{aaron2}
}
where we used the fixed point values for $\beta=0$ given in table~\ref{FPbeta0}.
Note that although we have used the different gauge fixing and ghost actions from~\cite{Eichhorn:2017eht}, the reason, why the critical exponent of the Yukawa coupling becomes positive, is same.
Since the dominant effects come from the transverse and traceless tensor mode which is independent of the gauge parameters, it is expected that there are no major differences. 
The values \eqref{aaron1} and \eqref{aaron2} are actually consistent with the region where the critical exponent of the Yukawa coupling become relevant; see~\cite{Eichhorn:2017eht}.

\section{Summary and Discussion}\label{Summary and Discussion}
In this paper we have studied higher derivative gravity coupled without and with matter fields.
In particular, for the matter theory, the Higgs-Yukawa model has been employed.
For pure gravity, the scalar--gravity and the Higgs-Yukawa system with Einstein-Hilbert (EH) term, we have obtained results which are consistent with previous studies.
In higher derivative gravity, we find several non-trivial fixed points.
For the EH$+R^2$ truncation one of them is consistent with the previous studies~\cite{Lauscher:2002sq,Falls:2013bv}.
For the EH$+R^2+R_{\mu\nu}^2$ truncation one features only positive critical exponents and the other three positive and one negative critical exponent.
In the ``Full theory space" (``Full" truncation) spanned by eight couplings, several possible fixed points have been found.

To see the stability of the critical exponents, we have investigated the gauge dependence.
We have found that in the Landau gauge ($\alpha=0$) the beta functions do not depend on $\rho_1$ and $\rho_2$. 
In the systems with the EH truncation, the values of the critical exponent are stable under variation of  the gauge parameters.
In contrast, for higher derivative gravity we find a fixed point at which all critical exponents become stable and positive.
We could conclude that pure asymptotically safe gravity has four relevant directions although this is not in agreement with the previous study~\cite{Benedetti:2009rx}.
However, the study using the vertex expansion suggests that the higher derivative gravity would have two relevant directions~\cite{Christiansen:2016sjn}.
These facts indicate that the order of the truncation has to be improved.
For the full truncation, there is a fixed point where all coupling constants become relevant.
Since this result is stable under variation of the gauge parameters, we conclude that this fixed point appears to be reliable.
However, there is no irrelevant direction and $\theta_4$ is large.
We have to include higher dimensional operators and investigate the stability of critical exponents.

In the ``Full" truncation, the Yukawa coupling becomes relevant.
This is a desired result for the asymptotic safety scenario.
The previous studies~\cite{Zanusso:2009bs,Oda:2015sma,Eichhorn:2016esv} have reported that the Yukawa coupling becomes irrelevant at the non-trivial fixed point of gravitational couplings.
This means, however, that fermions cannot have masses at low energy since the Yukawa interaction is prohibited by chiral symmetry.
It is inconsistent with our universe.
In order to confirm the relevance of the Yukawa coupling, the theory space has to be extended.

Finally, we would like to comment on the gauge hierarchy problem since asymptotically safe gravity could solve this problem~\cite{Wetterich:2016uxm}.
First of all, let us consider the meaning of the quadratic divergence $k^2$.
In the renormalization procedure the quadratic divergence strongly depends on the cutoff scheme.
When dimensional regularization is used, the quadratic divergence actually does not appear.
Even if another regularization which generates the quadratic divergence is used, it is subtracted by renormalization.
Therefore, we may be able to conclude that the quadratic divergence is meaningless.\footnote{
In viewpoint of renormalization group, the quadratic divergence determines the position of the second-order phase boundary~\cite{Wetterich:1981ir,Wetterich:1983bi,Wetterich:1987az,Wetterich:2011aa,Aoki:2012xs}.
}

In this point of view, the dimensionless scalar mass ${\bar m}=m/k$ is given by
\al{
{\bar m}^2\fn{k}={\bar m}^2\fn{M_\text{pl}}\left( \frac{k}{M_\text{pl}} \right)^{-\theta_m},
}
where ${\bar m}^2\fn{M_\text{pl}}$ is the scalar mass given at $M_\text{pl}$ is a scale such as the Planck scale and we defined the critical exponent $\theta_m:=2-\gamma_m$ with the anomalous dimension of the mass $\gamma_m$ which is obtained as $\gamma_m\approx 0.027$ in the standard model.
In order to obtain ${\bar m}^2\fn{k_\text{EW}}\approx 1$ at the electroweak scale, ${\bar m}^2\fn{M_\text{pl}}\approx 10^{-33}$ is required.
This is the gauge hierarchy problem.
In ref.~\cite{Wetterich:2016uxm}, it is discussed that asymptotically safe gravity could solve this problem.
If one can obtain large anomalous dimension $\gamma_m>2$ (equivalently $\theta_m<0$) around the non-trivial fixed point of quantum gravity above the Planck scale which means that the scalar mass is irrelevant, the mass given above the Planck scale decreases towards the Planck scale by lowering the cutoff scale.
Then the tiny mass at the Planck scale ${\bar m}^2\fn{M_\text{pl}}\approx 10^{-33}$ is automatically realized.
One calls this mechanism ``self-tuned criticality".
Since the gravitational effects decouple below the Planck scale, the critical exponent of the scalar mass turns to positive $\theta_m>0$ and the scalar mass realizes ${\bar m}^2\fn{k_\text{EW}}\approx 1$.
This mechanism is called ``resurgence mechanism".
In this work, we unfortunately have not obtained a solid result that the critical exponent of the scalar mass becomes negative as reported in \cite{Oda:2015sma}.
However, our work show that the ferminonic fluctuations make the critical exponent of the scalar mass small.
Its dependence on the number of flavors of fermion, scalar and gauge fields should be investigated in future works.
Note that in a viewpoint of string theory, there might be a physical cutoff above the Planck scale, and the problem of quadratic divergence becomes real issue.
In this case, it might be interesting to investigate the Veltman condition~\cite{Veltman:1980mj}.
As speculated by Veltman, the scale where Veltman condition would be related to the restoration of supersymmetry~\cite{Veltman:1980mj,Hamada:2012bp,Masina:2013wja}, and we can explore the scale of supersymmetry by numerical calculation.

\subsection*{Acknowledgement}
We thank Astrid Eichhorn and Aaron Held for helpful discussions on \cite{Eichhorn:2017eht} and carefully reading this manuscript.
We also thank Nicolai Christiansen, Tobias Denz, Kevin Falls, Hikaru Kawai, Stefan Lippoldt, Nobuyoshi Ohta, Jan M. Pawlowski, Manuel Reichert, Alberto Salvio, Frank Saueressig, Henry Tye and Christof Wetterich for useful discussions and comments.
M.\,Y. thanks all members in quantum gravity group of ITP Universit\"at Heidelberg for daily discussions.
Y.\,H. thanks the member of the Institute for Advanced Study of the Hong Kong University of Science and Technology for hospitality during his visit. 
The work of Y.\,H. is supported by the Grant-in-Aid for Japan Society for the Promotion of Science Fellows, No.\,16J06151.
The work of M.\,Y. is supported by the DFG Collaborative Research Centre SFB 1225 (ISOQUANT).
\begin{appendix}
\section{Functional renormalization group}\label{frgtreatment}
The FRG is a method to analyze systems without relying on a perturbative expansion.
Therefore, it is also called the non-perturbative renormalization group.
In the FRG, the effective action $\Gamma_k$ is defined by integrating out the high momentum modes $k < |p| <\Lambda$, where $\Lambda$ is the initial cutoff scale at which the bare action is given.
Integrating out quantum fluctuations with the ``shell momentum mode" $k-\delta k < |p| <k$, the effective action $\Gamma_{k-\delta k}$ is generated and the rate of variability is defined:
\al{
\frac{\Gamma_k-\Gamma_{k-\delta k}}{\delta k}:= \beta.
}
This functional differential equation with the initial condition $\Gamma_{\Lambda}=S$  at $k=\Lambda$ is the FRG equation where the right-hand side is called the beta function.
Its explicit form is known as the Wetterich equation~\cite{Wetterich:1992yh,Morris:1993qb}, which reads
\begin{align}
{\p\over\p k } \Gamma_k
	&=	\frac{1}{2}\STr \left[ \left( \frac{\overrightarrow \delta}{\delta \Phi\fn{x}} \Gamma _k\frac{\bara \delta}{\delta \Phi\fn{y}}+\mc R_k\fn{x,y}  \right) ^{-1}\cdot {\p\over\p k} \mc R_k\fn{y,x}\right],
	\label{wetterich}
\end{align}
where the field $\Phi$ is the superfield and ``STr" denotes the supertrace for the supermatrix.\footnote{
See~\cite{Oda:2015sma} for details on the manipulation for the supermatrix.
} 
The cutoff function $\mc R_k$ in momentum space behaves as 
\al{
\mc R_k\fn{p^2} \sim 
\begin{cases}
0 & |p|>k\\
k^2 & |p|<k.
\end{cases}
}
The cutoff function suppresses the fluctuations with lower momentum $|p|<k$.
Thanks to this profile, only the fluctuations with higher momentum $k<|p|<\Lambda$ are integrated out, that is, the path-integral is evaluated.

In order to calculate the beta functions in a system using equation \eqref{wetterich}, we rewrite it in a more explicit form.\footnote{
See e.g.,~\cite{Gies:2001nw,Braun:2011pp} for the deformation of the Wetterich equation.
}
Defining
\al{
\bmat{
\Gamma_\BB &\Gamma_\BF\\
\Gamma_\FB &\Gamma_\FF
}
	&:=	\frac{\overrightarrow \delta}{\delta \Phi\fn{x}} \Gamma _k\frac{\bara \delta}{\delta \Phi\fn{y}},	&
\bmat{
\mc R_\BB &0\\
0 &\mc R_\FF
}
	&:=	\mc R_k\fn{x,y},
}
and then
\al{
{\mc M}:=
\bmat{
{\mc M}_\BB &{\mc M}_\BF\\
{\mc M}_\FB &{\mc M}_\FF
}
:=
\frac{\overrightarrow \delta}{\delta \Phi\fn{x}} \Gamma _k\frac{\bara \delta}{\delta \Phi\fn{y}}+\mc R_k\fn{x,y}
	=	\bmat{
			\Gamma_\BB &\Gamma_\BF\\
			\Gamma_\FB &\Gamma_\FF
			}
		+
		\bmat{
			\mc R_\BB &0\\
			0 &\mc R_\FF
			},
			\label{curly M}
}
the Wetterich equation becomes
\al{
{\p\over\p k}\Gamma_k
	&=	{1\over2}\wt{\p\over\p k}\Tr\bigg[\ln {\mc M}_\BB\bigg]
			-{1\over2}\wt{\p\over\p k}\Tr\bigg[\ln\Big({\mc M}_\FF-{\mc M}_\FB {\mc M}_\BB^{-1}{\mc M}_\BF\Big)\bigg],
				\label{preWetterich}
}
where we have used the formula for the superdeterminant of supermatrix,
\al{
\text{SDet}\, {\mc M} = \frac{\text{det}{\mc M}_\BB}{\text{det}({\mc M}_\FF-{\mc M}_\FB{\mc M}_\BB^{-1}{\mc M}_\BF)}.
}
Here the derivative $\wt{\p/\p k}$ acts only on the cutoff scale in $\mc R_\BB$ and $\mc R_\FF$ and then we obtain
\al{
{\p\over\p k}\Gamma_k
	&=	{1\over2}\Tr\bigg[{\mc M}_\BB^{-1}{\p \mc R_\BB\over\p k}\bigg]\nn
	&\quad
		-{1\over2}\Tr\bigg[\Big({\mc M}_\FF-{\mc M}_\FB {\mc M}_\BB^{-1}{\mc M}_\BF\Big)^{-1}
					\paren{{\p \mc R_\FF\over\p k}+{\mc M}_\FB {\mc M}_\BB^{-1}\,
						{\p \mc R_\BB\over\p k}\,
						{\mc M}_\BB^{-1}{\mc M}_\BF}\bigg].
						\label{FRG_total}
}
If we expand the term $\Big({\mc M}_\FF-{\mc M}_\FB {\mc M}_\BB^{-1}{\mc M}_\BF\Big)^{-1}={\mc M}_\FF^{-1}\Big(1-{\mc M}_\FF^{-1}{\mc M}_\FB {\mc M}_\BB^{-1}{\mc M}_\BF\Big)^{-1}$ into the polynomials of ${\mc M}_\FF^{-1}{\mc M}_\FB {\mc M}_\BB^{-1}{\mc M}_\BF$, the equation is
\al{
{\p\over\p k}\Gamma_k
	&=	{1\over2}\Tr\bigg[{\mc M}_\BB^{-1}{\p \mc R_\BB\over\p k}\bigg]
	-{1\over2}\Tr\bigg[{\mc M}_\FF^{-1}{\p \mc R_\FF\over\p k}\bigg]\nn
	&\quad		-{1\over2}\Tr\bigg[{\mc M}_\FF^{-1}\,{\p \mc R_\FF\over\p k}\,{\mc M}_\FF^{-1}{\mc M}_\FB {\mc M}_\BB^{-1}{\mc M}_\BF\bigg] 
	 -{1\over2}\Tr\bigg[{\mc M}_\FF^{-1}{\mc M}_\FB {\mc M}_\BB^{-1}\,{\p \mc R_\BB\over\p k}\,{\mc M}_\BB^{-1}{\mc M}_\BF\bigg]\nn
		&\qquad +\cdots.
						\label{FRG_totalexpand}
}
The first and second terms on the right-hand side corresponds to the one-loop effects of boson and fermion, respectively.
The third and fourth terms are the one-loop corrections with both bosonic and fermionic fluctuations. 
This form is useful to obtain the RG equation for the Yukawa coupling constant since the vertex structure becomes clearer.

We now introduce the critical exponents which are central characteristic of an asymptotically safe fixed point.\footnote{
The discussions here are given in e.g.,~\cite{Aoki:2000wm,Niedermaier:2006wt,Reuter:2007rv,Oda:2015sma}.
}
Let us consider an effective action in $d$ dimensions,
\al{
\Gamma_k=\int \df^d x \sum_{i=0}^\infty \frac{\hat g_i}{k^{d_{O_i}-d}}\mathcal O_i,
\label{action general discussion}
}
where $\hat g_i$ is the dimensionless coupling constant and $d_{\mathcal O_i}$ is the dimension of the operator $\mathcal O_i$.
Note that one of the operators among \eqref{action general discussion} should be redundant, e.g., the kinetic term with the field renormalization factor.
We here assume that the operator $\mathcal O_0$ is redundant.
Using the Wetterich equation we obtain coupled RG equations of the coupling constants
\al{
k\frac{\p  g_i}{\p k}=\beta_i\fn{ g},
}
where $g_i={\hat g}_i/{\hat g}_0$ with a redundant coupling constant $g_0$ and $\bar g$ without the index stands for a set of coupling constants $\{  g_1,  g_2,\cdots \}$.
We assume that the system \eqref{action general discussion} has a fixed point $g^*$ at which the beta functions vanish $\beta_i\fn{ g^*}=0$ for all $i$.
The RG flows around the fixed point are governed by the linearized RG equations
\al{
k\frac{\p  g_i}{\p k} \simeq \sum_j \frac{\p\beta_i}{\p   g_j}\bigg|_{ g= g^*}( g_j- g_j^*).
}
We easily find their solution
\al{
 g_i = g_i^* +\sum_{j=0}^\infty C_i\zeta_{ij}\left( \frac{\Lambda}{k}\right)^{\theta_j},
}
where $C_i$ are constants of integration, $\zeta_{ij}$ is the matrix diagonalizing the stability matrix $M_{ij}:=-\frac{\p\beta_i}{\p g_j}|_{g=g^*}$ and $\theta_j$ being the eigenvalue of $M_{ij}$ is called the critical exponent.
We can classify the RG flow as being one of three types: 
\begin{enumerate}
\item relevant ($\text{Re}\fn{\theta_j}>0$);
\item marginal ($\text{Re}\fn{\theta_j}=0$);
\item irrelevant ($\text{Re}\fn{\theta_j}<0$).
\end{enumerate}
While lowering the cutoff scale $k\to 0$, the RG flow with the positive critical exponent grows and becomes dominant at low energy scales.
In contrast, the RG flow with the negative critical exponent shrinks towards the fixed point.
Low energy physics is determined by the relevant operators and their coupling constants become the free parameters of the system.
In other words, when fixing the physics at low energy the theory can asymptotically reach the fixed point in the limit $k\to \infty$.
Then, the theory is free from UV divergences.

More explicitly, the beta function of $\bar g_i$ is typically written as
\al{
k\frac{\p \bar g_i}{\p k}= (d_{\mathcal O_i}-d - \eta){\bar g}_i +f_i\fn{\bar g},
\label{explicit beta general}
}
where the first term in the right-hand side is the canonical dimension of the coupling constant ${\bar g}_i$ with $\eta:=-k \p_k \ln g_0$ on the anomalous dimension from the redundant operator $\mathcal O_0$
and $f_i\fn{g}$ encodes the loop effects.
Expanding \eqref{explicit beta general} around the fixed point $\bar g^*$, the RG equation becomes
\al{
k\frac{\p \bar g_i}{\p k}&= \sum_j\left((d_{\mathcal O_i}-d-\eta)\delta_{ij} +\frac{\p f_i\fn{\bar g}}{\p \bar g_j}\bigg|_{\bar g=\bar g^*} \right)(\bar g_j-\bar g_j^*) +{\mc O}\fn{(\bar g_i-\bar g^*_i)^2}\nn
&\simeq -\sum_jM_{ij}\big|_{\bar g=\bar g^*}(\bar g_j-\bar g_j^*).
\label{stabilitymatrix}
}
The critical exponent is given as the eigenvalue of the matrix $M_{ij}|_{\bar g=\bar g^*}$. 
If the off-diagonal part of $M_{ij}|_{\bar g=\bar g^*}$ is negligible, the critical exponent is given as
\al{
\theta_i&\simeq -(d_{\mathcal O_i}-d -\eta)\delta_{ii} -\frac{\p f_i\fn{\bar g}}{\p \bar g_i}\bigg|_{\bar g=\bar g^*}
&(\text{no summation for}~i).
}
We can see that the critical exponent is the ``effective" dimension around the fixed point and the loop correction $-\frac{\p f_i\fn{\bar g}}{\p \bar g_i}|_{\bar g=\bar g^*}$ corresponds to the anomalous dimension arising from the non-perturbative dynamics.
Note that since the canonical scaling term $-(d_{\mathcal O_i}-d-\eta)$ becomes dominant around the Gaussian (trivial) fixed point at which perturbation theory is valid and $\eta \approx 0$, we see that the value of critical exponent is found by the naive dimensional analysis.


\section{Variations}\label{variations of action}
The variations for the operators given in the action~\eqref{originaleffectiveaction} are calculated to derive the beta functions for the effective action.
To this end, the fields are split as given in \eqref{backgroundsplit}.
Here we assume that $\bar g_{\mu\nu}$ is an arbitrary background.
In this case, the results of first and second variations become \cite{Barth:1983hb}
\al{
\int \df^4x\, \delta  (\sqrt{g}{{\mathcal {O}}})&= \int _x\left[
\frac{1}{2}h{\bar{\mathcal O}}+\delta {\mathcal O}
\right],
\\
\int \df^4x\,  \delta  (\sqrt{g}F\fn{\phi^2}R)&= \int _x
\left[
\left\{\frac{\bar R}{2}h -\bar R^{\mu\nu}h_{\mu\nu} -\bar{{ \Box}} h + \bar\nabla_\mu \bar\nabla_\nu h^{\mu\nu}\right\}F
+\bar R\delta F
\right],
\\
\int \df^4x\,  \delta (\sqrt{g}R^2)&=\int _x
\left[\frac{1}{2}\bar R^2\bar g^{\mu\nu}-2\bar R\bar R^{\mu\nu}-2\bar\nabla_\alpha\bar\nabla^\alpha \bar R\bar g^{\mu\nu}+2\bar\nabla_\mu \bar\nabla_{\nu}\bar R\right]h_{\mu\nu},\\
\int _x \delta  (\sqrt{g}R^{\mu\nu}R_{\mu\nu})&=\int _x 
\bigg[
\frac{1}{2}\bar R^{\alpha\beta}\bar R_{\alpha\beta} \bar g^{\mu\nu} 
-\bar\nabla_\alpha \bar\nabla^\alpha \bar R^{\mu\nu}
-\frac{1}{2}\bar\nabla_\alpha\bar\nabla^\alpha \bar R \bar g^{\mu\nu}
+\bar\nabla_\alpha\bar\nabla^\beta \bar R 
-2\bar R^{\alpha \beta}\bar R_{\alpha~\beta}^{~\mu~\nu}
\bigg]h_{\mu\nu},
}
\al{
\int \df^4x\, \delta^2 (\sqrt{g} {\mathcal O})&= \int _x
\left[
\left\{-\frac{1}{2}h^{\mu\nu}h_{\mu\nu} +\frac{1}{4}h^2\right\} \bar {\mathcal O}
+h\delta {\mathcal O}+\delta^2{\mathcal O}
\right],\\
\int \df^4x\, \delta^2 (\sqrt{g}F\fn{\phi^2}R)&= \int _x
\bigg[
\bigg\{ \frac{1}{2}h^{\mu\nu}({\bar { \Box}} -\bar R)  h_{\mu\nu} +\bar R_{\alpha\mu\beta\nu}h^{\alpha\beta}h^{\mu\nu} -\frac{1}{2}h{\bar { \Box}} h +h{\bar \nabla}^\mu {\bar \nabla}^\nu h_{\mu\nu} \nn
&\quad+({\bar \nabla}^\mu h_{\mu\alpha} )({\bar \nabla}_\nu h^{\nu \alpha})
+h^{\alpha \beta}\bar R_\alpha^{~\rho}h_{\rho\beta} -h\bar R^{\alpha \beta}h_{\alpha \beta} +\frac{\bar R}{4}
h^2\bigg\} F\nn
&\qquad+ 2\left\{\frac{\bar R}{2}h -\bar R^{\mu\nu}h_{\mu\nu} -{\bar { \Box}} h + {\bar \nabla}_\mu {\bar \nabla}_\nu h^{\mu\nu}\right\} \delta F
 +\bar R \delta^2 F\bigg],\\
\int \df^4x\,\delta^2 (\sqrt{g}R^2)&= \int _x
\bigg[
2h\bigg\{ {\bar { \Box}}^2 + \frac{\bar R^2}{8}\bigg\} h +\bar Rh^{\alpha \beta}{\bar {\Box}} h_{\alpha \beta} +2\bar R \bar R_{\alpha \mu\beta \nu}h^{\alpha \beta}h^{\mu\nu}\nn
&\quad+2({\bar \nabla}_\mu{\bar \nabla}_\nu h^{\mu\nu})^2
+({\bar \nabla} _\alpha {\bar \nabla}_\beta h^{\alpha \beta})\bigg\{ -4{\bar { \Box}} -2\bar R \bigg\} h\nn
&\qquad-2\bar R {\bar \nabla}^\mu h_{\mu\nu} {\bar \nabla}_\alpha h^{\alpha \nu}
-\bar Rh{\bar { \Box}} h
+4\bar R^{\alpha \beta}h_{\alpha \beta}{\bar { \Box}} h
 -4\bar R^{\alpha \beta}h_{\alpha \beta}{\bar \nabla}_{\mu}{\bar \nabla}_{\nu}h^{\mu \nu}\nn
&\qquad -4\bar Rh^{\alpha \beta} {\bar \nabla}_\beta {\bar \nabla}^\mu h_{\alpha \mu}
-4\bar R{\bar \nabla} _\mu h^{\mu \nu}{\bar \nabla}_\nu h
-2h^{\alpha \beta}{\bar \nabla}_\alpha {\bar \nabla}^\mu \bar R h_{\mu\beta}
-\frac{1}{2}h({\bar { \Box}} \bar R) h \nn
&\qquad+4h({\bar \nabla}^\alpha {\bar \nabla}^\beta \bar R) h_{\alpha\beta}
+\frac{3}{2}({\bar { \Box}} \bar R) h^{\alpha \beta}h_{\alpha \beta}
-2\bar R h\bar R^{\alpha \beta}h_{\alpha \beta}
-\frac{1}{2}\bar R^2 h^{\alpha \beta} h_{\alpha \beta}\nn
&\qquad +2\bar R^{\alpha \beta} h_{\alpha \beta} \bar R^{\mu\nu}h_{\mu\nu}
+2\bar R h^{\alpha \beta}\bar R_\alpha ^{~\mu}h_{\mu\beta}
\bigg],\\
\int \df^4x\,\delta^2 (\sqrt{g}R_{\mu\nu}R^{\mu\nu})&=\int _x 
\bigg[
\frac{1}{2}h^{\alpha \beta}{\bar { \Box}}^2 h_{\alpha\beta} 
+\frac{1}{2}h{\bar { \Box}}^2 h - ({\bar \nabla}_\mu {\bar \nabla}_\nu h^{\mu\nu}){\bar { \Box}} h
+{\bar \nabla}_\mu h^{\mu\nu}{\bar { \Box}} {\bar \nabla}^\alpha h_{\alpha \nu}\nn
&\quad +({\bar \nabla}_\alpha {\bar \nabla}_\beta h^{\alpha \beta})^2
+2h_{\mu\nu}\bar R^{\alpha \mu \beta \nu}{\bar { \Box}} h_{\alpha \beta}
+2h_{\mu\nu}\bar R^{\alpha \mu \beta \nu}\bar R_{\alpha \rho \beta \tau}h^{\rho \tau}\nn
&\qquad+\frac{1}{4}\bar R^{\mu\nu}\bar R_{\mu\nu}h^2
-2h\bar R^{\mu\nu}\bar R_{\alpha \mu \beta \nu}h^{\alpha \beta}
-\frac{1}{2}h\bar R^{\alpha \beta} {\bar \nabla}_{\alpha}{\bar \nabla}_{\beta}h \nn
 &\qquad  +h\bar R^{\alpha \beta}{\bar \nabla}_\beta {\bar \nabla}_\mu h^{\mu}_{~\alpha} -\frac{1}{2}\bar R^{\mu\nu}\bar R_{\mu\nu}h^{\alpha \beta}h_{\alpha \beta} 
 -{\bar \nabla}_\alpha h^{\mu \alpha}\bar R_{\mu\nu}{\bar \nabla}^\nu h
 -2h^{\alpha \beta} {\bar \nabla}_\mu {\bar \nabla}^\rho h_{\alpha \rho}\bar R^\mu_{~\beta}\nn
&\qquad + h^{\alpha \beta}{\bar \nabla}_\alpha {\bar \nabla}^\nu h_{\nu\mu}\bar R^\mu_{~\beta}
+h^{\alpha \beta} \bar R_{\alpha \mu}\bar R_{\nu \beta} h^{\mu\nu} 
-h^{\alpha \beta}\bar R_{\alpha \mu}\bar R^{\mu \nu}h_{\beta \nu}
+2 h^{\alpha \rho}\bar R^{\mu\nu}\bar R_{\rho \mu~ \nu}^{~~\beta}h_{\alpha \beta}\nn
&\qquad +2h^{\mu\rho}\bar R_\rho^{~\nu}\bar R^{\alpha~\beta}_{~\mu~\nu}h_{\alpha\beta}
-2h^{\alpha \beta}\bar R^{\mu\nu}{\bar \nabla}_\nu {\bar \nabla}_\beta h_{\alpha \mu}
+h^{\alpha \beta}\bar R^{\mu\nu}{\bar \nabla}_\mu {\bar \nabla}_\nu h_{\alpha \beta}
-\frac{1}{8}h({\bar { \Box}} \bar R) h \nn
&\qquad  +h^{\alpha \mu}({\bar { \Box}} \bar R_{\mu}^{~\beta})h_{\alpha \beta}
+\frac{1}{4}({\bar { \Box}} \bar R) h^{\alpha \beta}h_{\alpha \beta}
\bigg],
}
where we used the shorthand notation, $\int_x=\int \df^4x \sqrt{\bar g}$, and $\mathcal O$ and $\bar {\mathcal O}$ are
\al{
\mathcal O&:=	V\fn{\Phi^2}
+\frac{1}{2} g^{\mu \nu}\,{\bar \p} _\mu{ \Phi}\,{\bar \p} _{\nu} \Phi 
		+{\bar \Psi}{\Slash {\bar D}}\Psi	
		+y  \Phi {\bar \Psi}\Psi,\\
\bar{\mathcal O}&:=	V\fn{ \phi^2}
		+y   \phi {\bar \psi}\psi,
}
respectively.
The variations for $F$ and $\mathcal O$ are given by
\al{
\int _x\delta F&=\int _x(2\phi F')\varphi,&\\
\int _x\delta^2 F&=\int _x\varphi(2F'+4\phi^2 F'')\varphi,&
\\
\int _x\delta \mathcal O
&=\int _x \bigg[(2\phi V')\varphi + y \varphi\bar \psi \psi + y \phi (\bar \chi \psi  + \bar \psi \chi)
+\bar \psi \big( (\delta \gamma^\mu)  {\bar D}_\mu+\gamma^\mu \delta {\bar D}_\mu\big)\psi
+\frac{1}{2}\bar \psi \Slash {\bar D} \chi
-\frac{1}{2}({\bar D}_\mu \bar \chi) \gamma^\mu \psi\bigg], 
&\\
\int _x \delta^2 \mathcal O&=\int _x \bigg[\varphi(-{\bar \Box}+2 V'+4\phi^2 V'')\varphi 
+\bar \chi( \Slash {\bar D} +y \phi )\chi
+ y \varphi (\bar \chi \psi  + \bar \psi \chi)\nn
&\quad +\frac{1}{2}\bigg\{\bar \psi \delta \gamma_\mu( {\bar D}^\mu \chi) 
- ({\bar D}_\mu \bar \chi) \delta \gamma^\mu \psi 
\bigg\} +\bar \psi \gamma_\mu (\delta {\bar D}^\mu) \chi
+\bar \chi \gamma_\mu (\delta {\bar D}^\mu) \psi\nn
&\qquad 
+\bar \psi (\delta^2 \gamma^\mu  {\bar D}_\mu+ 2\delta \gamma^\mu \delta {\bar D}_\mu + \gamma^\mu \delta^2 {\bar D}_\mu)\psi \bigg],
}
where the prime denotes the derivative with respect to $\phi^2$ and we have assumed that the background fields of scalar and fermion do not depend on the spacetime.
Here we evaluate the variations for the gamma matrix and the covariant derivative of fermion. 
To this end, we follow the literatures \cite{Gies:2013noa,Gies:2015cka}, where the local spin-based formalism has performed in four dimension.\footnote{The local spin-based formalism in arbitrary dimensions is discussed in \cite{Lippoldt:2015cea}.} 
The variation of the spin connection is given by
\al{
\hat \Gamma_\mu=
\bar{\hat \Gamma}_\mu +
\frac{1}{8}[\gamma^\kappa,\gamma^\sigma]\bigg[
\delta^{\alpha \beta}_{\mu[\kappa}\delta^\nu_{\sigma]}+h_{\rho\lambda}\bigg( \omega^{\alpha\beta\rho\lambda}_{~~~~~[\kappa\sigma]}\delta^\nu_\mu
-2\omega^{\alpha\beta\rho\lambda}_{~~~~~\mu[\kappa}\delta^\nu_{\sigma]}
-\frac{1}{2}\delta^{\alpha\beta}_{\mu[\kappa}\delta^{\rho\lambda}_{\sigma]\chi}\bar g^{\chi\nu}\bigg)\bigg]{\bar \nabla}_\nu h_{\alpha \beta}+{\mc O}\fn{h^3},
}
where we defined the tensors
\al{
\omega^{\rho\lambda\alpha\beta}_{~~~~~\mu\nu}
=\omega^{\alpha\beta\rho\lambda}_{~~~~~\nu\mu}
=\frac{1}{4}\delta^{\rho\lambda}_{\mu\kappa}\bar g^{\kappa\sigma}\delta^{\alpha\beta}_{\sigma\nu},
}
and $\delta^{\rho\lambda}_{\mu\nu}=\frac{1}{2}(\delta^\rho_\mu \delta^\lambda_\nu+\delta^\rho_\nu \delta^\lambda_\mu)$, and the indices with square brackets are antisymmetric, i.e., $T_{[\mu\nu]}=\frac{1}{2}(T_{\mu\nu}-T_{\nu\mu})$.
Then the variations for the covariant derivative are
\al{
\delta {D}_\mu
&=\frac{1}{8}[\gamma^\alpha, \gamma^\beta]{\bar \nabla}_\beta h_{\alpha \mu},
\label{delD1}
\\
\delta^2 {D}_\mu,
&=\frac{1}{8}[\gamma^\alpha, \gamma^\beta]\bigg( 
-h^\rho _{~\alpha} {\bar \nabla}_\beta h_{\mu \rho}
-h^\rho_{~\beta}{\bar \nabla}_\rho h_{\mu \alpha}
-\frac{1}{2}h_{\rho \alpha}{\bar \nabla}_\mu h^{\rho}_{~\beta}
\bigg).
\label{delD2}
}
Note that the gamma matrix $\gamma^\mu$ in \eqref{delD1} and \eqref{delD2} is defined on the background metric, namely, $\gamma^\mu\fn{\bar g}$.
For the gamma matrix, its expansion is
\al{
\gamma_\mu\fn{\bar g+h}=\gamma_\mu\fn{\bar g}
+\frac{\p \gamma_\mu}{\p g_{\rho\lambda}}\bigg|_{g=\bar g}h_{\rho\lambda}
+\frac{1}{2}\frac{\p^2 \gamma_\mu}{\p g_{\alpha \beta}\p g_{\rho\lambda}}\bigg|_{g=\bar g}h_{\alpha \beta}h_{\rho\lambda} +{\mc O}\fn{h^3},
}
where
\al{
\frac{\p \gamma_\mu}{\p g_{\rho\lambda}}\bigg|_{g=\bar g} 
&=\frac{1}{2}\delta^{\rho\lambda}_{\mu\nu}\gamma^\nu\fn{\bar g},&
\frac{\p^2 \gamma_\mu}{\p g_{\alpha \beta}\p g_{\rho\lambda}}\bigg|_{g=\bar g}
&=-\omega^{\alpha\beta\rho\lambda}_{~~~~~[\mu\nu]}\gamma^\nu\fn{\bar g}.&
}
Then the variations of the gamma matrix are
\al{
\delta \gamma_\mu&= \frac{1}{2}\gamma^\nu h_{\mu\nu},&
\delta^2 \gamma_{\mu} &= -\frac{1}{4}h^{\lambda}_{~\mu}h_{\lambda \nu}\gamma^\nu.&
} 
Using the results of the variations given above, the variation of ${\mc O}$ becomes
\al{
\int_x \delta \mathcal O&=\int_x \bigg[ (2\phi V')\varphi + y \varphi\bar \psi \psi + y \phi (\bar \chi \psi  + \bar \psi \chi)
+\frac{1}{8}\bar \psi \left(\gamma^\mu [\gamma^\alpha,\gamma^\beta]{\bar \nabla}_\beta h_{\alpha \mu}\right)\psi +\frac{1}{2}\bar \psi \Slash {\bar D} \chi
-\frac{1}{2}({\bar D}_\mu \bar \chi) \gamma^\mu \psi\bigg], 
&\\
\int_x \delta^2 \mathcal O&=\int_x \bigg[ \varphi(-{\bar \Box}+2 V'+4\phi^2 V'')\varphi 
+\bar \chi( \Slash {\bar D} +y \phi )\chi
+ y \varphi (\bar \chi \psi  + \bar \psi \chi)\nn
&\quad +\frac{1}{4}\left[\bar \psi \gamma^\nu( {\bar D}^\mu \chi) - ({\bar D}_\mu \bar \chi) \gamma^\nu \psi \right] h_{\mu\nu}
+\frac{1}{8}\left(
\bar \chi \gamma^\mu[\gamma^\alpha,\gamma^\beta] \psi +\bar \psi \gamma^\mu [\gamma^\alpha,\gamma^\beta] \chi
\right){\bar \nabla}_{\beta}h_{\alpha \mu}\nn
&\qquad -\frac{1}{16}(h_{\rho \alpha}{\bar \nabla}_\mu h^{\rho}_{~\beta})\bar \psi (\gamma^\mu[\gamma^\alpha, \gamma^\beta])\psi\bigg].
}
Note that we have used the fact that using the integration by part the covariant derivative is given so that
\al{
\int_x (\bar \psi \gamma^\nu {\bar D}^\mu  \chi) h_{\mu\nu}=
-\int_x(\bar \psi \gamma^\nu \chi) {\bar \nabla}^\mu h_{\mu\nu},
}
where ${\bar D}_\mu$ and ${\bar \nabla}_\mu$ are the covariant derivatives for the fermion and the graviton, respectively.
Note that the terms ${\bar D}_\mu \bar \psi$ and ${\bar D}_\mu \gamma^\nu$ are omitted.

We now assume that the background metric $\bar g_{\mu\nu}$ is the Einstein metric, i.e., $\bar R_{\mu\nu}=\frac{\bar R}{4}\bar g_{\mu\nu}$.
In this case, the second variations of the operators are reduced as follow:
\al{
\int \df^4x\, \delta^2 (\sqrt{g} {\mathcal O})&= \int _x
\left[
\left\{-\frac{1}{2}h^{\mu\nu}h_{\mu\nu} +\frac{1}{4}h^2\right\} \bar {\mathcal O}
+h\delta {\mathcal O}+\delta^2{\mathcal O}
\right],\\
\int \df^4x\,\delta^2 (\sqrt{g}F\fn{\phi^2}R)&=\int _x 
\bigg[
\bigg\{ -\frac{1}{2}h^{\mu\nu}\bigg( -\delta_{\mu\alpha}\delta_{\nu\beta}{\bar \Box} +\delta_{\mu\alpha}\delta_{\nu\beta}\frac{\bar R}{2} -2\bar R_{\alpha\mu\beta\nu}\bigg) h^{\alpha\beta}
 -\frac{1}{2}h{\bar \Box} h 
+h{\bar \nabla}^\mu {\bar \nabla}^\nu h_{\mu\nu} \nn
&\quad+({\bar \nabla}^\mu h_{\mu\alpha} )({\bar \nabla}_\nu h^{\nu \alpha})\bigg\} F
+2\bigg\{\frac{\bar R}{4}h  -{\bar \Box} h + {\bar \nabla}_\mu {\bar \nabla}_\nu h^{\mu\nu} \bigg\}
 \delta F
 +\bar R \delta^2 F\bigg]\nn
&=\int _x
\bigg[
\bigg\{ -\frac{1}{2}h^{\mu\nu}\Delta_{L2} h_{\mu\nu}
 -\frac{1}{2}h{\bar \Box} h 
+h{\bar \nabla}^\mu {\bar \nabla}^\nu h_{\mu\nu} \nn
&\quad+({\bar \nabla}^\mu h_{\mu\alpha} )({\bar \nabla}_\nu h^{\nu \alpha})\bigg\} F
+2\bigg\{\frac{\bar R}{4}h -{\bar \Box} h + {\bar \nabla}_\mu {\bar \nabla}_\nu h^{\mu\nu} \bigg\}
 \delta F
 +\bar R \delta^2 F\bigg],\\
\int \df^4x\, \delta^2 (\sqrt{g}R^2)&= \int _x
\bigg[
2h\bigg\{ {\bar \Box}^2 - \frac{\bar R^2}{16}\bigg\} h
 -\bar R h^{\mu\nu}\left(-\delta_{\mu\alpha}\delta_{\nu\beta}{\bar \Box}+\delta_{\mu\alpha}\delta_{\nu\beta}\frac{\bar R}{2}   
 -2 \bar R_{\alpha \mu\beta \nu}\right)h^{\alpha \beta}\nn
&\quad  +2({\bar \nabla}_\mu{\bar \nabla}_\nu h^{\mu\nu})^2
+\frac{\bar R^2}{2}h_{\mu\nu}h^{\mu\nu}   +({\bar \nabla} _\alpha {\bar \nabla}_\beta h^{\alpha \beta})\bigg\{ -4{\bar \Box} +\bar R \bigg\} h
+2\bar R {\bar \nabla}^\mu h_{\mu\nu} {\bar \nabla}_\alpha h^{\alpha \nu}
\bigg]\nn
&=
\int _x
\bigg[
2h\bigg\{ {\bar \Box}^2 - \frac{\bar R^2}{16}\bigg\} h
 -\bar R h_{\mu\nu}\Delta_{L2}h^{\mu \nu}
 +2({\bar \nabla}_\mu{\bar \nabla}_\nu h^{\mu\nu})^2+\frac{\bar R^2}{2}h_{\mu\nu}h^{\mu\nu}\nn
&\quad +({\bar \nabla} _\alpha {\bar \nabla}_\beta h^{\alpha \beta})\bigg\{ -4{\bar \Box} +\bar R \bigg\} h
+2\bar R {\bar \nabla}^\mu h_{\mu\nu} {\bar \nabla}_\alpha h^{\alpha \nu}
\bigg],\\
\int \df^4x\, 
\delta^2 (\sqrt{g}R_{\mu\nu}R^{\mu\nu})&=\int _x
\bigg[
\frac{1}{2}h^{\alpha \beta}\left( \delta_{\alpha\beta}\delta_{\lambda\sigma}\left(-{\bar \Box}+\frac{\bar R}{2}\right) - 2\bar R_{\alpha \lambda \beta \sigma}\right)
\left(\delta^{\lambda\sigma}\delta^{\mu\nu}\left(-{\bar \Box}+\frac{\bar R}{2}\right) - 2\bar R^{\lambda\mu\sigma\nu }\right) h_{\mu\nu}\nn
&+\frac{1}{2}h\bigg( {\bar \Box}^2 -\frac{\bar R}{4}{\bar \Box} -\frac{\bar R^2 }{8}\bigg)h 
- ({\bar \nabla}_\mu {\bar \nabla}_\nu h^{\mu\nu}) \bigg({\bar \Box} -\frac{\bar R}{2} \bigg)h
+{\bar \nabla}_\mu h^{\mu\nu}\bigg({\bar \Box} +\frac{3\bar R}{4} \bigg) {\bar \nabla}^\alpha h_{\alpha \nu}\nn
& +({\bar \nabla}_\alpha {\bar \nabla}_\beta h^{\alpha \beta})^2
-\frac{3\bar R}{4}h^{\mu\nu}\left(\delta_{\mu\alpha}\delta_{\nu\beta}(-{\bar \Box}+\frac{\bar R}{2}) - 2 \bar R_{\mu\alpha\nu\beta}\right)h^{\alpha\beta}
+\frac{\bar R^2}{4}h_{\mu\nu}h^{\mu\nu}
\bigg]\nn
&= \int_x
\bigg[
\frac{1}{2}h^{\alpha \beta}\Delta_{L2}^2 h_{\alpha\beta} 
-\frac{3\bar R}{4}h^{\alpha \beta}\Delta_{L2} h_{\alpha\beta}
+\frac{1}{2}h\bigg( {\bar \Box}^2 -\frac{\bar R}{4}{\bar \Box} -\frac{\bar R^2 }{8}\bigg)h \nn
&\quad- ({\bar \nabla}_\mu {\bar \nabla}_\nu h^{\mu\nu}) \bigg({\bar \Box} -\frac{\bar R}{2} \bigg)h
+{\bar \nabla}_\mu h^{\mu\nu}\bigg({\bar \Box} +\frac{3\bar R}{4} \bigg) {\bar \nabla}^\alpha h_{\alpha \nu}\nn
&\quad +({\bar \nabla}_\alpha {\bar \nabla}_\beta h^{\alpha \beta})^2
+\frac{\bar R^2}{4}h_{\mu\nu}h^{\mu\nu}
\bigg],
}
where we have neglected the terms which do not contribute to the beta functions and have defined the Lichnerowicz Laplacians in 4 dimension:
\al{
{\bar \Delta}_{L0}S&=-{\bar \Box} S,
\label{scalar Lich}\\
{\bar \Delta}_{L\frac{1}{2}}\psi &= \left( -{\bar D}^2  + \frac{\bar R}{4}\right)\psi,
\label{spinor Lich}\\
{\bar \Delta}_{L1}\xi_\mu&=\left(-{\bar \Box} \delta_{\mu}^{\nu} +\bar R_{\mu}^{~\nu}\right)\xi_\nu,
\label{vector Lich}\\
{\bar \Delta}_{L2}h_{\mu\nu}&=\left( -{\bar \Box} \delta_{\mu}^{\alpha}\delta_{\nu}^{\beta}
+ \bar R_{\mu}^{~\alpha}\delta_{\nu}^{\beta}
+ \bar R_{\nu}^{~\beta}\delta_{\mu}^{\alpha}
- 2\bar R_{\mu~\nu}^{~\alpha~\beta}\right)h_{\alpha \beta}.
\label{tensor Lich}
}
These Laplacians satisfy
\al{
{\bar \Delta}_{L1}{\bar \nabla}_\mu S &= {\bar \nabla}_\mu {\bar \Delta}_{L0}S,
\label{LL1}\\
{\bar \nabla}_\mu{\bar \Delta}_{L1} A^\mu &= {\bar \Delta}_{L0}{\bar \nabla}_\mu A^\mu,
\label{LL2}\\
{\bar \Delta}_{L2}({\bar \nabla}_\mu{\bar \nabla}_\nu S )&= {\bar \nabla}_\mu{\bar \nabla}_\nu {\bar \Delta}_{L0}S,
\label{LL3}\\
{\bar \Delta}_{L2}({\bar \nabla}_\mu A_\nu + {\bar \nabla}_\nu A_\mu)&= {\bar \nabla}_\mu {\bar \Delta}_{L1}A_\nu + {\bar \nabla}_\nu {\bar \Delta}_{L1}A_\mu,
\label{LL4}\\
{\bar \nabla}_\mu {\bar \Delta}_{L2}T^{\mu\nu}&={\bar \Delta}_{L2}{\bar \nabla}_\mu T^{\mu\nu}, 
\label{LL5}\\
{\bar \Delta}_{L2}{\bar g}_{\mu\nu} S&={\bar g}_{\mu\nu}{\bar \Delta}_{L0} S.
\label{LL6}
}
Employing the York decomposition for the metric
\al{\label{metricdecomposition in ap}
h_{\mu\nu}	=	h_{\mu\nu}^\perp		+{\bar \nabla} _\mu\ft\xi_{\nu}	+{\bar \nabla} _{\nu}\ft\xi_\mu	
+\paren{{\bar \nabla}_\mu {\bar \nabla}_{\nu}	-\frac{1}{4} \bar g_{\mu\nu}{\bar \Box}}\ft\sigma
 	+\frac{1}{4} \bar g_{\mu\nu}h,
}
and using the relationships \eqref{LL1}--\eqref{LL6}, we have
\al{
\int_x\,h_{\mu\nu}h^{\mu\nu}
&=\int_x\, \bigg[ h^\perp_{\mu\nu}h^{\perp}{}^{\mu\nu} 
 +2\xi^\mu \xi_\mu
 +\frac{3}{4}\sigma^2
  +\frac{1}{4}h^2 \bigg],
 \\
\int_x\,h^{\mu\nu}{\bar \Delta}_{L2} h_{\mu\nu}&= \int_x\,\bigg[ h^\perp_{\mu\nu}{\bar \Delta}_{L2}h^\perp{}^{\mu\nu} +2 \xi_\mu {\bar \Delta}_{L1}\xi^\mu +\frac{3}{4}\sigma {\bar \Delta}_{L0}\sigma + \frac{1}{4}h{\bar \Delta}_{L0}h\bigg] ,\\
\int_x\, h^{\mu\nu}{\bar \Delta}_{L2}^2 h_{\mu\nu}&=\int_x\, \bigg[ h^\perp_{\mu\nu}{\bar \Delta}_{L2}^2h^\perp{}^{\mu\nu} +2 \xi_\mu {\bar \Delta}_{L1}^2\xi^\mu +\frac{3}{4}\sigma {\bar \Delta}_{L0}^2\sigma + \frac{1}{4}h{\bar \Delta}_{L0}^2h \bigg],\\
{\bar \nabla}_\mu {\bar \nabla}_\nu h^{\mu\nu}
&=\bigg[ \frac{3}{4}{\bar \Box} \left( {\bar \Box} +\frac{\bar R}{3}\right) \tilde \sigma+\frac{1}{4}{\bar \Box} h\bigg],\\
{\bar \nabla}^\mu h_{\mu\alpha} 
&=\bigg[ \left( {\bar \Box} +\frac{\bar R}{4} \right) \tilde\xi_\alpha 
+\frac{3}{4}{\bar \nabla}_\alpha \left( {\bar \Box} +\frac{\bar R}{3} \right) \tilde \sigma + \frac{1}{4}{\bar \nabla}_\alpha h\bigg],
}
where we redefined the fields
\al{
\xi_\mu&=\sqrt{-{\bar \Box}-\frac{\bar R}{4}}\tilde \xi_\mu=\sqrt{{\bar \Delta}_{L0}-\frac{\bar R}{4}}\tilde \xi_\mu,&
\sigma&=\sqrt{-{\bar \Box}}\sqrt{-{\bar \Box}-\frac{\bar R}{3}}\tilde \sigma
=\sqrt{{\bar \Delta}_{L0}}\sqrt{{\bar \Delta}_{L0}-\frac{\bar R}{3}}\tilde \sigma.
}
Then, we obtain the second variations the operators with the Einstein metric and the York decomposition:
\al{
\int \df^4x\, \delta^2 (\sqrt{g} {\mathcal O})&= \int_x
\bigg[
\left\{
\frac{1}{8}h^2 -\frac{1}{2}h^\perp_{\mu\nu}h^\perp{}^{\mu\nu}- \xi_\mu\xi^\mu
-\frac{3}{8} \sigma^2
\right\} \bar {\mathcal O} 
+h\delta {\mathcal O}+\delta^2{\mathcal O}
\bigg]\\
\int \df^4x\,\delta^2 (\sqrt{g}F\fn{\phi^2}R)&= \int_x
\bigg[
\bigg\{
\frac{3}{16}h{\bar \Delta}_{L0}h  -\frac{1}{2}h^\perp_{\mu\nu} {\bar \Delta}_{L2} h^\perp{}^{\mu\nu}
-\frac{\bar R}{2}\xi_\mu \xi^\mu\nn
&\quad +\frac{3}{8}h\sqrt{\left( {\bar \Delta}_{L0}-\frac{\bar R}{3}\right) {\bar \Delta}_{L0}}\sigma
+\frac{3}{16} \sigma ({\bar \Delta}_{L0} -\bar R)\sigma
\bigg\}F\fn{\phi^2} \nn
&\quad + 2\delta F\left\{ \frac{1}{4} (3{\bar \Delta}_{L0} + \bar R) h  + \frac{3}{4}\sqrt{{\bar \Delta}_{L0}}\sqrt{{\bar \Delta}_{L0}-\frac{\bar R}{3}}\sigma \right\} +\bar R \delta^2F
\bigg],\\
\int \df^4x\,\delta^2(\sqrt{g}R^2)&=\int_x
\bigg[
\frac{9}{8}h\left({\bar \Delta}_{L0}-\frac{\bar R}{3}\right){\bar \Delta}_{L0}h 
-\bar R h^\perp_{\mu\nu}\Bigg({\bar \Delta}_{L2}  -\frac{\bar R}{2}\Bigg)h^\perp{}^{\mu\nu}
+\frac{9}{8}\sigma {\bar \Delta}_{L0}^2\sigma\nn
&\quad + \frac{9}{4} h {\bar \Delta}_{L0}\sqrt{{\bar \Delta}_{L0}}\sqrt{{\bar \Delta}_{L0}-\frac{\bar R}{3}}\sigma
\bigg],\\
\int \df^4x\,\delta^2 (\sqrt{g}R_{\mu\nu}R^{\mu\nu})&= \int_x
\bigg[
\frac{3}{8}h\left({\bar \Delta}_{L0}-\frac{\bar R}{3}\right){\bar \Delta}_{L0}h
+\frac{1}{2}h^\perp_{\mu\nu}\left({\bar \Delta}_{L2}^2-\frac{3\bar R}{2}{\bar \Delta}_{L2} +\frac{\bar R^2}{2}\right)h^\perp{}^{\mu\nu}\nn
&\quad +\frac{3}{8}\sigma {\bar \Delta}_{L0}^2\sigma
+\frac{3}{4}h{\bar \Delta}_{L0}\sqrt{{\bar \Delta}_{L0}}\sqrt{{\bar \Delta}_{L0}-\frac{\bar R}{3}}\sigma
\bigg].
}

In the same manner, for the gauge fixing and the ghost terms \eqref{gaugefixedaction} we have
\al{
S_{{\rm gf}}+S_{{\rm gh}}	
&=\int \df^4x \sqrt{\bar g}
\Bigg[
-\frac{\alpha}{2}{\hat B}_\mu Y^{\mu\nu}{\hat B}_\nu +\frac{1}{2\alpha}\Sigma_\mu Y^{\mu\nu} \Sigma_\nu
+\bar C_\mu Y^{\mu\rho} {\bar \Delta}^\text{ghost}_{\rho\nu}C^\nu
\bigg]\nn
&=\int \df^4x \sqrt{\bar g}\bigg[
-\frac{\alpha}{2}
\bigg\{
B^\perp_\mu\left( {\bar \Delta}_{L1} -\frac{1-\rho_2}{4} {\bar R} \right)B^\perp{}^\mu
+ B\left[ (1+\rho_1-\rho_2) {\bar \Delta}_{L0} -\frac{1-\rho_2}{4}{\bar R}
\right]B
\bigg\}\nn
&\quad +\frac{1}{2\alpha} 
\bigg\{
\xi_\mu \left( {\bar \Delta}_{L1}-\frac{\bar R}{2} \right)\left( {\bar \Delta}_{L1}-\frac{1-\rho_2}{4}{\bar R} \right) \xi^\mu\nn
&\quad+\frac{9}{16} \sigma \left( {\bar \Delta}_{L0} -\frac{\bar R}{3} \right)
\left(
(1+\rho_1-\rho_2){\bar \Delta}_{L0} -\frac{1-\rho_2}{4}{\bar R}
\right) \sigma\nn
&\quad+ \frac{3\beta}{8} \sigma\sqrt{{\bar \Delta}_{L0}\left({\bar \Delta}_{L0}-\frac{\bar R}{3}\right)}
\left(
(1+\rho_1-\rho_2){\bar \Delta}_{L0} -\frac{1-\rho_2}{4}{\bar R}
\right)  h\nn
&\quad+ \frac{\beta^2}{16} h{\bar \Delta}_{L0}\left(
(1+\rho_1-\rho_2){\bar \Delta}_{L0} -\frac{1-\rho_2}{4}{\bar R}
\right) h
\bigg\}\nn
&\quad
+\bar C^\perp_\mu \left( {\bar \Delta}_{L1} -\frac{1-\rho_2}{4}{\bar R} \right) 
\left( {\bar \Delta}_{L1} -\frac{\bar R}{2}  \right)C^\perp{}^\mu
\nn
&\quad
+\frac{3-\beta}{2} \bar C \left( {\bar \Delta}_{L0} -\frac{\bar R}{3-\beta} \right)\left\{ (1+\rho_1-\rho_2){\bar \Delta}_{L0}   -\frac{1-\rho_2}{4}{\bar R} \right\} C
\Bigg].
\label{SgfSgfexpl}
}
Here defining
\al{
{\mathcal X}\fn{{\bar \Delta}_{L1}}&:=  {\bar \Delta}_{L1}-\frac{1-\rho_2}{4}{\bar R},
\label{xmath}
\\
{\mathcal Y}\fn{{\bar \Delta}_{L0}}&:= (1+\rho_1-\rho_2){\bar \Delta}_{L0} -\frac{1-\rho_2}{4}{\bar R},
\label{ymath}
}
\eqref{SgfSgfexpl} can be written as
\al{
S_{{\rm gf}}+S_{{\rm gh}}	
&=\int \df^4x \sqrt{\bar g}\bigg[
-\frac{1}{2}
\bigg\{
B^\perp_\mu {\mathcal X}\fn{{\bar \Delta}_{L1}} B^\perp{}^\mu
+ B{\mathcal Y}\fn{{\bar \Delta}_{L0}}B
\bigg\}\nn
&+\frac{1}{2\alpha} 
\Bigg\{
\xi_\mu {\mathcal X}\fn{{\bar \Delta}_{L1}}\left( {\bar \Delta}_{L1}-\frac{\bar R}{2} \right) \xi^\mu
+\frac{1}{16}\left[3{\mathcal Y}^{1/2}\fn{{\bar \Delta}_{L0}}\sqrt{{\bar \Delta}_{L0} -\frac{\bar R}{3}} \sigma + \beta {\mathcal Y}^{1/2}\fn{{\bar \Delta}_{L0}}\sqrt{{\bar \Delta}_{L0}} h\right]^2
\Bigg\}
 \nn
&
+\bar C^\perp_\mu {\mathcal X}\fn{{\bar \Delta}_{L1}}\left( {\bar \Delta}_{L1} -\frac{\bar R}{2}  \right)C^\perp{}^\mu
+\frac{3-\beta}{2} \bar C {\mathcal Y}\fn{{\bar \Delta}_{L0}} \left( {\bar \Delta}_{L0} -\frac{\bar R}{3-\beta} \right) C
\Bigg],
\label{actionrewritten}
}
where $\alpha$ in front of $B$ action is absorbed into $B$ fields.

\section{Formula of heat kernel trace}\label{hkeapp}
In this appendix, we present the formula of heat kernel expansion.
Let us consider the trace for a function $W\fn{{\bar \Delta}_s}$ where ${\bar \Delta}_s=-D^2+{\bf Q}_s$ is the differential operator and the subscript $s$ denotes the spin of field on which ${\bar \Delta}_s$ acts.
By the Laplace transformation, we have
\al{
{\rm Tr}[W\fn{{\bar \Delta}_s}]=\int^\infty_{-\infty}\df t \,\tilde W\fn{t}{\rm Tr}[e^{-t{\bar \Delta}_s}].
\label{trace formula}
}
The trance for $e^{-t{\bar \Delta}_s}$ is expanded as
\al{
{\rm Tr}[e^{-t{\bar \Delta}_s}]=\left( \frac{1}{4\pi t}\right)^2 \int \df^4 x\sqrt{\bar g}
\Big[\tr_s {\bf b}_0 + t\tr_s{\bf b}_2 + t^2 \tr_s{\bf b}_4+\cdots \Big],
\label{general heat kernel expansion}
}
with the following heat kernel coefficients~\cite{Gilkey:1995mj,Codello:2008vh}:
\al{
{\bf b}_0&={\bf 1},~~~
{\bf b}_2=\frac{\bar R}{6}{\bf 1} -{\bf Q}_s,&\nn
{\bf b}_4
&=\frac{1}{180}\left( {\bar R}_{\mu\nu\alpha\beta}{\bar R}^{\mu\nu\alpha\beta} -{\bar R}_{\mu\nu}{\bar R}^{\mu\nu} 
+\frac{5}{2}{\bar R}^2
+6D^2 {\bar R}  \right){\bf 1}
+\frac{1}{12}{\bf\Omega}_{\mu\nu}{\bf \Omega}^{\mu\nu}
-\frac{\bar R}{6} {\bf Q}_s
+\frac{1}{2}{\bf Q}_s^2
-\frac{1}{6}D^2{\bf Q}_s.&
\label{heat kernel general}
}
Here ${\bf 1}$ is an unity in the space of the field acted the differential operator and the traces for these coefficients in the spin $\frac{1}{2}$, 1 and 2 fields are defined as
\al{
\tr_{\frac{1}{2}}[{\bf b}_{2l}]&=\delta^{ij}[{\bf b}_{2l}]_{(ij)},&
\tr_1[{\bf b}_{2l}]&={\bar g}^{\mu\nu}[{\bf b}_{2l}]_{(\mu\nu)},&
\tr_2[{\bf b}_{2l}]&={\bar g}^{\mu\alpha}{\bar g}^{\nu\beta}[{\bf b}_{2l}]_{(\mu\nu)(\alpha\beta)},&
}
where indices $i$, $j$ in the trace for spin $\frac{1}{2}$ stand for the Dirac spinor indices.
Using these definitions the traces for the unity matrices become
\al{
\tr_0[{\bf 1}]&=1,&
\tr_{\frac{1}{2}}[{\bf 1}]&=4,&
\tr_1[{\bf 1}]&=4,&
\tr_2[{\bf 1}]&=10.&
\label{trace of unity}
}
${\bf \Omega}_{\mu\nu}=[{\bar \nabla}_\mu,{\bar \nabla}_\nu]$ is the commutator of the covariant derivative and its square acting on vector and tensor fields becomes
\al{
[{\bf\Omega}_{\mu\nu}{\bf \Omega}^{\mu\nu}]_{ij}&=-\frac{1}{4}{\bar R}_{\mu\nu}^{~~ab}{\bar R}^{\mu\nu cd}(J_{ab}J_{cd})_{ij}
\\
[{\bf\Omega}_{\mu\nu}{\bf \Omega}^{\mu\nu}]_{\alpha\beta}&=-{\bar R}_{\mu\nu\gamma\alpha}{\bar R}^{\mu\nu\gamma}_{~~~\beta},
\label{omega2 vector}
\\
[{\bf\Omega}_{\mu\nu}{\bf \Omega}^{\mu\nu}]_{\alpha\beta\rho\sigma}
&=-{\bar R}_{\mu\nu\gamma\alpha}{\bar R}^{\mu\nu\gamma}_{~~~\rho}{\bar g}_{\beta\sigma}
-{\bar R}_{\mu\nu\gamma\beta}{\bar R}^{\mu\nu\gamma}_{~~~\sigma}{\bar g}_{\alpha\rho}
+2{\bar R}_{\mu\nu\alpha\rho}{\bar R}^{\mu\nu}_{~~\beta\sigma},
\label{omega2 tensor}
}
respectively, where $J^{ab}=\frac{i}{4}[\gamma^a,\gamma^b]$ is the generator of Lorentz transformation in Dirac spinor space.
Their traces become
\al{
\delta^{ij}[{\bf\Omega}_{\mu\nu}{\bf \Omega}^{\mu\nu}]_{ij}
&=-\frac{1}{4}{\bar R}_{\mu\nu}^{~~ab}{\bar R}^{\mu\nu cd}  \left({\bar g}^{ac}{\bar g}^{bd}-{\bar g}^{ad}{\bar g}^{bc}\right)
=-\frac{1}{2}{\bar R}_{\mu\nu \rho \sigma}{\bar R}^{\mu\nu \rho \sigma},\\
{\bar g}^{\alpha\beta}[{\bf\Omega}_{\mu\nu}{\bf \Omega}^{\mu\nu}]_{\alpha\beta}
&=-{\bar R}_{\mu\nu\alpha\beta}{\bar R}^{\mu\nu\alpha\beta},\\
{\bar g}^{\mu\alpha}{\bar g}^{\nu\beta}[{\bf\Omega}_{\mu\nu}{\bf \Omega}^{\mu\nu}]_{\alpha\beta\rho\sigma}
&=-8{\bar R}_{\mu\nu\rho\sigma}{\bar R}^{\mu\nu\rho\sigma} +2{\bar R}_{\alpha\beta}{\bar R}^{\alpha \beta}.
}

Consider here the case where ${\bar \Delta}_s$ is the Lichnerowicz Laplacians with the Einstein metric, that is, 
\al{
{\bf Q}_0&=0,&
[{\bf Q}_{\frac{1}{2}}]_{ij}&=\frac{\bar R}{4}\delta_{ij},&
[{\bf Q}_1]_{\mu\nu}&=\frac{{\bar R}}{4}{\bar g}_{\mu\nu},&
[{\bf Q}_2]_{\mu\nu\rho\sigma}&=\frac{\bar R}{2}{\bar g}_{\mu\rho}{\bar g}_{\nu\sigma}- 2\bar R_{\mu\rho\nu\sigma}.&
}
Substituting them with \eqref{trace of unity}, \eqref{omega2 vector} and \eqref{omega2 tensor} for \eqref{heat kernel general} the heat kernel traces are evaluated as
\al{
\tr_0[{\bf b}_0]&=1,&
\tr_0[{\bf b}_2]&=\frac{1}{6}\bar R,&
\tr_0[{\bf b}_4]&=\frac{1}{180}\bar R_{\mu\nu\rho\sigma}\bar R^{\mu\nu\rho\sigma}+\frac{1}{80}\bar R^2,&
\label{heat00}
\\
\tr_{\frac{1}{2}}[{\bf b}_0]&=4,&
\tr_{\frac{1}{2}}[{\bf b}_2]&=-\frac{1}{3}\bar R,&
\tr_{\frac{1}{2}}[{\bf b}_4]&=-\frac{11}{180}\bar R_{\mu\nu\rho\sigma}\bar R^{\mu\nu\rho\sigma}+\frac{1}{120}\bar R^2,&
\label{heat120}
\\
\tr_1[{\bf b}_0]&=4,&
\tr_1[{\bf b}_2]&=-\frac{1}{3}{\bar R},&
\tr_1[{\bf b}_4]&=-\frac{7}{360}\bar R_{\mu\nu\rho\sigma}\bar R^{\mu\nu\rho\sigma}+\frac{1}{120}\bar R^2,&
\label{heat10}
\\
\tr_2[{\bf b}_0]&=10,&
\tr_2[{\bf b}_2]&=-\frac{13}{3}\bar R,&
\tr_2[{\bf b}_4]&=\frac{19}{18}\bar R_{\mu\nu\rho\sigma}\bar R^{\mu\nu\rho\sigma} +\frac{7}{8}\bar R^2.&
\label{heat20}
} 

Let us next derive the heat kernel coefficients for the fields after the York decomposition.
Using the facts that a spin 1 vector field $A_\mu$ is decomposed as
\al{
A_\mu&=A_\mu^\perp+{\bar \nabla}_\mu {\mathcal A},&
{\bar \nabla}^\mu A_\mu^\perp&=0,&
}
and a field ${\bar \nabla}_\mu{\mathcal A}$ satisfies \eqref{LL1}, the trace for a spin 1 field is decomposed as
\al{
\tr_{1}\big[e^{-t{\bar \Delta}_{L1}}\big]
=\tr_{1\perp}\big[e^{-t{\bar \Delta}_{L1}}\big]+\tr_{0}\big[e^{-t{\bar \Delta}_{L0}}\big].
\label{trace relation vector}
}

Next, we derive the trace for a spin 2 tensor field $h_{\mu\nu}$ which is expanded as \eqref{metricdecomposition in ap}.
The trace is decomposed as
\al{
\tr_{2}\big[e^{-t{\bar \Delta}_{L2}}\big]
=\tr_{2\perp}\big[e^{-t{\bar \Delta}_{L2}}\big]+\tr_{1}\big[e^{-t{\bar \Delta}_{L1}}\big] +\tr_{0}\big[e^{-t{\bar \Delta}_{L0}}\big]-\sum_{l=0}^{n_\text{zero}}e^{-t \lambda_l},
\label{trace relation tensor}
}
where the last term corresponds to subtracting zero modes,
\al{
\sum_{l=0}^{n_\text{zero}}e^{-t \lambda_l}= n_\text{KV} +n_\text{CKV}e^{-t \frac{\bar R}{6}}.
}
Here we have written the number of Killing vectors and conformal one as $n_\text{KV}$ and $n_\text{CKV}$, respectively.
In this paper, we choose the Einstein metric such that $n_\text{KV}=n_\text{CKV}=0$.

Using \eqref{heat00}--\eqref{heat20} and the formulas \eqref{trace relation vector} and \eqref{trace relation tensor} the heat kernel coefficients for spin 0 scalar field, spin $\frac{1}{2}$ spinor field, transverse spin 1 vector field and transverse spin 2 tensor field become
\al{
\tr_0[{\bf b}_0]&=1,&
\tr_0[{\bf b}_2]&=\frac{1}{6}\bar R,&
\tr_0[{\bf b}_4]&=\frac{1}{180}\bar R_{\mu\nu\rho\sigma}\bar R^{\mu\nu\rho\sigma}+\frac{1}{80}\bar R^2,&\\
\tr_{\frac{1}{2}}[{\bf b}_0]&=4,&
\tr_{\frac{1}{2}}[{\bf b}_2]&=-\frac{1}{3}\bar R,&
\tr_{\frac{1}{2}}[{\bf b}_4]&=-\frac{7}{360}\bar R_{\mu\nu\rho\sigma}\bar R^{\mu\nu\rho\sigma}+\frac{1}{120}\bar R^2,&\\
\tr_1[{\bf b}_0]&=3,&
\tr_1[{\bf b}_2]&=-\frac{1}{2}\bar R,&
\tr_1[{\bf b}_4]&=-\frac{1}{15}\bar R_{\mu\nu\rho\sigma}\bar R^{\mu\nu\rho\sigma}-\frac{1}{240}\bar R^2,&\\
\tr_2[{\bf b}_0]&=5,&
\tr_2[{\bf b}_2]&=-\frac{25}{6}\bar R,&
\tr_2[{\bf b}_4]&=\frac{10}{9}\bar R_{\mu\nu\rho\sigma}\bar R^{\mu\nu\rho\sigma}+\frac{41}{48}\bar R^2,&
} 
respectively.

The Mellin transformation for \eqref{trace formula} with \eqref{general heat kernel expansion} yields
\al{\label{traceexplict}
\Tr \,[W\fn{{\bar \Delta}_{Ls}}] = \frac{1}{(4\pi)^{2}}\int \df^4x \sqrt{\bar g} \bigg\{ Q_{2}[W]  \tr_s[{\bf b}_0]+ Q_{1}[W] \tr_s[ {\bf b}_2] + Q_0[W] \tr_s[ {\bf b}_4]+\cdots  \bigg\},
}
where
\al{
Q_0[W] &= W(0),&
Q_n[W] &= \frac{1}{\Gamma[n]}\int^\infty_0\df z\, z^{n-1}W[z].&
\label{mellinQ}
}
This formula is used to derive the beta functions in next section.

\section{Derivation of beta function}\label{explicit derivation of beta functions}
In this appendix, we present the explicit calculation of each term appeared in Eq.~\eqref{FRG_totalexpand}.
To evaluate this, we need the derivatives of the cutoff function $\mathcal{R}$, which are given by
{\footnotesize
\al{
&{\partial\mathcal R_\BB\over \p k}
=\pmat{
{\p_k\mathcal R_k^{(h^\perp h^\perp)}}& 0 & 0 & 0 &0 \\
0 & {\p_k\mathcal R_k^{(\xi \xi)}} & 0 &0 & 0 \\
0 & 0 & {\p_k\mathcal R_k^{(SS)}} &0 & 0\\
0 & 0 & 0 & {\p_k\mathcal R_k^{(B^\perp B^\perp)}} & 0\\
0 & 0 & 0 & 0 & {\p_k\mathcal R_k^{(BB)}}
},&
\mathcal R_\FF
&=\pmat{
{\mathcal R_k^{(\bar \chi \chi)} } & 0\\
0 & {\mathcal R_k^{\text{ghost}}}
},
}
\al{
{\p_k\mathcal R_k^{(h^\perp h^\perp)}\over\theta\paren{k^2-{\bar \Delta}_{L2}}}&=
\frac{1}{2} k  \left\{2 F+b  \left(2 k ^2+2
   {\bar \Delta} _{\text{L2}}-3 R\right)-4 R a \right\}
   +{k^2-{\bar \Delta} _{L2}\over4} \left\{4 b  k +b '
   \left(2 k ^2+2 {\bar \Delta} _{L2}-3 R\right)-4 R a
   '+2 \partial _{k }F\right\}
   ,\nn\nn
{\p_k\mathcal R_k^{(\xi \xi)}\over\theta\paren{k^2-{\bar \Delta}_{L1}}}&=   
-\frac{k  \left\{8 k ^2+\left(\rho _2-3\right) R\right\}}{2 \alpha }
    ,\nn\nn
{\p_k\mathcal R_k^{(\sigma\sigma)}\over\theta\paren{k^2-{\bar \Delta}_{L0}}  }
&=
   {3\over 32 \alpha}
   \bigg[16 \alpha  b  k ^3-4 \alpha  F k +4 \alpha 
   \left\{12 k ^3 a +\left(b '+3 a '\right) \left(k
   ^4-{\bar \Delta} _{\text{L0}}^2\right)\right\}+2 \alpha  \partial _{k }F
   \left({\bar \Delta} _{\text{L0}}-k ^2\right)
   \nn
   &\qquad\quad+k  \left\{\left(4 \rho
   _1-7 \rho _2+7\right) R-24 k ^2 \left(\rho _1-\rho
   _2+1\right)\right\}\bigg]
   ,\nn\nn
{\p_k\mathcal R_k^{(\sigma h)}\over\theta\paren{k^2-{\bar \Delta}_{L0}}  }
&=
\frac{1}{64 \alpha  \sqrt{k^2-\frac{R}{3}}}  
\bigg[24 \alpha  k ^5 b '+72 \alpha  k ^5 a '+288 \alpha k ^4 a -48 \beta  k ^4 \rho _1+48 \beta  k ^4 \rho_2-48 \beta  k ^4+4 \alpha  F \left(R-6 k ^2\right)
   \nn
   &-8 \alpha 
   b ' {\bar \Delta} _{\text{L0}} \sqrt{3 k ^2-R} \sqrt{{\bar \Delta}_{\text{L0}} \left\{-\left(R-3 {\bar \Delta} _{\text{L0}}\right)\right\}}-24 \alpha 
   {\bar \Delta} _{\text{L0}} \sqrt{3 k ^2-R} a ' \sqrt{{\bar \Delta}
   _{\text{L0}} \left\{-\left(R-3 {\bar \Delta} _{\text{L0}}\right)\right\}}
   \nn
   &+4 \alpha 
   \partial _{k }F \sqrt{3 k ^2-R} \sqrt{{\bar \Delta} _{\text{L0}}
   \left\{-\left(R-3 {\bar \Delta} _{\text{L0}}\right)\right\}}+\beta  \rho _2
   R^2-\beta  R^2-8 \alpha  k ^3 R b '-24 \alpha  b  k
   ^2 \left(R-4 k ^2\right)
   \nn
   &-24 \alpha  k ^3 R a '-72 \alpha 
   k ^2 R a +12 \beta  k ^2 \rho _1 R-18 \beta  k ^2
   \rho _2 R+18 \beta  k ^2 R+4 \alpha  k  R \partial _{k
   }F-12 \alpha  k ^3 \partial _{k }F\bigg]
   ,\nn\nn
{\p_k\mathcal R_k^{(\sigma \varphi)}\over\theta\paren{k^2-{\bar \Delta}_{L0}}  }
&=  
\phi\frac{F' \left(R-6 k ^2\right)+\partial _{k
   }F' \left(-3 k ^3+\sqrt{3 k ^2-R} \sqrt{{\bar \Delta}
   _{\text{L0}} \left\{-\left(R-3 {\bar \Delta} _{\text{L0}}\right)\right\}}+k 
   R\right)}{2 \sqrt{k ^2-\frac{R}{3}}}
   ,\nn\nn
{\p_k\mathcal R_k^{(hh)}\over\theta\paren{k^2-{\bar \Delta}_{L0}}  }
&=   
\frac{1}{32\alpha }
\bigg[-12 \alpha  F k +4 \alpha  \left\{-\left(b '+3 a '\right)
   \left(k ^2-{\bar \Delta} _{\text{L0}}\right) \left(-3 k ^2-3 {\bar \Delta}
   _{\text{L0}}+R\right)-6 k  a  \left(R-6 k ^2\right)\right\}
   \nn
   &\qquad\quad+6
   \alpha  \partial _{k }F \left({\bar \Delta} _{\text{L0}}-k ^2\right)+8
   \alpha  b  k  \left(6 k ^2-R\right)+\beta ^2 k 
   \left\{-8 k ^2 \left(\rho _1-\rho _2+1\right)-\rho _2 R+R\right\}\bigg]
,\nn\nn
{\p_k\mathcal R_k^{(h\varphi)}\over\theta\paren{k^2-{\bar \Delta}_{L0}}}
&= 
-\frac{3}{2} \phi  \left\{2 F' k +\partial _{k }F'
   \left(k ^2-{\bar \Delta} _{\text{L0}}\right)\right\}
   ,\nn\nn   
{\p_k\mathcal R_k^{(\varphi\varphi)}\over\theta\paren{k^2-{\bar \Delta}_{L0}}}
&= 2k.
\label{Eq: dRdLambda}
}
}
The prime denotes the derivative with respect to $\phi^2$ and $k$ for $V, F$ and $a, b$ respectively.

Then, let us calculate Eq.~\eqref{FRG_totalexpand}.
Eq.~\eqref{FRG_totalexpand} consists of three contributions.
First and second terms correspond to the contributions from loops of bosonic particle and fermionic particle, respectively.
Third term comes from boson fermion mixed diagrams.
\subsection{Loop of bosonic particle}
The first term corresponds to the loop of particles which obey bosonic statics.
The transverse traceless tensor part gives
{\footnotesize
\al{
&\frac{1}{2}{\rm Tr}\sqbr{ \mathcal{M}_{BB}^{-1}\p_t{\mathcal R}_k}\bigg|_{h^\perp h^\perp}
=
\frac{1}{2}{\rm Tr}\sqbr{{1\over\Gamma_k ^{(\perp\perp)}+\mathcal R_k^{(h^\perp h^\perp)}}\p_k\mathcal R_k^{(h^\perp h^\perp)}}
\nn
					&=					
\frac{1}{2}{\rm Tr}
\bigg[
\frac{k\,\theta(k^2-{\bar \Delta}_{L2})}
{2 F k ^2+\left(R-2 k ^2\right)\br{b  \left(R-k ^2\right)+2 R a}-2 V-2 Y}\times
   \nn&
\br{3
   k ^2 R b '-4 F k -2 k ^4 b '+b  \left(6 k  R-8
   k ^3\right)+4 k ^2 R a '+8 k  R
   a -3 R {\bar \Delta}_{L2} b '-4 R {\bar \Delta}_{L2} a '+2 \partial
   _{k }F \left({\bar \Delta}_{L2}-k ^2\right)+2 {\bar \Delta}_{L2}^2 b'}
\bigg]
\nn
&=
{1\over(4\pi)^2}\int_x\sqbr{A_0^{h^\perp}k^4+A_2^{h^\perp}\phi^2k^2+A_4^{h^\perp}\phi^4+B_0^{h^\perp}\,Rk^2+B_2^{h^\perp}\,R\phi^2+C^{h^\perp} R^2+D^{h^\perp} R_{\mu\nu\rho\sigma}R^{\mu\nu\rho\sigma}+E^{h^\perp} Y},
\nn
A_0^{h^\perp}&=
{A^{h^\perp}\over k^4}
\bigg|_{\phi=0}
,\nn
A_2^{h^\perp}&=
{\partial\over \partial\phi^2}
{A^{h^\perp}\over k^2}
\bigg|_{\phi=0}
,\nn
A_4^{h^\perp}&=
{1\over2}{\partial^2\over \partial(\phi^2)^2}
A^{h^\perp}
\bigg|_{\phi=0}
,\nn
B_0^{h^\perp}&=
{B^{h^\perp}\over k^2}
   \bigg|_{\phi=0}
,\nn
B_2^{h^\perp}&= 
{\partial\over \partial\phi^2}  
B^{h^\perp}
   \bigg|_{\phi=0}
,\nn
C^{h^\perp}&=
\frac{k ^2 }{288 \br{V-k ^2\left(F+b  k ^2\right)}^3}
\bigg[
123 \br{V-k ^2 \left(F+bk ^2\right)}^2 
\br{2 F+k ^3 b'+k  (4 b  k +\partial _{k}F)}
   \nn&
+50 k ^2 \br{V-k ^2\left(F+b  k ^2\right)} 
\bigg\{
3 \sqbr{
	4k  a' \br{V-k ^2 \left(F+b k ^2\right)}
	+(3 b +4 a )\br{k ^3 (4 b  k +\partial _{k}F)+4 V}
	}
   \nn&
   +b' \br{-9 F k^3+k ^5 (3 b +16 a )+9 k V}
\bigg\}
-15 k ^4 
\bigg\{
2 k ^2 (3b +4 a ) \left(F k ^2+b  k^4-V\right) 
\left(3 k  b '+18 b +4k  a '+24 a \right)
   \nn&
-\sqbr{2 \br{6 F+k (12 b  k +\partial _{k }F)}+3 k ^3b'}
\sqbr{k ^2 \br{k ^2 \left(7b ^2+20 b  a +16 a ^2\right)-2 F(b +2 a )}
+2 V (b +2 a)}
\bigg\}\bigg]      
   \bigg|_{\phi=0}
,\nn
D^{h^\perp}&=-\frac{5 k ^2 \br{2 F+k ^3 b '+k 
   (4 b  k +\partial _{k }F)}}{9
   \left(F k ^2+b  k ^4-V\right)}  
\bigg|_{\phi=0}
,\nn
E^{h^\perp}&=-\frac{5 k ^6 
\sqbr{2 \br{6 F+k  (12 b k +\partial _{k }F)}+3 k ^3 b'}}
{24 \br{V-k ^2 \left(F+b  k^2\right)}^2}  
\bigg|_{\phi=0},
}
}
where $A^{h^\perp}$ and $B^{h^\perp}$ are given by
{\footnotesize
\al{
A^{h^\perp}&=\frac{5 k ^6}{24}
    \frac{2 \br{6 F+k  (12 b  k
   +\partial _{k }F)}+3 k ^3 b'}
   {V-k ^2 \left(F+b  k^2\right)},
   \nn
B^{h^\perp}&=
\frac{5 k ^4}{144 \left(F k ^2+b  k^4-V\right)^2}
\bigg[6 \bigg\{20 F^2 k ^2
   +5 F\br{k ^3 (12 b  k +\partial _{k}F)-4 V}
   +4 k ^3 a' \left(Fk ^2+b  k ^4-V\right)
   \nn
   &\qquad\qquad\qquad\qquad\qquad\quad
   +k  
   \sqbr{22b ^2 k ^5+2 b  k ^4 \partial_{k }F
   -4 k  a  \br{k ^3 (6b  k +\partial _{k }F)+6 V}-5
   \partial _{k }F V-58 b  k 
   V}
   \bigg\}
   \nn
   &\qquad\qquad\qquad\qquad\qquad\quad
   +b ' \br{58 F k ^5+k ^7(31 b -36 a )-58 k ^3V}\bigg].
}
}

The $\xi$ component gives
{\footnotesize
\al{
&\frac{1}{2}{\rm Tr}\sqbr{ \mathcal{M}_{BB}^{-1}\p_t{\mathcal R}_k}\bigg|_{\xi\xi}
=
\frac{1}{2}{\rm Tr}\sqbr{{1\over\Gamma_k ^{(\xi\xi)}+\mathcal R_k^{(\xi\xi)}}\p_k\mathcal R_k^{(\xi\xi)}}
\nn
					&=
\frac{1}{2}{\rm Tr}
\bigg[
-\frac{4 k ^2 \br{8 k ^2+\left(\rho _2-3\right) R}}
{-4\alpha  F R-\left(R-2 k ^2\right) \br{4 k ^2+\left(\rho_2-1\right) R}
+8 \alpha  V+8 \alpha  Y}
   \theta \left(k ^2-{\bar \Delta}_{L1}\right)
\bigg]
\nn
&=
{1\over(4\pi)^2}\int_x\sqbr{A_0^{\xi}k^4+A_2^{\xi}\phi^2k^2+A_4^{\xi}\phi^4+B_0^{\xi}\,Rk^2+B_2^{\xi}\,R\phi^2+C^{\xi} R^2+D^{\xi} R_{\mu\nu\rho\sigma}R^{\mu\nu\rho\sigma}+E^{\xi} Y},
}
\al{
A_0^{\xi}&={A^\xi\over k^4}
\bigg|_{\phi=0}
,\nn
A_2^{\xi}&={\p\over \p \phi^2}{A^\xi\over k^2}
\bigg|_{\phi=0}
,\nn
A_4^{\xi}&={1\over2}{\p^2\over \p (\phi^2)^2}A^\xi
\bigg|_{\phi=0}
,\nn
B_0^{\xi}&={B^\xi\over k^2}
\bigg|_{\phi=0}   
,\nn
B_2^{\xi}&={\p\over \p \phi^2}B^\xi
\bigg|_{\phi=0}
,\nn
C^{\xi}&=\frac{k ^4}{480 \left(k ^4+\alpha  V\right)^3}
\bigg[ \alpha  k ^4 \left(593 V-360 \alpha 
   F^2\right)-570 \alpha  F k ^6+30 \rho _2 \br{9 \alpha  F k
   ^6-3 \alpha  k ^2 V \left(\alpha  F+5 k ^2\right)+k ^8+2
   \alpha ^2 V^2}
   \nn&\qquad\qquad\qquad\qquad
   +510 \alpha ^2 F k ^2 V-41 k ^8-176 \alpha
   ^2 V^2+45 \rho _2^2 \left(\alpha  k ^4 V-k
   ^8\right)\bigg]      
\bigg|_{\phi=0}
,\nn
D^{\xi}&=\frac{2 k ^4}{15 \left(k ^4+\alpha  V\right)}  
\bigg|_{\phi=0}
,\nn
E^{\xi}&=\frac{3 \alpha  k ^8}{\left(k ^4+\alpha  V\right)^2}  
\bigg|_{\phi=0},
}
}
where $A^{\xi}$ and $B^{\xi}$ are given by
{\footnotesize
\al{
A^\xi&=-\frac{3 k ^8}{k ^4+\alpha  V},
\nn
B^\xi&=
\frac{k ^6}{8}
\frac{-k ^2 \left(12 \alpha  F+k^2\right)+3 \rho _2 \left(k ^4-\alpha  V\right)+17 \alpha 
   V}{ \left(k ^4+\alpha  V\right)^2}.
}
}

The loop of $\sigma, h$ and $\varphi$ contributes as
{\footnotesize
\al{
&\frac{1}{2}{\rm Tr}\sqbr{ \mathcal{M}_{BB}^{-1}\p_t{\mathcal R}_k}\bigg|_{\xi\xi}
=
\frac{1}{2}{\rm Tr}\sqbr{{1\over\Gamma_k ^{(SS)}+\mathcal R_k^{(SS)}}\p_k\mathcal R_k^{(SS)}}
\nn
&=
{1\over(4\pi)^2}\sqbr{A_0^S\int_xk^4+A_2^S\int_x\phi^2k^2+A_4^S\int_x\phi^4+B_0^S\int_x\,Rk^2+B_2^S\int_x\,R\phi^2+C^S\int_x R^2+D^S \int_x R_{\mu\nu\rho\sigma}R^{\mu\nu\rho\sigma}+E^S \int_x Y}
, 
}
\al{
A_0^S&\to{A^S\over k^4}\bigg|_{\phi=0}
,\nn
A_2^S&\to{\partial\over \partial\phi^2}{A^S\over k^2}\bigg|_{\phi=0}
,\nn
A_4^S&\to{1\over2}{\p^2\over \p (\phi^2)^2}A^S\bigg|_{\phi=0}
,\nn
B_0^S&\to {B^S\over k^2}\bigg|_{\phi=0}
,\nn
B_2^S&\to {\partial\over \partial\phi^2}{B^S}\bigg|_{\phi=0}
\nonumber
}
\al{
C^S&\to
-\frac{1}{2880 \br{2 (b +3 a )k ^4-F k ^2+V}^3 \left(k ^2+2 V'\right)^3}
\bigg[4 (b +3 a )^2 
\bigg\{
	\br{131 k ^2\left(b '+3 a '\right)-128 \partial _{k }F}\left(k ^2+2 V'\right)^3
   \nn&
   +24 b  k  \left(480F'^2 k ^6+\left(k ^2+2 V'\right)^2 
   \left(115k ^2+224 V'\right)
   +80 F' \left(k ^6+2 V' k^4\right)\right)
   \nn&
   +72 k  a  \left(480 F'^2
   k ^6+\left(k ^2+2 V'\right)^2 \left(115 k ^2+224
   V'\right)+80 F' \left(k ^6+2 V' k ^4\right)\right)\bigg\}
   k ^{11}
   \nn&
   -4 V (b +3 a ) \bigg\{\br{69 \left(b '+3a '\right) k ^2+38 \partial _{k }F}
   \left(k ^2+2 V'\right)^3-36 b  \big[480 F'^2k ^7
   +\left(k ^2+2 V'\right)^2 \left(53 k ^2+100V'\right) k 
   \nn&
   +80 F' \left(k ^7+2 V' k^5\right)\big]-108 a  \left(480 F'^2 k
   ^7+\left(k ^2+2 V'\right)^2 \left(53 k ^2+100 V'\right)
   k +80 F' \left(k ^7+2 V' k ^5\right)\right)\bigg\}
   k ^7
   \nn&
   -24 F^3 \left(240 F'^2 k ^6+\left(k
   ^2+2 V'\right)^2 \left(31 k ^2+59 V'\right)+40 F' \left(k
   ^6+2 V' k ^4\right)\right) k ^6
   \nn&
   +F^2 
   \bigg\{\bigg(
   -\br{89\left(b '+3 a '\right) k ^2+8 \partial _{k}F} \left(k ^2+2 V'\right)^3
   +72 b  k  \big\{480F'^2 k ^6+3 \left(k ^2+2 V'\right)^2 \left(21k ^2+40 V'\right)
   \nn&
   +80 F' \left(k ^6+2 V' k^4\right)\big\}
   +216 k  a  \left(480 F'^2k ^6+3 \left(k ^2+2 V'\right)^2 \left(21 k ^2+40
   V'\right)+80 F' \left(k ^6+2 V' k ^4\right)\right)
   \bigg)
   k ^3
   \nn&
   +36 V \left(480 F'^2 k ^6+\left(k^2+2 V'\right)^2 \left(61 k ^2+116 V'\right)
   +80 F' \left(k^6+2 V' k ^4\right)\right)\bigg\} k ^4
   \nn&
   +V^2 \bigg\{-\left(89\left(b '+3 a '\right) k ^2
   +8 \partial _{k}F\right) \left(k ^2+2 V'\right)^3
   +24 b  k 
   \big[1440 F'^2 k ^6+\left(k ^2+2 V'\right)^2
   \left(163 k ^2+308 V'\right)
   \nn&
   +240 F' \left(k ^6+2 V'
   k ^4\right)\big]+72 k  a  \left(1440
   F'^2 k ^6+\left(k ^2+2 V'\right)^2 \left(163
   k ^2+308 V'\right)+240 F' \left(k ^6+2 V' k
   ^4\right)\right)\bigg\} k ^3
   \nn&
   +2 F \bigg\{2 (b +3 a )
   \big[
   \br{69 \left(b '+3 a '\right) k ^2+38 \partial_{k}F} \left(k ^2+2 V'\right)^3
   -24 b 
   \big\{720 F'^2 k ^7+\left(k ^2+2 V'\right)^2
   \left(106 k ^2+203 V'\right) k
   \nn&
    +120 F' \left(k ^7+2V' k ^5\right)
   \big\}
   -72 a  \left(720 F'^2k ^7+\left(k ^2+2 V'\right)^2 \left(106 k ^2+203
   V'\right) k +120 F' \left(k ^7+2 V' k^5\right)\right)\big] k ^7
   \nn&
   +V \big[\br{89 \left(b '+3a '\right) k ^2+8 \partial _{k }F}
   \left(k ^2+2 V'\right)^3
   -48 b 
   \big\{720 F'^2
   k ^7+\left(k ^2+2 V'\right)^2 \left(88 k ^2+167
   V'\right) k 
   \nn&
   +120 F' \left(k ^7+2 V' k^5\right)
   \big\}
   -144 a  \left(720 F'^2 k
   ^7+\left(k ^2+2 V'\right)^2 \left(88 k ^2+167 V'\right)
   k +120 F' \left(k ^7+2 V' k ^5\right)\right)\big]k ^3
   \nn&
   -36 V^2 \left(240 F'^2 k ^6+3
   \left(k ^2+2 V'\right)^2 \left(10 k ^2+19 V'\right)+40 F'
   \left(k ^6+2 V' k ^4\right)\right)\bigg\} k ^2
   \nn&
   +12
   V^3 \left(480 F'^2 k ^6+\left(k ^2+2
   V'\right)^2 \left(59 k ^2+112 V'\right)+80 F' \left(k ^6+2
   V' k ^4\right)\right)\bigg]
,\nonumber
}
\al{
D^S&\to
\frac{1}{360} k ^2 \left(\frac{6 F k ^2+k ^3
   \br{\partial _{k }F-2 k  \left(k  b '+8
   b +3 k  a '+24 a \right)}-4 V}{k ^2
   \br{-F k ^2+2 k ^4 (b +3 a)+V}}-\frac{2}{k ^2+2 V'}\right)
,\nn
E^S&\to
\frac{k ^6}
{12 \left(k ^2+2 V'\right)^2\br{-F k ^2+2 k ^4 (b +3 a )+V}^2}\times
   \nn&
\bigg[36 k ^7 F' b '+288 b  k ^6
   F'+108 k ^7 F' a '+864 k ^6 a  F'-72 F k
   ^4 F'-12 k ^5 \partial _{k }F F'-24 k ^3 \partial
   _{k }F F' V'+72 k ^5 F' b ' V'
   \nn&
   +48 b  k ^4
   \left(6 F'-7\right) V'+216 k ^5 F' V' a '+864 k ^4
   a  F' V'-144 V F' V'-6 F k ^4
   \nn&
   -12 k  \partial
   _{k }F' \left(k ^2+2 V'\right) \br{-F k
   ^2+2 k ^4 (b +3 a )+V}+120 F V' \left(k
   ^2+V'\right)+3 k ^7 b '+24 b  k ^6+9 k ^7
   a '+72 k ^6 a 
   \nn&
   -k ^5 \partial _{k }F+8
   k ^3 \partial _{k }F V'+20 k  \partial _{k }F
   V'^2-24 k ^5 b ' V'-60 k ^3 b '
   V'^2-480 b  k ^2 V'^2-72 k
   ^5 V' a '-1008 k ^4 a  V'
   \nn&
   -180 k ^3
   V'^2 a '-1440 k ^2 a 
   V'^2-72 V V'\bigg],
}
}
where $A^S$ and $B^S$ are given by
{\footnotesize
\al{
&A^S=
{k ^4\over12}
\big[
-12 k ^4 \phi ^2F'^2
+24 k ^2 \phi ^2 F' V'
-F k ^2\left(k ^2+4 \phi ^2 V''+2 V'\right)
+2 b  k ^6
+6k ^6 a 
+4 \phi ^2 V'' \br{2 k ^4 (b +3 a)+V}
\nn&
+4 b  k ^4 V'+12 k ^4 a  V'-12 \phi ^2V'^2+2 V V'+k ^2 V\big]^{-1}\times
\nn&
\bigg[
288 k ^4 \phi ^2 F'^2
+24k ^5 \partial _{k }F' \phi ^2 F'
-432 k ^2\phi ^2 F' V'
+12 F k ^2 \left(2 k ^2+6 \phi ^2 V''+3V'\right)
-3 k ^7 b '-60 b  k ^6-9 k ^7a '-180 k ^6 a
\nn& 
+k ^5 \partial _{k }F
-4k ^3 \phi ^2 V'' 
\br{3 k  \left(k  b '+16b +3 k  a '+48 a \right)-\partial _{k}F}
+2 k ^3 \partial _{k }F V'-24 k ^3 \partial_{k }F' \phi ^2 V'-6 k ^5 b ' V'-96 b k ^4 V'
   \nn&
   -18 k ^5 V' a '-288 k ^4 a  V'+144
   \phi ^2 V'^2-6 V \left(3 k ^2+8 \phi ^2 V''+4
   V'\right)\bigg],
}
\al{
&B^S=
-\frac{k ^2}{72 \left(k ^2+2V'\right)^2 \br{-F k ^2+2 k ^4 (b +3 a)+V}^2}\times
   \nn&
   \bigg[ k ^3 V 
   \bigg\{
   12 b  \br{k ^5\left(24 F'+19\right)+68 k ^3 V'+60 k V'^2}
   +36 a \br{k ^5 \left(24F'+19\right)+68 k ^3 V'+60 k V'^2}
   \nn&
   +\left(k ^2+2 V'\right)^2 \br{2k ^2 \left(b '+3 a '\right)-3 \partial _{k}F}
   \bigg\}
   +2 k ^7 (b +3 a ) 
   \bigg\{48 b 
   \br{k ^5 \left(3 F'+4\right)+15 k ^3 V'+14 k V'^2}
   \nn&
   +144 a \br{k ^5 \left(3F'+4\right)+15 k ^3 V'+14 k V'^2}
   +\left(k ^2+2 V'\right)^2 
   \br{11k ^2 \left(b '+3 a '\right)-6 \partial _{k}F}
   \bigg\}
   \nn&
   -F k ^2 \bigg\{k ^3 
   \bigg(
   96 b \br{3 k ^5 \left(F'+1\right)+11 k ^3 V'+10 k V'^2}
   +288 a \br{3 k ^5\left(F'+1\right)+11 k ^3 V'+10 k V'^2}
   \nn&
   +\left(k ^2+2 V'\right)^2 
   \br{2k ^2 \left(b '+3 a '\right)-3 \partial _{k}F}
   \bigg)
   +24 V \br{k ^4 \left(6 F'+5\right)+18 k^2 V'+16 V'^2}\bigg\}
   \nn&
   +6 V^2 \br{3 k ^4\left(4 F'+3\right)+32 k ^2 V'+28 V'^2}
   +6 F^2\br{k ^8 \left(12 F'+11\right)+40 k ^6 V'+36 k ^4V'^2}\bigg]
\nn
&
-\phi^2
\frac{k ^2}
{6 \br{2(b +3 a ) k ^4-F k ^2+V}^3 
\left(k^2+2 V'\right)^3}\times
   \nn&
\bigg[
4 F^3 \br{\left(12 F' k ^2+k ^2+2V'\right) V''-3 \left(k ^4+2 V' k ^2\right) F''} k ^8
+2 (b +3 a ) 
\bigg\{
   16 b ^2 
   \bigg(
   3\left(k ^4+2 V' k ^2\right) F''
   \nn&
   -\left(12 F' k^2+k ^2+2 V'\right) V''
   \bigg) k ^7
   +144 a ^2 \bigg(3\left(k ^4+2 V' k ^2\right) F''-\left(12 F' k^2+k ^2+2 V'\right) V''\bigg) k ^7
   \nn&
   -6 \partial _{k}F' a  \left(k ^2+2 V'\right) 
   \bigg(F' \left(24F'+19\right) k ^4+\left(14 F'-13\right) V' k ^2-26V'^2\bigg) k ^4
   \nn&
   +72 a  
   	\bigg(
   F'^2\left(12 F'+11\right) k ^8-F' \left(4 F'+25\right) V' k
   ^6+\br{11-2 F' \left(2 F'+29\right)} V'^2 k
   ^4
   \nn&
   +2 \left(19-20 F'\right) V'^3 k ^2+32V'^4
   	\bigg) k 
   +2 b  \bigg(144F'^3 k ^8-12 F'^2 \big\{(2 \partial_{k }F' k -11) k^8
   \nn&
   +4 (\partial _{k}F' k +1) V' k ^6+4 V'^2 k^4\big\}
   +\left(k ^2+2 V'\right) 
   \big\{48 a  \left(3k ^2 F''-V''\right) k ^6+13 \partial _{k }F' V' k ^5
   \nn&
   +2 (13 \partial _{k }F' k +66)
   V'^2 k ^2+192 V'^3\big\}
   -F'\big\{\left(19 \partial _{k }F' k +576 a V''\right) k ^8+4 (13 \partial _{k }F' k +75)V' k ^6
   \nn&
   +4 (7 \partial _{k }F' k +174)V'^2 k ^4
   +480 V'^3 k^2\big\}\bigg) k 
   +\left(k ^2 F'-V'\right) \left(k^2+2 V'\right) 
   \bigg(F' \big\{\left(36 F'+41\right) b ' k^2
   +3 \left(36 F'+41\right) a ' k ^2
   \nn&
   -4 \partial _{k }F
   \left(3 F'+4\right)\big\} k ^4
   +\left(2 F'-1\right) V' \br{23k ^2 \left(b '+3 a '\right)-10 \partial _{k}F} k ^2
   -2 V'^2 \br{23 k ^2\left(b '+3 a '\right)-10 \partial _{k}F}\bigg)
\bigg\} k ^7
   \nn&
   +F^2 \bigg(72 \left(k ^8-2k ^6 V'\right) F'^3
   -6 \br{(4 \partial _{k}F' k -7) k ^8
   +2 (4 \partial _{k }F'k +19) V' k ^6+8 V'^2 k ^4}F'^2
   \nn&
   -\br{13 \partial _{k }F' k ^9+4
   (7 \partial _{k }F' k +30) V' k ^6+4
   (\partial _{k }F' k +30) V'^2 k
   ^4+144 \left(2 (b +3 a ) k ^4+V\right) V'' k
   ^4+48 V'^3 k ^2} F'
   \nn&
   +\left(k ^2+2V'\right) \br{7 \partial _{k }F' V' k ^5+14
   (\partial _{k }F' k +3) V'^2 k
   ^2+12 \left(2 (b +3 a ) k ^4+V\right) \left(3 k
   ^2 F''-V''\right) k ^2+48 V'^3}\bigg)
   k ^4
   \nn&
   +V \bigg\{48 b ^2 \bigg(3 \left(k ^4+2 V' k^2\right) F''
   -\left(12 F' k ^2+k ^2+2 V'\right) V''\bigg)k ^7
   \nn&
   +432 a ^2 k ^7\bigg(3 \left(k ^4+2 V' k^2\right) F''
   -\left(12 F' k ^2+k ^2+2 V'\right) V''\bigg)
   \nn&
   +24 a  \bigg(12 F'^2 \left(3 F'+1\right)
   k ^8-3 F' \left(F' \left(24 F'+53\right)+17\right) V' k
   ^6-3 \left(26 F'^2+2 F'-7\right) V'^2
   k ^4
   \nn&
   -\partial _{k }F' \left(k ^2+2 V'\right)
   \br{4 F' \left(3 F'+2\right) k ^4+\left(4 F'-5\right) V'
   k ^2-10 V'^2} k ^3
   +3 \left(16F'+19\right) V'^3 k ^2+30 V'^4\bigg)
   k
   \nonumber
   }
\al{   
   &+8 b \bigg(36 \left(k ^8-2 k ^6 V'\right)
   F'^3
   -3 \br{4 (\partial _{k }F' k -1)k ^8
   +(8 \partial _{k }F' k +53) V' k^6
   +26 V'^2 k ^4} F'^2
   \nn&
   -\br{8\left(\partial _{k }F' k +54 a  V''\right)k^8
   +(20 \partial _{k }F' k +51) V' k^6
   +2 (4 \partial _{k }F' k +3) V'^2k^4
   -48 V'^3 k ^2} F'
   \nn&
   +\left(k^2+2 V'\right) 
   \br{36 a  \left(3 k ^2 F''-V''\right)k ^6+5 \partial _{k }F' V' k ^5+(10 \partial_{k }F' k +21) V'^2 k ^2+15V'^3}\bigg) k 
   \nn&
   +\left(k ^2 F'-V'\right)
   \left(k ^2+2 V'\right) 
   \bigg(F' \br{\left(36 F'+23\right)
   b ' k ^2+3 \left(36 F'+23\right) a ' k ^2-2
   \partial _{k }F \left(6 F'+5\right)} k ^4
   \nn&
   +\left(2F'-1\right) V' 
   \br{5 k ^2 \left(b '+3 a '\right)-4\partial _{k }F} k ^2
   -2 V'^2 \br{5k ^2 \left(b '+3 a '\right)-4 \partial _{k}F}\bigg)
   \bigg\} k ^3
   \nn&
   +F \bigg\{
   \bigg(48 b ^2
   \br{\left(12 F' k ^2+k ^2+2 V'\right) V''-3\left(k ^4+2 V' k ^2\right) F''} k ^7
   +432a ^2 \big\{\left(12 F' k ^2+k ^2+2 V'\right) V''
   \nn&
   -3\left(k ^4+2 V' k ^2\right) F''\big\} k ^7
   +12a  \big[-3 F'^2 \left(36 F'+23\right) k ^8
   +6F' \br{F' \left(12 F'+35\right)+29} V' k^6
   \nn&
   +3 \br{8 F'\left(5 F'+11\right)-23} V'^2 k ^4
   +2 \partial_{k }F' \left(k ^2+2 V'\right) \br{4 F' \left(3F'+2\right) k ^4+\left(4 F'-5\right) V' k ^2-10V'^2} k ^3
   \nn&
   +6 \left(20 F'-37\right)V'^3 k ^2-168 V'^4\big] k 
   -4b \big[-3 F'^2 \br{(8 \partial _{k}F' k -23) k ^4+2 (8 \partial _{k }F'
   k +35) V' k ^2+40 V'^2} k ^4
   \nn&
   +36F'^3 \left(3 k ^8-2 k ^6V'\right)
   +\left(k ^2+2 V'\right) \br{72 a  \left(3
   k ^2 F''-V''\right) k ^6+10 \partial _{k }F'
   V' k ^5+(20 \partial _{k }F' k +69)
   V'^2 k ^2+84 V'^3}
   \nn&
   -2 F' \br{8\left(\partial _{k }F' k +54 a  V''\right)k ^8
   +(20 \partial _{k }F' k +87) V' k^6
   +4 (2 \partial _{k }F' k +33) V'^2k^4
   +60 V'^3 k ^2}\big] k
   \nn&
   -\left(k ^2 F'-V'\right) \left(k ^2+2 V'\right) 
   \big[F'\br{\left(36 F'+23\right) b ' k ^2+3 \left(36F'+23\right) a ' k ^2-2 \partial _{k }F \left(6F'+5\right)} k ^4
   \nn&
   +\left(2 F'-1\right) V' \br{5 k^2 \left(b '+3 a '\right)-4 \partial _{k }F}k ^2
   -2 V'^2 \br{5 k ^2 \left(b '+3a '\right)-4 \partial _{k }F}\big]
   \bigg) k^3
   -12 V^2 \bigg(3 \left(k ^6+2 V' k ^4\right) F''
   \nn&
   -k^2 \left(12 F' k ^2+k ^2+2 V'\right) V''\bigg)
   -2 V\bigg(36 \left(k ^8-6 k ^6 V'\right) F'^3
   \nn&
   -3\br{(8 \partial _{k }F' k -5) k ^8+8 (2
   \partial _{k }F' k +11) V' k ^6+4
   V'^2 k ^4} F'^2
   -\big[\br{13\partial _{k }F' k +288 (b +3 a )V''} k ^8
   \nn&
   +28 (\partial _{k }F' k +3) V'k ^6
   +4 (\partial _{k }F' k +3)V'^2 k ^4
   -24 V'^3 k ^2\big]F'
   +\left(k ^2+2 V'\right) \big[24 (b +3 a ) \left(3
   k ^2 F''-V''\right) k ^6
   \nn&
   +7 \partial _{k }F'V' k ^5
   +(14 \partial _{k }F' k +33)V'^2 k ^2+30 V'^3\big]\bigg)\bigg\}k ^2
   +4 V^3 \br{3 \left(k ^6+2 V' k ^4\right)F''
   -k ^2 \left(12 F' k ^2+k ^2+2 V'\right)V''}
   \nn&
   +V^2 \bigg\{-288 F'^3 V' k ^6
   -12F'^2 \bigg(2 \partial _{k }F' k
   ^9+k ^8+(4 \partial _{k }F' k +25) V' k
   ^6-2 V'^2 k ^4\bigg)
   \nn&
   +\left(k ^2+2 V'\right)
   \bigg(24 (b +3 a ) \left(3 k ^2 F''-V''\right) k^6+7 \partial _{k }F' V' k ^5+2 (7 \partial_{k }F' k +12) V'^2 k ^2+12V'^3\bigg)
   \nn&
   -F' \bigg(\br{13 \partial _{k}F' k +288 (b +3 a ) V''} k ^8
   +4 (7\partial _{k }F' k +12) V' k ^6+4 (\partial_{k }F' k -24) V'^2 k ^4
   -96V'^3 k ^2\bigg)\bigg\}
\bigg].
}
}
Here we have provided the result for Landau gauge for simplicity.

Finally, the contribution from $B$ ghost is
{\footnotesize
\al{
&\frac{1}{2}{\rm Tr}\left. \frac{\p_t{\mathcal R}_k}{\Gamma _k^{(B^\perp B^\perp)}
									+{\mathcal R}_k}\right|_{ B^\perp B^\perp}
								+\frac{1}{2}{\rm Tr}\left. \frac{\p_t{\mathcal R}_k}{\Gamma _k^{(BB)}
								+{\mathcal R}_k}\right|_{BB}\nn
							&
=\frac{1}{2}\Tr
\sqbr{
\frac{-2 \alpha  k ^2 \theta \left(k^2-{\bar \Delta}_{L1}\right)}
   {\frac{1}{4} \alpha  \left(\rho _2-1\right)R
   +\alpha  \br{\left(k ^2-z\right) \theta\left(k ^2-{\bar \Delta}_{L1}\right)+{\bar \Delta}_{L1}}   }
}
   \nn&
   +\frac{1}{2}\Tr
   \sqbr{\frac{-8 k ^2 \left(\rho _1-\rho _2+1\right) \theta
   \left(k ^2-{\bar \Delta}_{L0}\right)}{\left(\rho _2-1\right) R
   +4\left(\rho _1-\rho _2+1\right) \br{\left(k^2-z\right) \theta \left(k ^2-{\bar \Delta}_{L0}\right)+{\bar \Delta}_{L0}}}}
   \nn
&=   
{1\over(4\pi)^2}
\int_x\sqbr{A_0^{B}k^4+A_2^{B}\phi^2k^2+A_4^{B}\phi^4+B_0^{B}\,Rk^2+B_2^{B}\,R\phi^2+C^{B} R^2+D^{B} R_{\mu\nu\rho\sigma}R^{\mu\nu\rho\sigma}+E^{B} Y},
}
}
where
{\footnotesize
\al{&
A_0^{B}=-2
,&&
A_2^{B}=0
,&&
A_4^{B}=0
,\nn
&
B_0^{B}=\frac{1}{24} \left(\frac{3 \rho _1}{\rho _1-\rho _2+1}+9 \rho
   _2-4\right)
,&&
B_2^{B}=0   
,&&
C^{B}=\frac{1}{480} \left(\frac{5 \rho _1 \left(7 \rho _1-10 \rho
   _2+10\right)}{\left(\rho _1-\rho _2+1\right){}^2}+15
   \left(2-3 \rho _2\right) \rho _2-24\right)      
,\nn
&
D^{B}=  \frac{11}{180}
,&&
E^{B}=0.  
\label{Eq:B contribution}
}
}

\subsection{Loop of fermionic particle}
Next, the functional traces corresponding to loops of fermionic statics particle are shown.
The contribution from $\chi$ particle is
{
\al{
&-{\rm Tr}\sqbr{ \mathcal{M}_{FF}^{-1}\p_t{\mathcal R}_k}\bigg|_{\bar{\chi}\chi}
=
-{1\over2}{\rm Tr}\sqbr{\p_t\mathcal{R}_k\over\Gamma_k^{(\chi\chi)}+\mathcal{R}_k }
=
-{1\over2}{\rm Tr}\sqbr{{-2k^2\over k^2 + y^2\phi^2}\theta(k^2-{\bar \Delta}_{L0})}\nn
&=
{1\over(4\pi)^2}
\int_x\sqbr{A_0^{\chi}k^4+A_2^{\chi}\phi^2k^2+A_4^{\chi}\phi^4+B_0^{\chi}\,Rk^2+B_2^{\chi}\,R\phi^2+C^{\chi} R^2+D^{\chi} R_{\mu\nu\rho\sigma}R^{\mu\nu\rho\sigma}+E^{\chi} Y},
}
}
where
{
\al{&
A_0^{\chi}=2
,&&
A_2^{\chi}=-2y^2
,&&
A_4^{\chi}=2y^4
,\nn
&
B_0^{\chi}=-{1\over3}
,&&
B_2^{\chi}={y^2\over3}   
,&&
C^{\chi}={1\over120}      
,\nn
&
D^{\chi}=-{7\over360}  
,&&
E^{\chi}=0.  
}
}

The contribution from $C$ ghost is
{\footnotesize
\al{
							&-\left. \Tr \frac{\p_t {\mathcal R} _k}{\Gamma_k^{(C^\perp C^\perp)}+{\mathcal R}_{k}}\right|_{{\bar C}^\perp C}
							-\left. \Tr \frac{\p_t {\mathcal R} _k}{\Gamma_k^{(CC)}+{\mathcal R}_{k}}\right|_{\bar C C}
\nn
&=
-\Tr\sqbr{8 k ^2 \left(\frac{2}{-4 k ^2-\rho _2
   R+R}+\frac{1}{R-2 k ^2}\right)\theta(k^2-{\bar \Delta}_{L1})} 
   \nn&
-\Tr\sqbr{4 k ^2 \left(\frac{3-\beta }{(\beta -3) k
   ^2+R}+\frac{4}{-4 k ^2-\rho _2 R+R}\right)\theta(k^2-{\bar \Delta}_{L0})}   
\nn
&=
{1\over(4\pi)^2}
\int_x\sqbr{A_0^{C}k^4+A_2^{C}\phi^2k^2+A_4^{C}\phi^4+B_0^{C}\,Rk^2+B_2^{C}\,R\phi^2+C^{C} R^2+D^{C} R_{\mu\nu\rho\sigma}R^{\mu\nu\rho\sigma}+E^{C} Y},
}		
}
where
{\footnotesize
\al{
&
A_0^{C}=8
,\quad
A_2^{C}=0
,\quad
A_4^{C}=0
,\nn
&
B_0^{C}=\frac{1}{3-\beta }-\frac{\rho _1}{4 \left(\rho _1-\rho
   _2+1\right)}-\frac{3 \rho _2}{4}+\frac{7}{6}
,\quad
B_2^{C}=0
,\nn
&C^{C}=
   \frac{1}{240} \left(\frac{8 \br{\beta  (11 \beta -76)+159}}{(\beta-3)^2}
   +\frac{15 \rho _1^2}{\left(\rho _1-\rho_2+1\right){}^2}
   -\frac{50 \rho _1}{\rho _1-\rho _2+1}
   +45\rho _2^2-30 \rho _2\right)      
,\nn
&
D^{C}=  -{11\over45}
,\quad
E^{C}=0.  
\label{Eq:C contribution}
}
}
					
\subsection{Mixed diagrams}
Finally, we consider the third and fourth term in Eq.~\eqref{FRG_totalexpand}.
These functional trace corresponds to the contribution of the diagrams VII--XIV in Fig.~\ref{feynman diagrams of yukawa vertex}:
\al{\label{Eq:Yukawa}
-{1\over2}{\rm Tr}
	\sqbr{
	\mathcal{M}_{FF}^{-1}
	\paren{\p_k \mathcal R_k^{(FF)}}
	\mathcal{M}_{FF}^{-1}\mathcal{M}_{FB}\mathcal{M}_{BB}^{-1}\mathcal{M}_{BF}
	}
-{1\over2}{\rm Tr}
	\sqbr{
	\mathcal{M}_{FF}^{-1}\mathcal{M}_{FB}\mathcal{M}_{BB}^{-1}
	\paren{\p_k \mathcal R_k^{(BB)}}\mathcal{M}_{BB}^{-1}\mathcal{M}_{BF}
	}
	\Bigg|_{R=0}	.
}
Since $\mathcal{M}_{FB}$ and $\mathcal{M}_{BF}$ contain one background fermion, only the Yukawa coupling is corrected by Eq.~\eqref{Eq:Yukawa}.
Hence we can safely put $R=0$ in Eq.~\eqref{Eq:Yukawa}.
We expand the matrices by the power of $\phi$,
\al{
\mathcal{M}_{FF}^{-1}\bigg|_{R=0}&=\mathcal{M}_{FF(0)}^{-1}+\mathcal{M}_{FF(1)}^{-1}\phi+...,
\nn
\mathcal{M}_{BB}^{-1}\bigg|_{R=0}&=\mathcal{M}_{BB(0)}^{-1}+\mathcal{M}_{BB(1)}^{-1}\phi+...,
\nn
\p_k \mathcal R_k^{(BB)}\bigg|_{R=0}&=\p_k \mathcal R_{k(0)}^{(BB)}+\p_k \mathcal R_{k(0)}^{(BB)}\phi+...,
\nn
\mathcal{M}_{BF}^{-1}\bigg|_{R=0}&=\mathcal{M}_{BF(0)}^{-1}+\mathcal{M}_{BF(1)}^{-1}\phi+...,
\nn
\mathcal{M}_{FB}^{-1}\bigg|_{R=0}&=\mathcal{M}_{FB(0)}^{-1}+\mathcal{M}_{FB(1)}^{-1}\phi+...,
}
where $...$ represents $\mathcal{O}(\phi^2)$, which does not contribute to the truncated effective action. 
In the following, we show the explicit formula employing Landau gauge.

The contribution from diagrams VII--X is
{\footnotesize
\al{\label{Eq:diagram6-9}
&-{1\over2}{\rm Tr}
	\sqbr{
	\mathcal{M}_{FF(1)}^{-1}
	\paren{\p_k \mathcal R_k^{(FF)}}
	\mathcal{M}_{FF(0)}^{-1}\mathcal{M}_{FB(0)}\mathcal{M}_{BB(0)}^{-1}\mathcal{M}_{BF(0)}
	+
	\mathcal{M}_{FF(0)}^{-1}
	\paren{\p_k \mathcal R_k^{(FF)}}
	\mathcal{M}_{FF(1)}^{-1}\mathcal{M}_{FB(0)}\mathcal{M}_{BB(0)}^{-1}\mathcal{M}_{BF(0)}
	}\phi	
	\nn
&-{1\over2}{\rm Tr}
	\sqbr{
	\mathcal{M}_{FF(1)}^{-1}\mathcal{M}_{FB(0)}\mathcal{M}_{BB(0)}^{-1}
	\paren{\p_k \mathcal R_{k(0)}^{(BB)}}\mathcal{M}_{BB(0)}^{-1}\mathcal{M}_{BF(0)}
	}\phi
\nn
&\to
{Y\over(4\pi)^2}
\frac{1}{320} k ^2 
\left(
\frac{k ^2 
\bigg\{
-80 F k^2+k ^3 
\sqbr{16 k  \br{k  \left(b '+3 a'\right)+15 (b +3 a )}-5 \partial _{k}F}
+40 V
\bigg\}
   }
   {\br{-F k ^2+2 k ^4 (b +3a )+V}^2}
   -\frac{640 y^2 \left(k^2+V'\right)}{\left(k ^2+2 V'\right)^2}
\right)\Bigg|_{\phi=0}.
}
}

The contribution from diagram XI and XII is
{\footnotesize
\al{\label{Eq:diagram10 and 11}
&-{1\over2}{\rm Tr}
	\sqbr{
	\mathcal{M}_{FF(0)}^{-1}
	\paren{\p_k \mathcal R_k^{(FF)}}
	\mathcal{M}_{FF(0)}^{-1}\mathcal{M}_{FB(1)}\mathcal{M}_{BB(0)}^{-1}\mathcal{M}_{BF(0)}
	+
	\mathcal{M}_{FF(0)}^{-1}
	\paren{\p_k \mathcal R_k^{(FF)}}
	\mathcal{M}_{FF(0)}^{-1}\mathcal{M}_{FB(1)}\mathcal{M}_{BB(0)}^{-1}\mathcal{M}_{BF(1)}
	}\phi
	\nn
&-{1\over2}{\rm Tr}
	\sqbr{
	\mathcal{M}_{FF(0)}^{-1}\mathcal{M}_{FB(1)}\mathcal{M}_{BB(0)}^{-1}
	\paren{\p_k \mathcal R_{k(0)}^{(BB)}}\mathcal{M}_{BB(0)}^{-1}\mathcal{M}_{BF(0)}
	+
	\mathcal{M}_{FF(0)}^{-1}\mathcal{M}_{FB(0)}\mathcal{M}_{BB(0)}^{-1}
	\paren{\p_k \mathcal R_{k(0)}^{(BB)}}\mathcal{M}_{BB(0)}^{-1}\mathcal{M}_{BF(1)}
}\phi
\nn
&\to
{Y\over(4\pi)^2}
\frac{189 F k ^6+18 k ^7 \partial _{k }F-7 k ^4
   \br{8 k ^5 \left(b '+3 a '\right)+90 k ^4(b +3 a )+9 V}}
   {105 \br{-F k ^2+2 k ^4(b +3 a )+V}^2}
   \Bigg|_{\phi=0}.
}
}

Finally, the contribution from diagram XIII and XIV is
{\footnotesize
\al{\label{Eq:diagram12 and 13}
&-{1\over2}{\rm Tr}
	\sqbr{
	\mathcal{M}_{FF(0)}^{-1}
	\paren{\p_k \mathcal R_k^{(FF)}}
	\mathcal{M}_{FF(0)}^{-1}\mathcal{M}_{FB(0)}\mathcal{M}_{BB(1)}^{-1}\mathcal{M}_{BF(0)}
	+
	\mathcal{M}_{FF(0)}^{-1}\mathcal{M}_{FB(0)}\mathcal{M}_{BB(0)}^{-1}
	\paren{\p_k \mathcal R_{k(1)}^{(BB)}}\mathcal{M}_{BB(0)}^{-1}\mathcal{M}_{BF(0)}
	}\phi
	\nn
&-{1\over2}{\rm Tr}
	\sqbr{
	\mathcal{M}_{FF(0)}^{-1}\mathcal{M}_{FB(0)}\mathcal{M}_{BB(1)}^{-1}
	\paren{\p_k \mathcal R_{k(0)}^{(BB)}}\mathcal{M}_{BB(0)}^{-1}\mathcal{M}_{BF(0)}
	+
	\mathcal{M}_{FF(0)}^{-1}\mathcal{M}_{FB(0)}\mathcal{M}_{BB(0)}^{-1}
	\paren{\p_k \mathcal R_{k(0)}^{(BB)}}\mathcal{M}_{BB(1)}^{-1}\mathcal{M}_{BF(1)}
}\phi
\nn
&\to
{Y\over(4\pi)^2}
\frac{2 k ^4 }{105\left(k ^2+2 V'\right)^2 
\br{-F k ^2+2 k ^4(b +3 a )+V}^2}\times
\nn&\bigg[
-56 k ^9 F' \left(b '+3 a'\right)
+18 k ^7 \partial _{k }F F'
+36 k ^5 \partial_{k }F F' V'
+18 k ^4 (b +3 a) 
		\bigg\{
-35 k^4 F'+7 k ^2 \left(7-6 F'\right) V'
   \nn&
+2 k ^3 \partial_{k }F' \left(k ^2+2 V'\right)+70V'^2
   		\bigg\}
   +9 F k^2 
   \bigg\{
   7 \br{3 k ^4F'+k ^2 \left(2 F'-5\right) V'-6 V'^2}
   -2k ^3 \partial _{k }F' 
   \left(k ^2+2V'\right)
   \bigg\}
   \nn&
   -112 k ^7 F' V' \left(b '+3 a
   '\right)+126 k ^2 V F' V'-63 k ^4 V F'-18 k ^5
   \partial _{k }F V'-36 k ^3 \partial _{k }F
   V'^2+18 k ^3 \partial _{k }F' V
   \left(k ^2+2 V'\right)
   \nn&
   +56 k ^7 V' \left(b '+3 a
   '\right)+112 k ^5 V'^2 \left(b '+3 a
   '\right)+189 k ^2 V V'+126 V V'^2\bigg]
\Bigg|_{\phi=0}.
}
}

\section{Fixed point and critical exponent}\label{FPCX}
Here, we list fixed points and critical exponents.
\begin{table}[H]
  \begin{center}
    \begin{tabular}{|c||c|c|c|c|c|} \hline
Truncation & $\tilde \xi_0^*\times 10^{2}$ & $\tilde \lambda_0^*\times 10^{3}$ & $\tilde a^*\times 10^{2}$ & $\tilde b^*\times 10^{2}$ & ${\tilde \lambda}_0^*/ {\tilde \xi}_0^*{}^2$  \\ \hline \hline
EH & $2.502$ & $4.450$ & ---  & --- & $7.11$ \\
EH $+R^2$ (i) & $2.123$ & $2.733$ & $0.2438$  & --- & $6.06$ \\
EH $+R^2$ (ii) & $2.195$ & $6.435$ & $0.4389$  & --- & $13.4$ \\
EH $+R^2$ (iii) & $9.331$ & $87.05$ & $0.5173$  & --- & $10.0$ \\
EH $+R^2+R_{\mu\nu}^2$ (i) & $0.8250$ & $3.546$ & $0.3564$ & $-0.8080$ & $52.1$ \\
EH $+R^2+R_{\mu\nu}^2$ (ii) & $166.3$ & $2491$ & $-21.28$ & $63.93$ & $0.90$ \\
EH $+R^2+R_{\mu\nu}^2$ (iii) & $2.701$ & $-9.402$ & $2.951$ & $-2.731$ & $-12.9$ \\
EH--scalar & $2.571$ & $5.611$ & ---  & --- & $8.49$ \\
EH--HY & $2.041$ & $1.147$ & ---  & --- & $2.75$ \\ 
Full (i)  & $0.3996$ & $1.537$ & $0.1737$ & $-0.3913$ & $96.3$ \\ 
Full (ii)  & $165.9$ & $2486$ & $-21.23$ & $63.77$ & $0.90$ \\
Full (iii)  & $2.112$ & $-9.91$ & $2.521$ & $-2.154$ & $-22.2$ \\ \hline
    \end{tabular}
  \end{center}
\caption{The values of fixed points for $\alpha=0$ and $\beta=0$.}
\label{FPbeta0}
\end{table}

\begin{table}[H]
  \begin{center}
    \scalebox{0.9}[0.9]{
    \begin{tabular}{|c||c|c|c|c|c|c|c|c|} \hline
      Truncation & $\theta_1$ & $\theta_2$  & $\theta_3$ & $\theta_4$ & $\theta_5$  & $\theta_6$ & $\theta_7$ & $\theta_8$ \\ \hline \hline
EH & $2.79+1.45i$ & $\theta_1^*$ & ---  & --- &  ---   &  ---   & ---  &  ---   \\
EH $+R^2$ (i)  & $2.58+1.39i$ & $\theta_1^*$ & $19.0$ & --- & --- & --- & --- & --- \\
EH $+R^2$ (ii)  & $2.54+1.62i$ & $\theta_1^*$ & $-38.6$ & --- & --- & --- & --- & --- \\
EH $+R^2$ (iii)  & $12.6$ &  $-180+28.6i$ & $\theta_2^*$ & --- & --- & --- & --- & --- \\
EH $+R^2+R_{\mu\nu}^2$ (i) & $3.41+0.772i$  & $\theta_1^*$ & $9.00+2.47i$ & $\theta_3^*$ &  --- & --- & --- & ---\\
EH $+R^2+R_{\mu\nu}^2$ (ii) & $22.6$  & $-0.644$ & $-3.22\times10^3$ & $-2.56\times10^4$  &  --- & --- & --- & ---\\
EH $+R^2+R_{\mu\nu}^2$ (iii) & $3.61+ 4.75i$  & $\theta_1^*$ & $2.04$ & $42.5$ &  --- & --- & --- & ---\\
EH--scalar & $2.83+1.57i$ & $\theta_1^*$ & --- & --- & $0.829+1.57i$ & $\theta_5^*$ & $-1.49$ & ---\\
EH--HY & $2.67+1.28i$ & $\theta_1^*$ & --- & --- & $0.665+1.23i$ & $\theta_5^*$ & $-1.31$ & $-0.929$ \\
Full (i) & $4.02$ & $8.97$ & $-5.41+14.6i$ & $\theta_3^*$ & $47.5+82.0i$ & $\theta_5^*$  & $16.7$ & $20.4$ \\ 
Full (ii) & $22.6$  & $-0.644$ & $-3220$ & $-2.55\times10^4$ & $34.3$ & $1730$ & $-1.00\times10^4$ & $9650$ \\
Full (iii) & $4.07 + 3.86i$  & $\theta_1^*$ & $1.95$ & $33.1$ & $13.4$ & $0.474$ & $2.68$ & $2.54$ \\ \hline
\end{tabular}
}
\end{center}
\caption{The values of critical exponents for $\alpha=0$ and $\beta=0$.}
\label{criticalbeta0}
\end{table}

\begin{table}[H]
  \begin{center}
    \begin{tabular}{|c||c|c|c|c|c|} \hline
Truncation & $\tilde \xi_0^*\times 10^{2}$ & $\tilde \lambda_0^*\times 10^{3}$ & $\tilde a^*\times 10^{2}$ & $\tilde b^*\times 10^{2}$ & ${\tilde \lambda}_0^*/ {\tilde \xi}_0^*{}^2$  \\ \hline \hline
EH (i) & $2.466$ & $4.343$ & ---  & --- & $7.14$ \\
EH (ii) & $7.045\times10^{-2}$ & $2.434$ & ---  & --- & $4900$ \\
EH $+R^2$ (i) & $1.772$ & $0.4947$ & $0.2477$  & --- & $1.58$ \\
EH $+R^2$ (ii) & $2.298$ & $11.26$ & $0.3873$  & --- & $21.3$ \\
EH $+R^2$ (iii) & $2.456$ & $15.22$ & $0.3854$  & --- & $25.2$ \\
EH $+R^2$ (iv) & $0.05807$ & $1.803$ & $1.104\times10^{-3}$  & --- & $5350$ \\
EH $+R^2+R_{\mu\nu}^2$ (i) & $0.4096$ & $1.590$ & $0.1892$ & $-0.4040$ & $94.8$ \\
EH $+R^2+R_{\mu\nu}^2$ (ii) & $0.01033$ & $0.9478$ & $-2.143\times10^{-3}$ & $-6.509\times10^{-3}$ & $8.89\times10^4$ \\
EH $+R^2+R_{\mu\nu}^2$  (iii) & $1.512$ & $20.17$ & $-0.2749$ & $1.329$ & $88.2$ \\
EH $+R^2+R_{\mu\nu}^2$  (iv) & $2.723$ & $-9.477$ & $2.971$ & $-2.753$ & $-12.8$ \\
EH--scalar (i) & $2.535$ & $5.471$ & ---  & --- & $8.51$ \\
EH--scalar (ii) & $0.1011$ & $3.512$ & ---  & --- & $3440$ \\
EH--HY (i) & $2.003$ & $1.119$ & ---  & --- & $2.79$ \\ 
EH--HY (ii) & $0.01149$ & $0.3996$ & ---  & --- & $3.03\times10^4$ \\ 
EH--HY (iii) & $0.01310$ & $-5.320$ & ---  & --- & $-3.10\times10^5$ \\ 
Full (i)  & $1.339$ & $17.86$ & $-0.2423$ & $1.174$ & $99.6$ \\
Full (ii)  & $2.135$ & $-10.00$ & $2.543$ & $-2.177$ & $-22.0$ \\ \hline
    \end{tabular}
  \end{center}
\caption{The values of fixed points for $\alpha=0$ and $\beta=-1$.}
\label{FPbeta-1}
\end{table}

\begin{table}[H]
  \begin{center}
    \begin{tabular}{|c||c|c|c|c|c|c|c|c|} \hline
      Truncation & $\theta_1$ & $\theta_2$  & $\theta_3$ & $\theta_4$ & $\theta_5$  & $\theta_6$ & $\theta_7$ & $\theta_8$ \\ \hline \hline
EH (i) & $2.84+1.47i$ & $\theta_1^*$ & ---  & --- &  ---   &  ---   & ---  &  ---   \\
EH (ii) & $-4.20$ & $-93.4$ & ---  & --- &  ---   &  ---   & ---  &  ---   \\
EH $+R^2$ (i)  & $2.43+1.07i$ & $\theta_1^*$ & $38.7$ & --- & --- & --- & --- & --- \\
EH $+R^2$ (ii)  & $1.71+0.523i$ & $\theta_1^*$ & $-41.4$ & --- & --- & --- & --- & --- \\
EH $+R^2$ (iii)  &  $3.23$ & $-1.66$ &  $-48.0$ & --- & --- & --- & --- & --- \\
EH $+R^2$ (iv)  &  $39.1$ & $155$ & $-2.02$ & --- & --- & --- & --- & --- \\
EH $+R^2+R_{\mu\nu}^2$ (i) & $3.25$  & $8.58$ & $17.0+15.6i$ & $\theta_3^*$ &  --- & --- & --- & ---\\
EH $+R^2+R_{\mu\nu}^2$ (ii)  & $4.05$ & $15.8$ & $-28.0$  & $-1.13$ &  --- & --- & --- & ---\\
EH $+R^2+R_{\mu\nu}^2$ (iii) & $7.99+ 0.515i$ & $\theta_1^*$ & $59.6$ & $3380$ &  --- & --- & --- & ---\\
EH $+R^2+R_{\mu\nu}^2$ (iv) & $3.58+ 4.79i$  & $\theta_1^*$ & $2.04$ & $41.4$ &  --- & --- & --- & ---\\
EH--scalar (i) & $2.88+1.59i$ & $\theta_1^*$ & --- & --- & $0.880+1.59i$ & $\theta_5^*$ & $-1.40$ & ---\\
EH--scalar (ii) & $-20.1+3.76i$ & $\theta_1^*$ & --- & --- & $-22.1+3.76i$ & $\theta_5^*$ & $-63.7$ & ---\\
EH--HY (i) & $2.70+1.32i$ & $\theta_1^*$ & --- & --- & $0.697+1.32i$ & $\theta_5^*$ & $-1.24$ & $-0.980$ \\
EH--HY (ii) & $-0.374$ & $-1680$ & --- & --- & $-2.37$ & $-1680$ & $-567$ & $1490$ \\
EH--HY (iii) & $3.95$ & $-72.8$ & --- & --- & $1.95$ & $-74.8$ & $-0.158$ & $0.840$ \\
Full (i)  & $8.18$ & $10.3$ & $50.3$ & $2970$ & $-11.0$ & $0.901$ & $126$ & $-3.85\times10^4$ \\
Full (ii)  & $4.06+ 3.91i$ & $\theta_1^*$ & $1.95$ & $32.1$ & $13.2$ & $0.458$ & $2.62$ & $2.53$ \\ \hline
\end{tabular}
\end{center}
\caption{The values of critical exponents for $\alpha=0$ and $\beta=-1$.}
\label{criticalbeta-1}
\end{table}


\begin{table}[H]
  \begin{center}
    \begin{tabular}{|c||c|c|c|c|c|} \hline
Truncation & $\tilde \xi_0^*\times 10^{2}$ & $\tilde \lambda_0^*\times 10^{3}$ & $\tilde a^*\times 10^{2}$ & $\tilde b^*\times 10^{2}$ & ${\tilde \lambda}_0^*/ {\tilde \xi}_0^*{}^2$  \\ \hline \hline
EH & $2.2485$ & $3.896$ & ---  & --- & $7.71$ \\
EH (ii) & {$0.1964$} & {$-1.738$} & ---  & --- & {$-451$} \\
EH $+R^2$ (i) & $2.403$ & $2.158$ & $0.3143$  & --- & $3.74$ \\
EH $+R^2$ (ii) & $16.32$ & $16.36$ & $3.455$  & --- & ${0.614}$ \\
EH $+R^2+R_{\mu\nu}^2$ (i) & $0.3445$ & $2.437$ & $0.6195$ & $-0.3366$ & $205$ \\
EH $+R^2+R_{\mu\nu}^2$ (ii) & $1.086$ & $3.623$ & $0.5547$ & $-1.077$ & $30.7$ \\
EH $+R^2+R_{\mu\nu}^2$ (iii) & $39.78$ & $51.12$ & $21.67$ & $-39.42$ & $0.323$ \\
EH $+R^2+R_{\mu\nu}^2$ (iv) & $0.416$ & $-4.79$ & $0.4104$ & ${-0.4302}$ & $-276.6$ \\
EH $+R^2+R_{\mu\nu}^2$ (v) & ${1.20}$  & $-2.01$ & $0.343$ & $-0.632$ & $-14.1$ \\
EH $+R^2+R_{\mu\nu}^2$ (vi) & $2.65$  & $-9.54$ & $2.85$ & $-2.68$ & $-13.6$ \\
EH--scalar & $2.576$ & ${4.909}$ & ---  & --- & ${7.40}$ \\
EH--scalar (ii) & ${0.1088}$ & ${-0.9569}$ & ---  & --- & ${-809}$ \\
EH--HY & $1.870$ & $0.9842$ & ---  & --- & $2.81$ \\ 
EH--HY (ii) & ${0.4182}$ & ${-4.000}$ & ---  & --- & ${-2.29\times10^2}$ \\
Full (i) & $0.6101$ & $1.777$ & $0.3184$ & $-0.6037$ & ${47.8}$ \\ 
Full (ii) & $37.57$ & $48.12$ & $20.46$ & $-37.19$ & $0.341$  \\ 
Full (iii) & ${1.080\times10^7}$ & $1.801\times10^7$ & $5.402\times10^6$ & $-9.003\times10^6$ & ${1.54\times10^{-6}}$  \\ 
Full (iv) & ${0.423}$ & $-6.68$ & $0.482$ & ${-0.45}$ & $-372.8$ \\ 
Full (v) & $1.64$ & $-2.75$ & $0.346$ & $-0.493$ & $-10.3$ \\ 
Full (vi) & ${2.05}$ & $-10.1$ & $2.38$ & $-2.09$ & $-24.1$ \\ 
\hline
    \end{tabular}
  \end{center}
\caption{The values of fixed points for $\alpha=0$ and $\beta=2$.}
\label{FPbeta2}
\end{table}

\begin{table}[H]
  \begin{center}
  \scalebox{0.9}[0.9]{
    \begin{tabular}{|c||c|c|c|c|c|c|c|c|} \hline
      Truncation & $\theta_1$ & $\theta_2$  & $\theta_3$ & $\theta_4$ & $\theta_5$  & $\theta_6$ & $\theta_7$ & $\theta_8$ \\ \hline \hline
EH & ${2.94+1.67i}$ & $\theta_1^*$ & ---  & --- &  ---   &  ---   & ---  &  ---   \\
EH (ii) & ${2.70}$ &${63.5}$ & ---  & --- &  ---   &  ---   & ---  &  ---   \\
EH $+R^2$ (i)  & $2.65+0.800i$ & $\theta_1^*$ & $9.74$ & --- & --- & --- & --- & --- \\
EH $+R^2$ (ii)  & $0.881+4.50i$ & $\theta_1^*$ & $821$ & --- & --- & --- & --- & --- \\
EH $+R^2+R_{\mu\nu}^2$ (i) & $4.35$ & $5.75$ & $9.35$ & $-2.70$ &  --- & --- & --- & ---\\
EH $+R^2+R_{\mu\nu}^2$ (ii) & ${2.96+0.52i}$ & $\theta_3^*$ & $6.84$  & $18.8$  &  --- & --- & --- & ---\\
EH $+R^2+R_{\mu\nu}^2$ (iii) & ${59.3}$ & ${712}$ &  ${-0.623}$ &  ${-3.78}$  & --- & --- & --- & ---\\
EH $+R^2+R_{\mu\nu}^2$ (iv) & $2.15$  & $4.26$ & $6.75$ & $14.7$ &  --- & --- & --- & ---\\
EH $+R^2+R_{\mu\nu}^2$ (v) & $2.61$  & ${5.48}$ & $8.03$ & ${-1720}$ &  --- & --- & --- & ---\\
EH $+R^2+R_{\mu\nu}^2$ (vi) & $3.53 + 4.68i$  & $\theta_1^*$ & $2.04$ & $33.6$ &  --- & --- & --- & ---\\
EH--scalar & $3.00+1.71i$ & $\theta_1^*$ & --- & --- & $0.996+1.71i$ & $\theta_5^*$ & $-1.08$ & ---\\
EH--scalar (ii) & ${2.18}$ &${108}$ & ---  & --- &  ${0.175}$   &  ${106}$   & ${7.96}$  &  ---   \\
EH--HY & $2.58+2.01i$ & $\theta_1^*$ & --- & --- & $0.577+2.01i$ & $\theta_5^*$ & $-0.888$ & $0.253$ \\
EH--HY (ii) & ${3.25}$ &${25.4}$ & ---  & --- &  ${1.25}$   &  ${23.4}$   & ${1.10}$  &  ${4.64}$   \\
Full (i) & $5.15+1.08i$ & $\theta_1^*$ & $14.3+7.16i$ & $\theta_5^*$ & $-1.05$ & $41.0$  & $15.8$ & $5.50$ \\ 
Full (ii) & $65.8$ & $610$ & $-0.566$ & $-4.27$  &  $-129$ & $2.07$  & $-20.2$ & $-73.1$ \\ 
Full (iii) & $-2.83\times10^8$ & $-2.59\times10^7$ & $-1.17\times10^5$ & $0.00585$ & $4.41\times10^5$ & $5.20\times10^7$  & $-2.22\times10^7$ & $-7.44\times10^7$ \\ 
Full (iv) & $2.02$ & $4.34$ & ${6.45}$ & $30.8$ & ${-19.0}$ & $3.16$  & $1.03$ & $0.213$ \\ 
Full (v) & $2.72$ & $3.53$ & $8.02$ & $-5853$ & $11.7$ & $1.11$  & $134.7$ & ${-10000}$ \\ 
Full (vi) & $4.03 + 3.74i$ & $\theta_1^*$ & ${1.94}$ & $24.1$ & $11.5$ & $0.316$  & $2.34$ & $2.02$ \\ 
\hline
\end{tabular}
}
\end{center}
\caption{The values of critical exponents for $\alpha=0$ and $\beta=2$.}
\label{criticalbeta2}
\end{table}

\newpage
\begin{table}[H]
  \begin{center}
    \begin{tabular}{|c||c|c|c|c|c|} \hline
Truncation & $\tilde \xi_0^*\times 10^{2}$ & $\tilde \lambda_0^*\times 10^{3}$ & $\tilde a^*\times 10^{2}$ & $\tilde b^*\times 10^{2}$ & ${\tilde \lambda}_0^*/ {\tilde \xi}_0^*{}^2$  \\ \hline \hline
EH & $2.441$ & $4.271$ & ---  & --- & $7.17$ \\
EH (ii) & {$1.837\times10^{-2}$} & {$-4.211$} & ---  & --- & {$-1.25\times10^5$} \\
EH (iii) & {$1.328\times10^{-2}$} & {$-0.5620$} & ---  & --- & {$-3.19\times10^4$} \\
EH $+R^2$ (i) & $45.94$ & $87.02$ & $7.825$  & --- & $0.412$ \\
EH $+R^2$ (ii)& $0.0105$ & $-0.898$ & $-0.00681$  & --- & $-81256$ \\
EH $+R^2$ (iii)& $2.19$ & $-0.834$ & $0.321$  & --- & $-1.73$ \\
EH $+R^2+R_{\mu\nu}^2$ (i) & $52.00$ & $107.1$ & $26.03$ & $-51.52$ & $0.396$ \\
EH $+R^2+R_{\mu\nu}^2$ (ii) & ${0.0289}$ & ${-2.29}$ & $0.0352$ & $-0.0359$ & $-27406$ \\ 
EH $+R^2+R_{\mu\nu}^2$ (iii) & $0.0647$ & ${-4.44}$ & $0.030$ & $-0.0706$ & $-10604$ \\ 
EH $+R^2+R_{\mu\nu}^2$ (iv) & ${0.0906}$ & $-4.41$ & $0.0367$ & $-0.0809$ & $-5376.8$ \\ 
EH $+R^2+R_{\mu\nu}^2$ (v) & $2.74$ & $-9.54$ & $2.99$ & $-2.77$ & $-12.69$ \\ 
EH--scalar & $2.510$ & $5.380$ & ---  & --- & $8.54$ \\
EH--scalar (ii) & ${1.418\times10^{-2}}$ & ${-3.348}$ & ---  & --- & ${-1.66\times10^5}$ \\
EH--HY & $1.974$ & $1.098$ & ---  & --- & $2.82$ \\ 
EH--HY (ii) & ${2.426\times10^{-2}}$ & ${-6.320}$ & ---  & --- & ${-1.07\times10^5}$ \\ 
EH--HY (iii) & ${8.884\times10^{-2}}$ & ${-2.870}$ & ---  & --- & ${-3.64\times10^3}$ \\ 
Full (i) & $50.43$ & $103.4$ & $25.22$ & $-49.87$ & $0.47$ \\ 
Full (ii) & ${0.0245}$ & $-6.25$ & $0.0171$ & $-0.0465$ & $-104180$ \\ 
Full (iii) & $0.142$ & $-4.74$ & $0.0778$ & $-0.160$ & $-2348$ \\ 
Full (iv) & $0.142$ & $-6.30$ & $0.0428$ & $-0.0818$ & $-3119$ \\ 
Full (v) & $0.231$ & $-3.81$ & $0.121$ & $-0.245$ & $-717.0$ \\ 
Full (vi) & $2.15$ & $-10.1$ & $2.56$ & $-2.20$ & $-21.73$ \\ 
\hline
    \end{tabular}
  \end{center}
\caption{The values of fixed points for $\alpha=0$ and $\beta=-2$.}
\label{FPbeta-2}
\end{table}

\begin{table}[H]
  \begin{center}
    \begin{tabular}{|c||c|c|c|c|c|c|c|c|} \hline
      Truncation & $\theta_1$ & $\theta_2$  & $\theta_3$ & $\theta_4$ & $\theta_5$  & $\theta_6$ & $\theta_7$ & $\theta_8$ \\ \hline \hline
EH & $2.87+1.49i$ & $\theta_1^*$ & ---  & --- &  ---   &  ---   & ---  &  ---   \\
EH (ii) & ${4.16}$ &${24.4}$ & ---  & --- &  ---   &  ---   & ---  &  ---   \\
EH (iii) & ${1.20}$ &${-1360}$ & ---  & --- &  ---   &  ---   & ---  &  ---   \\
EH $+R^2$ (i) & ${1.34+4.65i}$ & $\theta_1^*$ & $1572.5$ & --- & --- & --- & --- & --- \\
EH $+R^2$ (ii) & ${1.70}$ & $5.57$ & $-116.7$ & --- & --- & --- & --- & --- \\
EH $+R^2$ (iii) & $2.85 + 0.59i$ & $\theta_1^*$ & $19.0$ & --- & --- & --- & --- & --- \\
EH $+R^2+R_{\mu\nu}^2$ (i) & $1.10$  & $2.88$ & $6.84$ & $1290$ &  --- & --- & --- & ---\\
EH $+R^2+R_{\mu\nu}^2$ (ii) & $2.83 + 0.109i$ & $\theta_1^*$ & $5.084$ & $11.26$ &  --- & --- & --- & ---\\
EH $+R^2+R_{\mu\nu}^2$ (iii) & $2.23$ & $4.02$ & $6.93$ & $12.1$ &  --- & --- & --- & ---\\
EH $+R^2+R_{\mu\nu}^2$ (iv) & $3.87$ & $6.72$ & $10.2$ & $-9.57$ &  --- & --- & --- & ---\\
EH $+R^2+R_{\mu\nu}^2$ (v) & $3.56+ 4.83i$ & $\theta_1^*$ & $2.03$ & $40.8$ &  --- & --- & --- & ---\\
EH--scalar & $2.91+1.61i$ & $\theta_1^*$ & --- & --- & $0.909+1.61i$ & $\theta_5^*$ & $-1.35$ & ---\\
EH--scalar (ii) & ${4.19}$ &${23.6}$ & ---  & --- &  ${2.19}$   &  ${21.6}$   & ${2.77}$  &  ---   \\
EH--HY & ${2.72+1.36i}$ & $\theta_1^*$ & --- & --- & $0.717+1.36i$ & $\theta_5^*$ & $-1.18$ & $-0.999$ \\
EH--HY (ii) & ${4.05}$ &${18.6}$ & ---  & --- &  ${2.05}$   &  ${16.6}$   & ${1.07}$  &  ${8.54}$   \\
EH--HY (iii) & ${2.69}$ &${-94.1}$ & ---  & --- &  ${0.693}$   &  ${-96.1}$   & ${-0.636}$  &  ${6.44}$   \\
Full (i) & $1.71+0.175i$ & $\theta_1^*$ & ${7.84}$ & $1220$ & $-522$ & $0.171$ & $-9.79$ & $-106$ \\
Full (ii) & $4.48 + 2.83i$ & $\theta_1^*$ & $4.00$ & $6.57$ & $3.96$ & $0.576$ & $0.545$ & $0.802$ \\
Full (iii) & $0.507$ & $3.38$ & ${7.79}$ & $-276.4$ & $-203.4$ & $1.34$ & $-0.458$ & $-15.7$ \\
Full (iv) & $3.89$ & $6.19$ & $8.36$ & $-161.7$ & $13.0$ & $2.13$ & $15.0$ & $-192.1$ \\
Full (v) & $2.90$ & $7.15$ & $1838$ & $-0.820$ & $0.44$ & $-840$ & $-1.26$ & $-35.1$ \\
Full (vi) & $4.06 + 3.95i$ & $\theta_1^*$ & $1.95$ & $31.5$ & $13.07$ & $0.444$ & $2.58$ & $2.51$ \\
 \hline
\end{tabular}
\end{center}
\caption{The values of critical exponents for $\alpha=0$ and $\beta=-2$.}
\label{criticalbeta-2}
\end{table}

\end{appendix}

\bibliographystyle{TitleAndArxiv}
\bibliography{refs}
\end{document}